\pdfoutput=1

\documentclass[preprint,aps,nofootinbib,superscriptaddress,floatfix,preprintnumbers]{revtex4-1}

\usepackage{bm}

\usepackage[hyperfootnotes=false]{hyperref}
\usepackage{url}

\usepackage{color}

\usepackage{amsmath}
\usepackage{amssymb}
\usepackage{graphicx}

\def\lsim{\mathrel{\raise.3ex\hbox{$<$\kern-.75em\lower1ex\hbox{$\sim$}}}}
\def\gsim{\mathrel{\raise.3ex\hbox{$>$\kern-.75em\lower1ex\hbox{$\sim$}}}}

\newcommand{\bea}{\begin{eqnarray}}
\newcommand{\eea}{\end{eqnarray}}

\newcommand{\be}{\begin{equation}}
\newcommand{\ee}{\end{equation}}

\def\sw{s_W}
\def\cw{c_W}

\begin{document}

\title{On NMSSM Higgs Search Strategies at the LHC \\ and the Mono-Higgs Signature in Particular}

\newcommand{\OKC}{\affiliation{The Oskar Klein Centre for Cosmoparticle Physics, Department of Physics, Stockholm University, Alba Nova, 10691 Stockholm, Sweden}}
\newcommand{\Nordita}{\affiliation{Nordita, KTH Royal Institute of Technology and Stockholm University, Roslagstullsbacken 23, 10691 Stockholm, Sweden}}
\newcommand{\UoM}{\affiliation{Michigan Center for Theoretical Physics, Department of Physics, University of Michigan, Ann Arbor, MI 48109, USA.}}
\newcommand{\WS}{\affiliation{Department of Physics \& Astronomy, \\Wayne State University, Detroit, Michigan 48201, USA}}
\newcommand{\UCin}{\affiliation{Department of Physics, University of Cincinnati, Cincinnati, OH 45221, USA}}

\author{Sebastian Baum}
\email{sbaum@fysik.su.se}
\OKC
\Nordita

\author{Katherine Freese}
\email{ktfreese@umich.edu}
\OKC
\Nordita
\UoM

\author{Nausheen R. Shah}
\email{nausheen.shah@wayne.edu}
\WS

\author{Bibhushan Shakya}
\email{bshakya@umich.edu}
\UoM
\UCin

\preprint{NORDITA-2017-24}
\preprint{MCTP-17-03}
\preprint{WSU-HEP-1702}

\begin{abstract}

We study the collider phenomenology of the extended Higgs sector of the Next-to-Minimal Supersymmetric Standard Model (NMSSM). The region of NMSSM parameter space favored by a 125\,GeV SM-like Higgs and naturalness generically features a light Higgs and neutralino spectrum as well as a large $\mathcal{O}(1)$ coupling between the Higgs doublets and the NMSSM singlet fields. In such regimes, the heavier Higgs bosons can decay dominantly into lighter Higgs bosons and neutralinos. We study the prospects of observing such decays at the 13\,TeV LHC, focusing on mono-Higgs signatures as probes of such regions of parameter space. We present results for the mono-Higgs reach in a framework easily applicable to other models featuring similar decay topologies. In the NMSSM, we find that the mono-Higgs channel can probe TeV scale Higgs bosons and has sensitivity even in the low tan$\beta$, large $m_A$ regime that is difficult to probe in the MSSM. Unlike for many conventional Higgs searches, the reach of the mono-Higgs channel will improve significantly with the increased luminosity expected to be collected at the LHC in the ongoing and upcoming runs.

\end{abstract}

%\pacs{}
%\keywords{}

\maketitle

\section{Introduction and Motivation}\label{sec:intro}

Weak scale supersymmetry (SUSY) as a solution to the hierarchy problem \cite{Witten:1981nf, Dimopoulos:1981zb, Witten:1981kv, Kaul:1981hi,Sakai:1981gr} has faced severe challenges from the observation of a Standard Model (SM) like 125\,GeV Higgs boson and the absence of signals of superpartners at the Large Hadron Collider (LHC). This is particularly serious in the Minimal Supersymmetric Standard Model (MSSM) (see e.g. Refs.~\cite{Nilles:1983ge, Martin:1997ns,Chung:2003fi} for reviews of the MSSM), where large radiative corrections are required to yield a 125\,GeV SM-like Higgs boson. In addition, the MSSM suffers from the so-called $\mu$-problem \cite{Kim:1983dt}, i.e. to generate proper electroweak symmetry breaking the dimensionful MSSM parameter $\mu$ that appears in the superpotential must be of the order of the electroweak scale rather than the expected cutoff scale of the theory (the GUT or Planck scale).

These problems can be alleviated in the Next-to-Minimal Supersymmetric Standard Model (NMSSM), see Refs.~\cite{Ellwanger:2009dp,Maniatis:2009re} for reviews, which augments the MSSM particle content with a chiral superfield $\widehat{S}$ uncharged under any of the SM gauge groups. In this paper, we consider the scale-invariant NMSSM, where all dimensionful parameters in the superpotential are set to zero, yielding an accidental $\mathbb{Z}_3$ symmetry under which all superfields transform by $e^{2\pi i/3}$. This singlet field leads to the following additional terms in the superpotential:  
\be
	W \supset \lambda \widehat{S} \widehat{H}_u \cdot \widehat{H}_d + \frac{\kappa}{3} \widehat{S}^3,
\ee
where $\widehat{H}_u, \widehat{H}_d$ are the up- and down-type Higgs doublets and $\lambda$ and $\kappa$ are dimensionless coefficients. The $\mu \widehat{H}_u \cdot \widehat{H}_d$ term of the MSSM is forbidden in the scale-invariant NMSSM; however, an effective $\mu$-term is generated when the scalar component of the field $\widehat{S}$ gets a vacuum expectation value (vev), $\mu=\lambda\langle S \rangle/\sqrt{2}$. If the vev of the singlet is induced by the breaking of supersymmetry, $\langle S \rangle$ is of the order of the supersymmetry breaking scale, thereby alleviating the $\mu$-problem for low-scale supersymmetry. 

Recall that in the MSSM, the tree-level mass term for the SM-like Higgs field is ${m_h^2 \approx m_Z^2 \cos^2 2\beta \lesssim (90\,{\rm GeV})^2}$. In the NMSSM, the $F$-term scalar potential leads to an additional tree-level mass term for the SM-like Higgs field proportional to $\lambda^2$,
\be
	m_h^2\approx  m_Z^2 \cos^2 2\beta + \frac{1}{2} \lambda^2 v^2 \sin^2 2\beta,
	\label{eq:higgsmass}
\ee
and hence the 125\,GeV Higgs mass can be obtained without significant fine-tuning (i.e. without large loop corrections from stops) for a sizable $\lambda\gsim 0.5$ and low values of $\tan\beta$. 

Even larger values of $\lambda\lsim2$ have been studied in the literature (also referred to as $\lambda$-SUSY); these are not perturbative up to the GUT scale but are nevertheless compatible with electroweak precision data for low values of $\tan\beta$ and can successfully incorporate a 125\,GeV SM-like Higgs \cite{Hall:2011aa,Agashe:2012zq,Gherghetta:2012gb,Farina:2013fsa,Gherghetta:2014xea, Cao:2008un, Cao:2014kya}. For such large values of $\lambda$, the sensitivity of the electroweak scale to the stop mass scale is reduced by a factor $\sim g^2/\lambda^2$ \cite{Barbieri:2006bg,Hall:2011aa,Perelstein:2012qg,Farina:2013fsa}, where $g \approx 0.5$. Given the current stringent bounds on stop masses from the LHC, such values of $\lambda$ are appealing because they allow for a higher scale of supersymmetry compatible with naturalness arguments. These considerations motivate the study of the NMSSM in the large singlet-doublet coupling regime $0.5 \lesssim \lambda \lesssim 2$. 

The scalar components of the additional NMSSM superfield $\widehat{S}$ give rise to a singlet scalar boson $H^S$ and a singlet pseudoscalar $A^S$, which mix with their corresponding Higgs-doublet counterparts. Likewise, the fermionic component of $\widehat{S}$ gives a neutralino, the singlino $\widetilde{S}$, which mixes with the other neutralinos, in particular the Higgsinos, whose masses are controlled by $\mu$. Therefore, both the Higgs and neutralino sectors in the NMSSM are larger than those of the MSSM, leading to significantly richer phenomenology. For some recent discussions of Higgs and neutralino phenomenology at the LHC in the NMSSM, see Refs.~\cite{Gherghetta:2012gb,Christensen:2013dra,Cheung:2014lqa,Dutta:2014hma,Ellwanger:2016sur} and references therein. It is worth pointing out here that the most interesting region of parameter space in the NMSSM lies at $\tan\beta\lsim 5$ (see Eq.\,\ref{eq:higgsmass} and subsequent discussion), which is a challenging region to probe at the LHC due to the heavy Higgs bosons decaying dominantly into $t\bar{t}$ \cite{Gori:2016zto, Carena:2016npr, Craig:2015jba}. The neutralino sector can also provide a viable dark matter candidate with interesting phenomenology (see e.g. Refs.~\cite{Perelstein:2012qg,Cheung:2014lqa,Ellwanger:2016sur}); the dark matter aspect of the NMSSM lies beyond the scope of this work.

In this paper, we aim to study the prospects of probing the Higgs sector of the NMSSM in the large singlet-doublet coupling regime $0.5 \lesssim \lambda \lesssim 2$. In Section \ref{sec:NMSSM} we review the electroweak sector of the NMSSM. We discuss the parameter regions that can accommodate a SM-like Higgs via alignment and show how significant interactions among the Higgs bosons and electroweakinos (charginos and neutralinos) arise from the term $\lambda \widehat{S} \widehat{H}_u \cdot \widehat{H}_d$. We present the details of our parameter scan in Section \ref{sec:scan} and constraints on our data set from direct Higgs searches at the LHC in Section \ref{sec:lhcconstraints}. Section \ref{sec:NMSSMsearch} contains a discussion of NMSSM specific LHC search strategies in the most interesting regions of parameter space. In Section \ref{sec:monoH} we focus on the mono-Higgs channel and present results of our collider simulation in a framework easily applicable to other models featuring similar decay topologies. These results are interpreted in the NMSSM framework in Section \ref{sec:NMSSMinterpretation}. We present our conclusions in Section \ref{sec:conclusions}. Tables of the trilinear Higgs couplings and figures for LHC constraints are presented in the Appendices. Our main results are contained in Figs.~\ref{fig:reach_Higgs bosons},~\ref{fig:reach_neutralinos} (model-independent framework) and Figs.~\ref{fig:detectable_ma-tb},~\ref{fig:detectable_ma-lam_lam-kap} (NMSSM parameter space).

\section{Electroweak Sector of the NMSSM}\label{sec:NMSSM}
We follow the notation of Refs.~\cite{Ellwanger:2009dp,Agashe:2012zq}. The superpotential of the $\mathbb{Z}_3$-invariant NMSSM reads
\be
	W \supset \lambda \widehat{S} \, \widehat{H}_u \cdot \widehat{H}_d + \frac{\kappa}{3} \widehat{S}^3,
\ee
where we employ the dot-product notation for SU(2) doublets,
\be
	\widehat{H}_u \cdot \widehat{H}_d = \epsilon_{ij} \widehat{H}_u^i \cdot \widehat{H}_d^j = \widehat{H}_u^+ \widehat{H}_d^- - \widehat{H}_u^0 \widehat{H}_d^0.
\ee
In the following, fields written without the hat represent the scalar component.
The soft supersymmetry-breaking terms involving only the Higgs scalar fields are
\be
	V_{\rm soft} = m_{H_u}^2 H_u^\dagger H_u + m_{H_d}^2 H_d^\dagger H_d + m_S^2 S^\dagger S + \left( \lambda A_\lambda H_u \cdot H_d \, S + \frac{\kappa}{3} A_\kappa S^3 + h.c. \right), \label{eq:Vsoft}
\ee
and the usual $F$- and $D$-terms contributing to the scalar potential are given by
\be \begin{split}
 	V_{\rm SUSY} = &\ \left|\lambda H_u \cdot H_d + \kappa S^2 \right|^2 + \lambda^2 S^\dagger S \left( H_u^\dagger H_u + H_d^\dagger H_d\right) 
 	\\ & + \frac{g_1^2 + g_2^2}{8}\left( H_u^\dagger H_u - H_d^\dagger H_d \right)^2 +  \frac{g_2^2}{2} \left| H_d^\dagger H_u \right|^2 . \label{eq:Vsusy}
\end{split} \ee
One obtains the physical Higgs fields by expanding $H_u$, $H_d$ and $S$ around their respective vevs $v_u$, $v_d$ and $s$, which can be obtained by minimizing the Higgs potential built from $V_{\rm SUSY}$, Eq.~\eqref{eq:Vsusy}, and $V_{\rm soft}$, Eq.~\eqref{eq:Vsoft}. 
Separating the complex scalar fields into real ($H_u^R, H_d^R, H^S$) and imaginary ($H_u^I, H_d^I, A^S$) components and choosing the vevs to lie along the neutral components of the Higgs fields\footnote{Here our notations differs from Ref.~\cite{Ellwanger:2009dp} where the vevs are defined without the $\sqrt{2}$ and there $v = \sqrt{v_u^2 + v_d^2} \simeq 174\,$GeV.}, 
\begin{equation}
	H_u^0 = \frac{v_u + H_u^R+iH_u^I}{\sqrt{2}}, \ H_d^0 = \frac{v_d + H_d^R+iH_d^I}{\sqrt{2}}, \ S = \frac{s+H^S+iA^S}{\sqrt{2}},
\end{equation}
one obtains three CP-even neutral Higgs bosons $H_u^R$, $H_d^R$ and $H^S$, two CP-odd neutral Higgs bosons\,\footnote{The superscript ``NSM'' stands for non-SM, to distinguish from the part of the doublet that is SM-like.} $A^{\rm NSM}$ (composed of $H_u^I$ and $H_d^I$) and $A^S$, and one charged Higgs $H^\pm$. The remaining degrees of freedom make up the longitudinal components of the $W$ and $Z$ bosons after electroweak symmetry breaking. Defining
\begin{equation}
	\tan\beta = \frac{v_u}{v_d},~~~~~	\mu = \lambda \left\langle S \right\rangle / \sqrt{2},
\end{equation}
and setting $v = \sqrt{v_u^2 + v_d^2} \simeq 246\,$GeV, the Higgs sector of the NMSSM contains six free parameters:
\be
	p_i = \{\lambda, \kappa, \tan\beta, \mu, A_\lambda, A_\kappa\}.
	\label{eq:parameters}
\ee
By definition, $\tan\beta$ is positive. Without loss of generality, one can choose $\lambda \geq 0$, while $\kappa$ and the dimensionful parameters $\mu$, $A_\lambda$ and $A_\kappa$ can have either sign.

It is useful to rotate the neutral CP-even Higgs bosons $\{H_u^R, H_d^R, H^S\}$ to the so-called Higgs basis \cite{Georgi:1978ri, Donoghue:1978cj, gunion2008higgs, Lavoura:1994fv, Botella:1994cs, Branco99, Gunion:2002zf, Carena:2015moc} $\{H^{\rm SM}, H^{\rm NSM}, H^S\}$, where the entire vev of the Higgs doublets lies along $H^{\rm SM}$. In the Higgs basis, the elements of the symmetric squared-mass matrix for the CP-even neutral Higgs bosons, including the leading one-loop stop corrections, are given by \cite{Carena:2015moc}\footnote{Note, that Ref.~\cite{Carena:2015moc} uses the parameter $\overline{M}_Z^2 \equiv m_Z^2 - \lambda^2 v^2/2.$}
\begin{eqnarray}
	\mathcal{M}_{S,11}^2 &=&  m_Z^2 c_{2\beta}^2 + \frac{1}{2} \lambda^2 v^2 s_{2\beta}^2 + \frac{3 v^2 s_\beta^4 h_t^4}{8 \pi^2} \left[ \ln\left(\frac{M_S^2}{m_t^2}\right) + \frac{X_t^2}{M_S^2}\left(1-\frac{X_t^2}{12 M_S^2}\right) \right] \label{eq:MS11}, 
	\\ \mathcal{M}_{S,22}^2 &=& M_A^2 + \left(m_Z^2 - \frac{1}{2}\lambda^2v^2\right)s_{2\beta}^2  + \frac{3 v^2 s_{2\beta}^2 h_t^4}{32 \pi^2} \left[ \ln\left(\frac{M_S^2}{m_t^2}\right) + \frac{X_t Y_t}{M_S^2}\left(1-\frac{X_t Y_t}{12 M_S^2}\right) \right] ,
	\\ \mathcal{M}_{S,33}^2 &=& \frac{1}{4} \lambda^2 v^2 s_{2\beta} \left(\frac{M_A^2}{2\mu^2}s_{2\beta}-\frac{\kappa}{\lambda}\right) + \frac{\kappa \mu}{\lambda}\left(A_\kappa + \frac{4 \kappa \mu}{\lambda}\right),
	\\ \mathcal{M}_{S,12}^2 &=& -\left(m_Z^2 - \frac{1}{2}\lambda^2 v^2\right)s_{2\beta} c_{2\beta} + \frac{3 v^2 s_\beta^2 s_{2\beta} h_t^4}{16 \pi^2} \left[ \ln\left(\frac{M_S^2}{m_t^2}\right) + \frac{X_t \left( X_t + Y_t \right)}{2 M_S^2} - \frac{X_t^3 Y_t}{12 M_S^4} \right] ,
	\\ \mathcal{M}_{S,13}^2 &=& \sqrt{2} \lambda v \mu \left(1-\frac{M_A^2}{4\mu^2}s_{2\beta}^2 - \frac{\kappa}{2\lambda} s_{2\beta}\right) ,
	\\ \mathcal{M}_{S,23}^2 &=& - \frac{1}{\sqrt{2}}\lambda v \mu c_{2\beta} \left(\frac{M_A^2}{2\mu^2}s_{2\beta} + \frac{\kappa}{\lambda}\right), \label{eq:MS23}
\end{eqnarray}
where $c_\beta \equiv \cos\beta, \ s_\beta \equiv \sin\beta$, $M_S$ is the geometric mean of the two stop mass eigenstates, $X_t = A_t - \mu \cot\beta$ and $Y_t = A_t + \mu \tan\beta$ parametrize the stop mixing, $h_t$ is the top Yukawa coupling, and we have introduced
\be
	M_A^2 \equiv \frac{\mu}{s_\beta c_\beta} \left(A_\lambda + \frac{\kappa \mu}{\lambda} \right). \label{eq:MA}
\ee

The tree-level squared-mass matrix for the CP-odd neutral Higgs bosons in the basis $\{A^{\rm NSM}, A^S\}$ is given by
\be
	\mathcal{M}_P^2 = \begin{pmatrix} M_A^2 & \frac{1}{\sqrt{2}} \lambda v \left(\frac{M_A^2}{2\mu}s_{2\beta} - \frac{3\kappa\mu}{\lambda}\right)
	\\ \frac{1}{\sqrt{2}} \lambda v \left(\frac{M_A^2}{2\mu}s_{2\beta} - \frac{3\kappa\mu}{\lambda}\right) \hspace{.5cm} & \frac{1}{2} \lambda^2 v^2 s_{2\beta} \left(\frac{M_A^2}{4\mu^2} s_{2\beta} + \frac{3\kappa}{2\lambda}\right) - \frac{3\kappa A_\kappa \mu}{\lambda} \end{pmatrix}. \label{eq:MP}
\ee
For completeness we record the mass of the charged Higgs,
\be
	m_{H^\pm}^2 = M_A^2 + m_W^2 - \frac{1}{2}\lambda^2 v^2.
\ee

In the basis $\{\widetilde{B}, \widetilde{W}^3, \widetilde{H}_d^0, \widetilde{H}_u^0, \widetilde{S}\}$, where $\widetilde{B}$ and $\widetilde{W}^3$ are the bino and the neutral wino respectively, $\widetilde{H}_d^0$ and $\widetilde{H}_u^0$ are the neutral Higgsinos belonging to the respective doublet superfields, and $\widetilde{S}$ is the singlino, the symmetric tree-level neutralino mass matrix reads
\be
	M_{\chi^0}= \begin{pmatrix} M_1 & 0 & -m_Z \sw c_\beta & m_Z \sw s_\beta & 0
	\\ & M_2 & m_Z \cw c_\beta & -m_Z \cw s_\beta & 0
	\\ & & 0 & -\mu & -\lambda v s_\beta
	\\ & & & 0 & -\lambda v c_\beta
	\\ & & & & 2\kappa \mu / \lambda \end{pmatrix}, \label{eq:neumassmatrix}
\ee
where $s_W \equiv \sin\theta_W,$ with $\theta_W$ the weak mixing angle.  In this paper, we decouple the gauginos from the collider phenomenology by taking $M_1, M_2 = 1\,$TeV.

\subsection{Higgs Couplings and Alignment} \label{sec:Hcouplings}
The couplings of the Higgs basis states to SM particles are given by
\bea
	H^{\rm NSM}({\rm down}, {\rm up}, {\rm V}) &=& \left( g_{\rm SM} \tan\beta, \ \frac{g_{\rm SM}}{\tan\beta}, \ 0 \right), \label{eq:Hcouplings1}
	\\ H^{\rm SM}({\rm down}, {\rm up}, {\rm V}) &=& \left( g_{\rm SM}, \ g_{\rm SM}, \ g_{\rm SM}\right), \label{eq:Hcouplings2}
	\\ H^S({\rm down}, {\rm up}, {\rm V}) &=& \left( 0,\ 0,\ 0 \right), \label{eq:Hcouplings3}
\eea
where ``down'' (``up'') stands for down-type (up-type) SM-fermions, ``V'' for vector bosons, and $g_{\rm SM}$ indicates the respective coupling of such particles to the SM Higgs. The couplings of the Higgs mass eigenstates $H_i$ and $A_i$ can be obtained from those of the Higgs basis eigenstates via
\bea
	H_i &=& S_{i1} H^{\rm SM} + S_{i2} H^{\rm NMS} + S_{i3} H^S, \label{eq:Hmassstates}
	\\ A_i &=& P_{i1} A^{\rm NMS} + P_{i2} A^S, \label{eq:Amassstates}
\eea
where the $S_{ij}$ and $P_{ij}$ are obtained by diagonalizing the respective mass matrices. Note that, for the Higgs sector, we denote interaction basis eigenstates with superscripts, e.g.\,$H^{\rm SM}$, while letters with subscripts or standalone letters denote mass eigenstates, e.g.\,$h_{\rm SM},H_i, A$. Likewise, an uppercase $M$ denotes quantities of mass dimension defined in terms of fundamental model parameters, while we use a lowercase $m$ for masses of physical particles; in particular, $m_A$ is the mass of the mostly doublet-like CP-odd neutral Higgs boson mass eigenstate $A$, while $M_A$ is the mass parameter defined in Eq.\,\eqref{eq:MA}.

By definition, $H^{\rm SM}$ has the same couplings to SM particles as the SM Higgs boson. Since the couplings of the 125\,GeV Higgs boson discovered at the LHC are bound to be within $\mathcal{O}(10\,\%)$ of the SM values \cite{Khachatryan:2016vau}, any NMSSM realization compatible with LHC bounds must have a Higgs mass eigenstate $h_{\rm SM}$ of mass $m_{h_{\rm SM}} \simeq 125\,$GeV approximately aligned with $H^{\rm SM}$. Recalling the CP-even mass matrix given in Eqs.~\eqref{eq:MS11}-\eqref{eq:MS23}, approximate alignment is realized when $\mathcal{M}^2_{S,12}$ and $\mathcal{M}^2_{S,13}$, parametrizing the mixing of $H^{\rm SM}$ with $H^{\rm NSM}$ and $H^{\rm S}$ respectively, are small compared to the diagonal entries. There are two ways to achieve this: either $\mathcal{M}^2_{S,22}$ and $\mathcal{M}^2_{S,33}$ can be large, i.e. $H^{\rm NMS}$ and $H^{\rm S}$ are heavy (the decoupling regime), or the NMSSM parameters conspire to cancel the $\mathcal{M}^2_{S,12}$ and $\mathcal{M}^2_{S,13}$ terms. The latter case is referred to as alignment without decoupling, see Ref.~\cite{Carena:2015moc} for an in-depth discussion; this case is particularly relevant for collider phenomenology since the additional NMSSM Higgs bosons can remain light and be accessible at the LHC.

Perfect alignment is achieved for $\mathcal{M}^2_{S,12}$ and $\mathcal{M}^2_{S,13}$ vanishing, yielding the following conditions on the NMSSM parameters \cite{Carena:2015moc}:
\begin{eqnarray}
	 \mathcal{M}^2_{S,12}=0 \implies && \lambda^2 = \frac{m_{h_{\rm SM}}^2 - m_Z^2 \cos(2\beta)}{v^2 \sin^2\beta}, \label{eq:align1}
	\\ 
	\mathcal{M}^2_{S,13}=0 \implies &&  \frac{M_A^2}{\mu^2} = \frac{4}{s_{2\beta}^2}\left( 1- \frac{\kappa}{2 \lambda} s_{2\beta} \right), \label{eq:align2} 
\end{eqnarray}
where the mass of the SM-like Higgs mass eigenstate is given by
\be
	m_{h_{\rm SM}}^2 = \mathcal{M}_{S,11}^2 = m_Z^2 c_{2\beta}^2 + \frac{1}{2} \lambda^2 v^2 s_{2\beta}^2 + \frac{3 v^2 s_\beta^4 h_t^4}{8 \pi^2} \left[ \ln\left(\frac{M_S^2}{m_t^2}\right) + \frac{X_t^2}{M_S^2}\left(1-\frac{X_t^2}{12 M_S^2}\right) \right]\;, \label{eq:mhsm}
\ee
and it is assumed that $\left|\mu\right| \ll M_S$.

For moderate values of $\tan\beta$, requiring $m_{h_{\rm SM}} \approx 125\,$GeV leads to $\lambda \approx 0.65$ in the alignment limit \cite{Carena:2015moc}. The remaining CP-even states $H^{\rm NSM}$ and $H^S$ mix to a mostly doublet-like $H$ and mostly singlet-like $h_S$ mass eigenstate. Similarly, the CP-odd states $A^{\rm NSM}$ and $A^S$ mix into a mostly doublet-like $A$ and mostly singlet-like $a_S$ mass eigenstate. In the alignment limit, the singlet-like mass eigenvalues, taking into account the first non-trivial corrections to $m_{h_S}^2 \sim \mathcal{M}_{S,33}^2$ and $m_{a_S}^2 \sim \mathcal{M}_{P,22}^2$, are \cite{Carena:2015moc}
\begin{eqnarray}
	m_{h_S}^2 &\simeq& \frac{\kappa \mu}{\lambda} \left( A_\kappa + \frac{4 \kappa \mu}{\lambda} \right) + \frac{\lambda^2 v^2 M_A^2}{8 \mu^2} s_{2\beta}^4 - \frac{1}{4}v^2 \kappa \lambda \left( 1+ 2 c_{2\beta}^2 \right) s_{2 \beta} - \frac{1}{2} v^2 \kappa^2 \frac{\mu^2}{M_A^2} c_{2\beta}^2, \label{eq:mhs}
	\\ m_{a_S}^2 &\simeq& 3 \kappa \left[ \frac{3}{4} \lambda v^2 s_{2\beta} - \mu \left( \frac{A_\kappa}{\lambda} + \frac{3 v^2 \kappa \mu}{2 M_A^2} \right) \right]. \label{eq:mas}
\end{eqnarray}
Such approximate formulae are useful to infer possible parameter combinations compatible with physical Higgs spectra; for instance, from the above equations for $m_{a_S}^2$ and $m_{h_S}^2$, one can infer that $\kappa < 0$ can lead to large negative contributions to $m_{a_S}^2$. In particular, contributions to $m_{a_S}^2$ linear only in $\kappa$ are significantly larger than those to $m_{h_S}^2$. Hence, prohibiting $a_S$ from becoming tachyonic when randomly sampling NMSSM parameters leads to a preference for positive values of $\kappa$. 

It is interesting to note the correlations between the Higgs and the neutralino masses due to the presence of a SM-like Higgs. Consider the region of parameters containing non-decoupled singlet Higgs bosons $\left|\kappa\right| \lesssim \lambda$, where approximate alignment must be fulfilled for consistent Higgs phenomenology. From Eq.~\eqref{eq:align2}, we see that $\mu$ is generically lighter than $M_A$ -- we find that typically $2 \lesssim M_A^2/\mu^2 \lesssim 8$. This leads to the singlet-like states $h_S$ and $a_S$ being lighter than the doublet-like $H$ and $A$, whose masses are mostly degenerate and controlled by $M_A$ (cf. the mass matrices Eqs. \eqref{eq:MS11}--\eqref{eq:MS23}, Eq.~\eqref{eq:MP} and Eqs. \eqref{eq:mhs}--\eqref{eq:mas} and discussion in Ref.~\cite{Carena:2015moc}). Furthermore, due to the relationship between $M_A$ and $\mu$, the singlinos ($m_{\widetilde{S}} \sim ~2 \kappa \mu /\lambda$) and Higgsinos ($m_{\widetilde{H}_u^0} = m_{\widetilde{H}_d^0} \sim \mu$) are also lighter than $A$ and $H$. However, we emphasize that while $M_A$ (controlling $m_A$ and $m_H$), $m_{h_S}$, $m_{\widetilde{S}}$, $m_{\widetilde{H}_u^0}$ and $m_{\widetilde{H}_d^0}$ are all strongly correlated with $|\mu|$, this is not necessarily the case for $m_{a_S}$: $A_\kappa$ can be used to vary $m_{a_S}$ independently of the value of $\mu$.
Hence the presence of a 125 GeV SM-like Higgs and light ($\lesssim$ 1 TeV) additional Higgs bosons folds the extended NMSSM parameter space such that the entire Higgs and neutralino mass spectrum is essentially driven by the two mass scales $\mu$~(or $M_A$) and $m_{a_S}$. 

The NMSSM parameters $\lambda$ and $\kappa$ induce additional couplings beyond the MSSM within the Higgs sector and between the Higgs bosons and neutralinos, which can change the Higgs collider phenomenology significantly. In particular, apart from decays into SM particles, the branching ratios of $\left(H_i \to H_j H_k / A_j A_k\right)$, $\left(A_i \to A_j H_k\right)$, $\left(H_i/A_i \to \chi_j \chi_k\right)$, $\left(H_i/A_i \to Z A_j / Z H_j\right)$ and $\left( \chi_i \to H_j \chi_k / A_j \chi_k\right)$ decays can be significant if kinematically allowed. 

The couplings
\begin{itemize}
	\item $\left(H^{\rm SM} H^{\rm SM} H^{\rm NSM}\right) \propto \mathcal{M}_{S,12}^2 \sim 0$,
	\item $\left(H^S H^{\rm SM} H^{\rm SM}\right) \propto \mathcal{M}_{S,13}^2 \sim 0$,
	\item $\left(H^{\rm NSM} A^{\rm NSM} A^S\right) =0$
\end{itemize}
are supressed close to the alignment limit, with the last one strictly vanishing. The couplings $\left(H^S H^{\rm SM} H^{\rm NSM}\right)$ and $\left(H^{\rm SM} A^{\rm NSM} A^S\right)$ are large for sizable values of $\lambda$ and $\kappa$ barring accidental cancellations. The singlet states $A^S$ and $H^S$ have no couplings to the gauge bosons at tree-level. By definition, $H^{\rm NSM}$ does not couple to pairs of gauge bosons. Its remaining coupling to neutral gauge bosons is given by 
\be
	g_{H^{\rm NSM} A^{\rm NSM} Z} = \frac{i}{2} \sqrt{g_1^2 + g_2^2} \left( p-p' \right)^\mu,
\ee
where $p$ ($p'$) is the incoming momentum of the $H^{\rm NMS}$ ($A^{\rm NMS}$). A complete list of the Higgs to Higgs couplings in the Higgs basis can be found in the Appendix of Ref.~\cite{Carena:2015moc}, and we tabulate them in Appendix \ref{app:triH} for the convenience of the reader.

The couplings of the Higgs basis states to the neutralino mass eigenstates $\chi_i$ are
\bea
	g_{(H^{\rm SM} \chi_i \chi_j)} &=& \frac{1}{\sqrt{2}} \left[ \lambda N_{i5} \left( N_{j3} s_\beta + N_{j4} c_\beta \right) + \left( i \leftrightarrow j \right) \right],
	\\ g_{(H^{\rm NSM} \chi_i \chi_j)} &=& \frac{1}{\sqrt{2}} \left[ \lambda N_{i5} \left( N_{j3} c_\beta - N_{j4} s_\beta \right) + \left( i \leftrightarrow j \right) \right],
	\\ g_{(A \chi_i \chi_j)} &=& \frac{i}{\sqrt{2}} \left[ \lambda N_{i5} \left( N_{j3} c_\beta + N_{j4} s_\beta \right) + \left( i \leftrightarrow j \right) \right],
	\\ g_{(H^S \chi_i \chi_j)} = i g_{(A^S \chi_i \chi_j)} &=& \frac{1}{\sqrt{2}} \left[ \left( \lambda N_{i4} N_{j3} - \kappa  N_{i5} N_{j5} \right) + \left( i \leftrightarrow j \right) \right],
\eea
where the neutralino mass eigenstates $\chi_i$ are related to the interaction eigenstates by
\be
	\chi_i = N_{i1} \widetilde{B} + N_{i2} \widetilde{W}^3 + N_{i3} \widetilde{H}_d^0 + N_{i4} \widetilde{H}_u^0 + N_{i5} \widetilde{S},
\ee
where the $N_{ij}$ are obtained by diagonalizing the neutralino mass matrix given in Eq.~(\ref{eq:neumassmatrix}),
and we take $N_{i1} \approx N_{i2} \approx 0$ since the bino and wino are decoupled from our analysis.

We stress that several of the above couplings are proportional to $\lambda$ as they contain the singlet-doublet-doublet structure, which originates from the $\lambda \widehat{S} \widehat{H}_u \cdot \widehat{H}_d$ term in the superpotential. Since a 125\,GeV SM-like Higgs and naturalness considerations favor large values of $\lambda$, these couplings are expected to be significant.

\section{Numerical Scan}\label{sec:scan}
We perform a random scan of the NMSSM parameter space with the program package \texttt{NMSSMTools\_4.9.3} \cite{NMSSMTools}, which includes \texttt{NMHDECAY} \cite{Ellwanger:2004xm, Ellwanger:2005dv} to compute masses, couplings, and decay widths of the Higgs bosons and \texttt{NMSDECAY} \cite{Das:2011dg, Muhlleitner:2003vg} to compute sparticle widths and branching ratios. We scan over a wide range of values of the parameter set from Eq.~\eqref{eq:parameters}, listed in Table~\ref{tab:scan_param}.  In addition to the ``standard'' scan, we also perform a second scan over a narrower range of parameters focused on producing lighter Higgs spectra accessible at the LHC, which we label the ``light subset''. The chosen range $1 \leq \tan\beta \leq 5$ is motivated by $m_{h_{\rm SM}} \simeq 125\,$GeV, as the crucial contribution $\frac{1}{2} \lambda^2 v^2 s_{2\beta}^2$ to $m_{h_{\rm SM}}^2$ Eq.~\eqref{eq:mhsm} is suppressed at larger values of $\tan\beta$. Note that we also scan over the stop mass parameter $M_{U_3} = M_{Q_3}$, since stops can give large radiative corrections to the mass of the SM-like Higgs. We set the stop and sbottom mixing parameters $X_t \equiv (A_t - \mu \cot\beta) = 0$ and $X_b \equiv (A_b - \mu\tan\beta) = 0$ since large third generation sfermion mixing is not necessary to obtain the correct Higgs mass in the NMSSM, and is thus irrelevant for Higgs phenomenology in our region of interest. The remaining supersymmetric particles are decoupled from our study: we set sfermion mass parameters (except $M_{U_3}, M_{Q_3}$) to $3\,$TeV, the bino and wino mass parameters to $M_1 = M_2 = 1\,$TeV, and the gluino mass to $M_3 = 2\,$TeV.

\begin{table}
	\begin{centering}
		\begin{tabular}{|c|c|c|}
			\hline
			& ``standard'' & ``light subset'' \\ \hline
			$\tan\beta$ & $ \left[ 1; 5 \right] $ & $ \left[ 1; 5 \right] $  \\ \hline
			$\lambda$ & $ \left[ 0.5; 2 \right] $ & $ \left[ 0.5; 1 \right] $  \\ \hline
			$\kappa$ & $ \left[ -1; +1 \right] $ & $ \left[ -0.5; +0.5 \right] $  \\ \hline
			$A_\lambda$ & $ \left[ -1; +1 \right] \,$TeV & $ \left[ -0.5; +0.5 \right] \,$TeV \\ \hline
			$A_\kappa$ & $ \left[ -1; +1 \right] \,$TeV & $ \left[ -0.5; +0.5 \right] \,$TeV \\ \hline
			$\mu$ & $ \left[-1; +1 \right] \,$TeV & $ \left[ -0.5; +0.5 \right] \,$TeV \\ \hline
			$M_{Q_3}$ & $ \left[ 1; 10 \right] \,$TeV & $  \left[ 1; 10 \right] \,$TeV \\ \hline
		\end{tabular}
		\caption{NMSSM parameter ranges used in \texttt{NMSSMTools} scans.}
		\label{tab:scan_param}
	\end{centering}
\end{table}

For each parameter set, we scan $10^8$ points randomly chosen from linear-flat distributions over the respective parameter ranges, imposing a subset of the constraints implemented in \texttt{NMSSMTools} (see Ref.~\cite{NMSSMTools} for details). Points are excluded if they have unphysical global minima, soft Higgs masses much larger than $M_{\rm SUSY}$, or if the lightest neutralino $\chi_1$ is not the lightest supersymmetric particle (LSP). We also require compatibility with constraints from the Large Electron-Positron Collider (LEP), Tevatron, and searches for sparticles and charged Higgs bosons \cite{ATLAS-CONF-2011-151} at the LHC as implemented in \texttt{NMSSMTools}. Finally, points are required to contain a SM-like Higgs boson with couplings to photons, massive gauge bosons, and $b$-quarks compatible with LHC bounds and with a mass of $125 \pm 3\,$GeV, where the width of this band is given by the theoretical uncertainty of the Higgs mass calculation\footnote{Higher-order loop corrections not taken into account in NMSSMTools can account for differences in the SM-like Higgs boson mass as large as 6\,GeV when compared to other spectrum generators \cite{Goodsell:2014pla, Staub:2015aea}. We scan over the stop mass parameter to allow for the required loop corrections to obtain a SM-like Higgs with mass $125\,$GeV. Taking into account higher order loop corrections to the Higgs mass would affect the value of the stop mass parameters for a given point, but not the allowed range of the parameters $\{\lambda, \kappa, \tan\beta, \mu, A_\lambda, A_\kappa\}$ relevant for the Higgs and neutralino sector phenomenology. \\ \vspace{.8cm}}~\cite{NMSSMTools, Goodsell:2014pla, Staub:2015aea}.

\begin{figure}
	\begin{center}
		\includegraphics[width=1\linewidth]{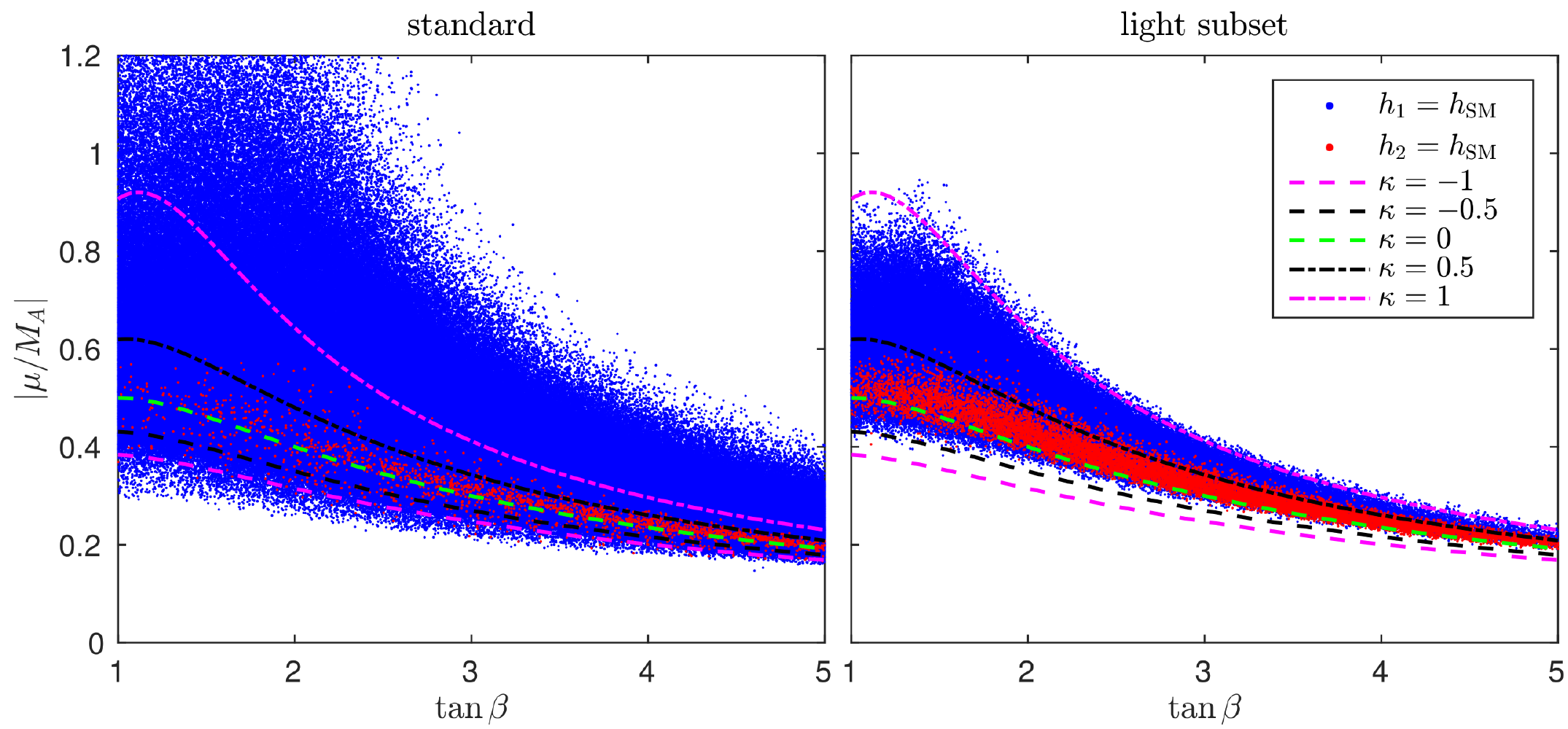}
		\caption{Distribution of $\left|\mu/M_A\right|$ vs. $\tan\beta$ obtained from our \texttt{NMSSMTools} scan for the ``standard'' (left panel, $M_A\lesssim3$~TeV) and ``light subset'' (right panel, $M_A \lesssim 1$~TeV). The dashed and dash-dotted lines display the values of $\left|\mu/M_A\right|$ in the alignment limit for different values of $\kappa$ as indicated in the legend, with $\lambda$ also set to the alignment value; see Eqs. \eqref{eq:align1} and \eqref{eq:align2}. For the ``light subset", which has lighter Higgs spectra, we find values close to the alignment limit; for the ``standard" set, compatibility with the 125\,GeV Higgs boson can also be achieved by decoupling, hence the points are more dispersed. The asymmetry of the distribution of points relative to the alignment limit contours reflects the preference for positive $\kappa$ to avoid tachyonic Higgs masses. See text for details.}
		\label{fig:mu_align}
	\end{center}
\end{figure}

We keep points violating direct Higgs search constraints from ${\left(H_i/A_i \to \tau\tau\right)}$, $\left(H_i/A_i \to \gamma\gamma\right)$, and ${\left(h_{\rm SM} \rightarrow A_i A_i \rightarrow 4 \mu\right)}$ in order to compare the implementation of the constraints in \texttt{NMSSMTools} against our own implementation of direct LHC constraints on NMSSM Higgs bosons (see Section\,\ref{sec:lhcconstraints}).
We also keep points violating the flavor physics constraints in \texttt{NMSSMTools}, as it is non-trivial to find combinations of NMSSM parameters simultaneously satisfying theoretical consistency of the spectrum, a neutralino LSP, and flavor constraints, with the justification that additional degrees of freedom in the flavor sector can generally be adjusted independently to achieve compatibility, see discussions in Ref.~\cite{Altmannshofer:2012ks}.

As anticipated in the previous section, we find that points satisfying these constraints, driven particularly by the requirement of a 125\,GeV SM-like Higgs, lie close to the alignment limit (Eqs.\,\eqref{eq:align1}, \eqref{eq:align2}). This pattern is illustrated in Fig.~\ref{fig:mu_align}, where we show the distribution of $\left|\mu/M_A\right|$ vs. $\tan\beta$ obtained from our scans together with contours of the alignment limit. How close the NMSSM parameters are to the alignment values is driven by the Higgs spectrum: lighter Higgs spectra have NMSSM parameters closer to the alignment limit than heavier spectra. This behavior is evident when comparing the two panels; for the ``light subset" (right panel), where $m_{A_2} \lesssim 1$\,TeV, the distribution obtained from our scan follows the alignment band closely, while for the ``standard" set (left panel) $m_{A_2}$ can be as large as $\sim 3$\,TeV and we see that the distribution is more dispersed since compatibility with the 125\,GeV Higgs boson can also be achieved by decoupling. We emphasize that the NMSSM parameters are not a priori set to be close to the alignment limit in our scan, but forced into this regime by the requirement of a CP-even Higgs mass eigenstate compatible with the 125\,GeV SM-like Higgs detected at the LHC. 

Points passing all our constraints typically have moderate values of $\lambda \gsim 0.6$ and $1 \lesssim \tan\beta \lesssim 3$. We also observe the preference for positive values of $\kappa$ to avoid tachyonic masses as discussed below Eq.~\eqref{eq:mas}. Compared to points where the lightest CP-even Higgs boson is SM-like, those where the second lightest CP-even Higgs is SM-like feature smaller $\lambda$, smaller $\left|\kappa\right|$, larger $\tan\beta$ and larger $\left|A_\lambda\right|$. This is because when $h_2=h_{\rm SM}$, the lightest CP-even Higgs must be almost exclusively singlet-like to be compatible with phenomenological constraints. Smaller values of $\lambda$, $\left|\kappa\right|$ and larger values of $\tan\beta$ lead to a lighter singlet mass (cf. Eq.~\eqref{eq:mhs}) and reduce the singlet-doublet mixing (cf. Eqs.~\eqref{eq:MS23} and \eqref{eq:align2}). Furthermore due to values of $\lambda$ smaller than those preferred by alignment (cf. Eq.~\eqref{eq:align1}), somewhat larger masses of $H$ are preferred, which are controlled by $A_\lambda$ ($M_A$).

In the following, we provide results based on the combined ``standard" and ``light subset" scans. We note that the most relevant LHC phenomenology is obtained for the region of parameter space corresponding to the ``light subset" as this tends to give lighter physical states. When referring to points from our scans, we denote the Higgs mass eigenstates by $\{h_1, h_2, H_3\}$ for the CP-even Higgs bosons and $\{A_1, A_2\}$ for the CP-odd Higgs bosons. The index denotes the mass hierarchy $m_{h_1} < m_{h_2} < m_{H_3}$ and $m_{A_1} < m_{A_2}$. One of the lighter CP-even Higgs eigenstates $h_1, h_2$ is identified with $h_{\rm SM}$, the 125\,GeV SM-like Higgs boson observed at the LHC. Of the remaining CP-even mass eigenstates $h_i$ and $H_3$, one is identified with the mostly singlet-like $h_S$ and the other with the mostly doublet-like $H$. Similarly, one of the CP-odd mass eigenstates $A_1, A_2$ is identified with the mostly singlet like $a_S$, and the other with the mostly doublet-like $A$.

\section{Constraints from direct Higgs searches at the LHC}
\label{sec:lhcconstraints}

\begin{table}
	\begin{center}
	\begin{tabular}{|c|c|c|c|}
		\hline
		decay channel & NMSSM Higgs & Reference & Reference \\ & tested & $\sqrt{s}=8\,$TeV & $\sqrt{s}=13\,$TeV \\ \hline 
		$H \rightarrow \tau^+\tau^-$ & $h_i, H_3, A_1, A_2$ & \cite{Khachatryan:2014wca, CMS-PAS-HIG-14-029, Aad:2014vgg} & \cite{ATLAS-CONF-2016-085, CMS-PAS-HIG-16-037} \\ \hline
		$H \rightarrow b\bar{b}$ & $h_1, H_3, A_1, A_2$ & -- & \cite{CMS-PAS-HIG-16-025} \\ \hline
		$H \rightarrow \gamma\gamma$ & $h_i, H_3, A_1, A_2$ & \cite{Khachatryan:2015qba, CMS-PAS-HIG-14-037, Aad:2014ioa} & \cite{ATLAS-CONF-2016-018, CMS-PAS-EXO-16-027, ATLAS-CONF-2016-059} \\ \hline
		$H \rightarrow ZZ$ & $h_1, H_3$ & \cite{Aad:2015kna}  & \cite{CMS-PAS-HIG-16-001, ATLAS-CONF-2016-012, ATLAS-CONF-2016-016, ATLAS-CONF-2015-071, CMS-PAS-HIG-16-033, ATLAS-CONF-2016-056, ATLAS-CONF-2016-079} \\ \hline
		$H \rightarrow WW$ & $h_i, H_3$ & \cite{Aad:2015agg, CMS-PAS-HIG-13-027, ATLAS-CONF-2013-067}  & \cite{ATLAS-CONF-2016-021, CMS-PAS-HIG-16-023, ATLAS-CONF-2016-074, ATLAS-CONF-2016-062} \\ \hline
		$H \rightarrow h_{\rm SM}h_{\rm SM} \rightarrow b\bar{b}\tau^+\tau^-$ & $h_i, H_3$ & \cite{Khachatryan:2015tha, CMS-PAS-HIG-15-013, Aad:2015xja}  & \cite{CMS-PAS-HIG-16-013, CMS-PAS-HIG-16-029} \\ \hline
		$H \rightarrow h_{\rm SM}h_{\rm SM} \rightarrow b\bar{b} \ell \nu_\ell \ell \nu_\ell $ & $h_i, H_3$ & -- & \cite{CMS-PAS-HIG-16-011} \\ \hline
		$H \rightarrow h_{\rm SM}h_{\rm SM} \rightarrow b\bar{b}b\bar{b} $ & $h_i, H_3$ & \cite{Khachatryan:2015yea, Aad:2015uka} & \cite{CMS-PAS-HIG-16-002, ATLAS-CONF-2016-017, ATLAS-CONF-2016-049} \\ \hline
		$H \rightarrow h_{\rm SM}h_{\rm SM} \rightarrow b\bar{b} \gamma\gamma $ & $h_i, H_3$ & \cite{Khachatryan:2016sey, Aad:2014yja} & \cite{CMS-PAS-HIG-16-032, ATLAS-CONF-2016-004} \\ \hline
		$A \rightarrow Z h_{\rm SM} \to Z b\bar{b} $ & $A_1, A_2$ & \cite{Aad:2015wra, Khachatryan:2015lba} &  \cite{ATLAS-CONF-2016-015} \\ \hline
		$A \rightarrow Zh_{\rm SM} \rightarrow Z \tau^+\tau^-$ & $A_1, A_2$ & \cite{Khachatryan:2015tha, Aad:2015wra} & -- \\ \hline
		$h_{\rm SM} \to A A \to \tau^+ \tau^- \tau^+ \tau^-$ & $A_1$, $A_2$ & \cite{Khachatryan:2240709} & --\\ \hline
		$h_{\rm SM} \to A A \to \mu^+ \mu^- b \bar{b}$ & $A_1$, $A_2$ & \cite{Khachatryan:2240709} & -- \\ \hline
		$h_{\rm SM} \to A A \to \mu^+ \mu^- \tau^+ \tau^-$ & $A_1$, $A_2$ & \cite{Khachatryan:2240709} & -- \\ \hline
		$h_{\rm SM} \to A A \to \mu^+ \mu^- \mu^+ \mu^-$ & $A_1$, $A_2$ & -- & \cite{CMS-PAS-HIG-16-035} \\ \hline
		$A/H \to Z h_i/A_1$ & $A_2/H_3$, $h_i/A_1$ & \cite{Khachatryan:2016are} & -- \\ \hline 
	\end{tabular}
	\caption{Direct Higgs searches at the LHC used for this work. $h_i = h_2$ ($h_1$) if the (second)~lightest  scalar is SM-like.}
	\label{tab:LHCSearches}
	\end{center}
\end{table}

We constrain our NMSSM data set with the null-results of a number of direct Higgs searches at the LHC, listed in Table~\ref{tab:LHCSearches}, by comparing the production cross section times branching ratio in the respective final state with the corresponding bound.

Over the range $1 \leq \tan\beta \leq 5$ the production cross section of all NMSSM Higgs bosons at the LHC is dominated by gluon fusion. \texttt{NMSSMTools} calculates the ratio of the coupling of the NMSSM (pseudo)~scalar Higgs bosons to gluons with respect to the coupling of a SM Higgs of the same mass at next-to-leading order (NLO) in QCD, $\kappa_{gg}^{A_i/H_i}$. We first approximate the gluon fusion production cross section for NMSSM Higgs bosons by
\begin{equation} \sigma(ggH_i/ggA_i) = \left(\kappa_{gg}^{H_i/A_i}\right)^2 \times \sigma^{\rm SM}_{ggH}, \end{equation}
where $\sigma^{\rm SM}_{ggh}$ is the gluon fusion production cross of the SM Higgs boson, which we calculate at NLO precision with the program \texttt{SusHi-1.5.0}\footnote{Our production cross section calculation for SM Higgs bosons agrees with those from the LHC Higgs Cross Section Working Group for $80\,{\rm GeV} \leq m_h \leq 1\,$TeV at next-to-next-to-leading log (NNLL) accuracy in QCD and NLO in electroweak (EW) corrections \cite{Heinemeyer:2013tqa} within theoretical uncertainties, taking into account that we compute our SM-like cross sections at the renormalization scale recommended by \texttt{SusHi-1.5.0} and that we do not take into account NNLL QCD corrections for consistency with the \texttt{NMSSMTools} calculation of $\kappa_{gg}^{A_i/H_i}$ at NLO QCD. See Ref.~\cite{deFlorian:2016spz} for a recent updated calculation of SM Higgs production cross section at NNLO+NNLL for $10\,{\rm GeV} \leq m_h \leq 3\,$TeV. \\ \vspace{.8cm}} \cite{Harlander:2012pb, Harlander:2002wh, Harlander:2005rq}. We validate the gluon fusion cross section thus obtained by comparing it with a sampling of the gluon fusion cross section computed directly from the NMSSM implementation in \texttt{SusHi}. We find agreement to better than 5\,\% in most cases, with deviations of up to 15\,\% in rare cases, particularly for CP-odd Higgs bosons with masses close to the top-resonance $m_{A_i} \simeq 2m_t$. We address such discrepancies by recalculating the gluon fusion production cross section with the NMSSM implementation of \texttt{SusHi} for points with $\sigma(ggH_i/ggA_i) \times {\rm BR}(H_i/A_i \rightarrow {\rm final\,\,state})$ within $\pm 20\,\%$ of the respective LHC exclusion limit.

LHC searches for additional Higgs bosons with pairs of leptons, quarks, or photons in the final states are applicable for the two CP-even neutral Higgs bosons not identified with the SM-like Higgs boson, and for both CP-odd neutral Higgs bosons. Searches for Higgs bosons decaying to a pair of vector bosons $\left(H \rightarrow ZZ/WW\right)$ are only checked for the CP-even Higgs bosons as this decay is forbidden for CP-odd scalars at tree-level. For similar reasons, searches for additional Higgs bosons decaying to a pair of SM-like Higgs bosons $\left(H \rightarrow h_{\rm SM} h_{\rm SM}\right)$ (a $Z$-boson and a SM-like Higgs $\left(A \rightarrow Z h_{\rm SM}\right)$) are only tested for CP-even (CP-odd) NMSSM Higgs bosons. 

We show the Higgs production cross section into various channels obtained for our scanned points together with the respective limits from LHC in Appendix \ref{app:constraints} to illustrate the constraining power of the respective searches. We find that the most constraining LHC searches are $\left(ggH/ggA \rightarrow \gamma\gamma\right)$, $\left(ggH/ggA \rightarrow \tau\tau\right)$, $\left(ggH \to ZZ\right)$, and $\left(ggA \rightarrow Z h_{\rm SM}\right)$. We also note that our implemented constraints are more stringent than the \texttt{NMSSMTools} implementation of direct Higgs searches, since we take many more searches into account. Generically, points excluded by the LHC tend to have larger $\left|\kappa\right|$ and smaller $\tan\beta$, $\lambda$, $\left|\mu\right|$, $\left|A_\lambda\right|$, and $\left|A_\kappa\right|$ compared to those that pass the constraints.

\begin{figure}
	\begin{center}
		\includegraphics[width = .5\linewidth]{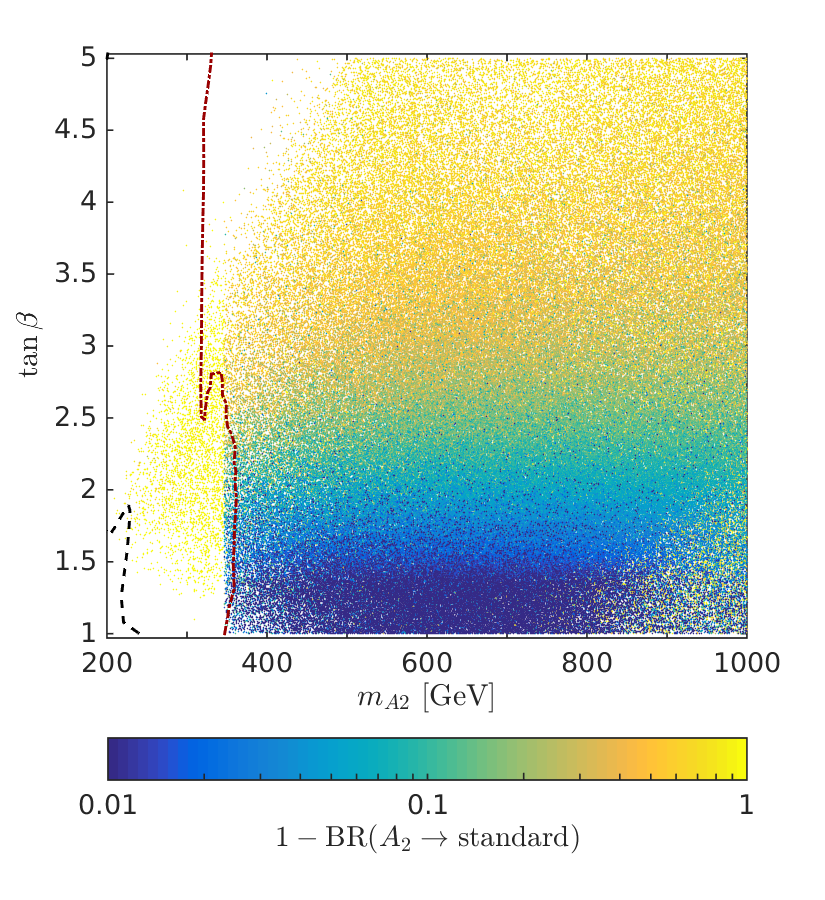}
		\caption{Points from our \texttt{NMSSMTools} scans passing all constraints listed in Table\,\ref{tab:LHCSearches} in the ${m_{A_2}-\tan\beta}$ plane. The color coding shows the branching ratios of $A_2$ into non-SM particles, see Eq.~\eqref{eq:BRAstandard}. The dash-dotted dark red (dashed black) line shows the limit from CMS after LHC8 for the hMSSM (MSSM $m_h^{\rm mod+}$) scenario \cite{CMS-PAS-HIG-16-007}, excluding the regions to the left. The allowed yellow points to the left of the red curve illustrate that large branching fractions into non-SM states in the NMSSM framework can enable light Higgs bosons to circumvent the traditional MSSM bounds.}
		\label{fig:ma-tb}
	\end{center}
\end{figure}

As discussed in Section \ref{sec:Hcouplings}, the phenomenology of the NMSSM Higgs sector can differ significantly from that of the MSSM framework. This is illustrated in Fig.~\ref{fig:ma-tb}, where we show points from our scans passing all constraints listed in Table\,\ref{tab:LHCSearches} in the $m_{A_2} - \tan\beta$ plane. The color coding shows the branching ratios of the heavy CP-odd Higgs boson $A_2$ to non-SM particles $\left[1 - {\rm BR}(A_2 \to {\rm standard})\right]$, where
\be \begin{split}
	{\rm BR}(A_2 \to {\rm standard}) = & \  {\rm BR}(A_2 \to t\bar{t}) +  {\rm BR}(A_2 \to b\bar{b}) 
	\\ & + {\rm BR}(A_2 \to \tau^+ \tau^-) + {\rm BR}(A_2 \to \mu^+ \mu^-) 
	\\ & + {\rm BR}(A_2 \to Z h_{\rm SM}) + {\rm BR}(A_2 \to \gamma \gamma).
	\label{eq:BRAstandard}
\end{split} \ee
These non-standard decays include channels such as ($A_2 \to Z h_i$) and ($A_2 \to \chi_j \chi_1$). The equivalent plot for the branching ratios of $H_3$ to non-SM particles looks very similar. The dash-dotted dark red (dashed black) curves denote the limits from CMS from 8\,TeV data for the hMSSM (MSSM $m_h^{\rm mod+}$) scenarios \cite{CMS-PAS-HIG-16-007}, which should be interpreted as lower limits on the heavy Higgs mass $m_A$. We see that such limits are not applicable to the NMSSM, as several allowed points~(mostly yellow, representing sizable branching ratios into non-SM particles) lie to the left of the red curve; the additional couplings and decay modes in the NMSSM can thus enable light Higgs bosons to evade MSSM-specific LHC bounds. 

%%%%%%%%%%%%%%%%%%%%%%%%%%%%%%%%%%%%%%%%%
\section{NMSSM specific search strategies at the LHC} \label{sec:NMSSMsearch}

\begin{figure}
	\begin{center}
		\includegraphics[width=1\linewidth]{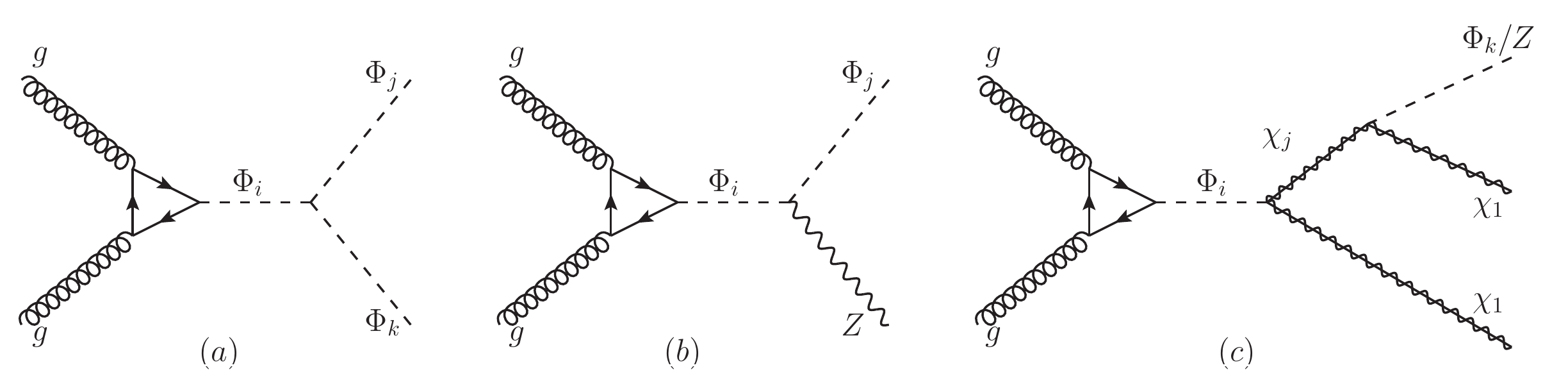}
		\caption{Illustration of NMSSM-specific Higgs decay topologies, where the $\Phi_i$ stand for one of the five NMSSM Higgs bosons. For channel $(a)$, either one or all three of the $\Phi_{i,j,k}$ must be CP-even. For channel $(b)$, if $\Phi_i$ is CP-even, $\Phi_j$ must be a CP-odd state, and vice-versa. For channel $(c)$, the final state can be $\chi_1 \chi_1 H_i$, $\chi_1 \chi_1 A_i$, or $\chi_1 \chi_1 Z$, and $\Phi_i$ can be CP-even or -odd.}
		\label{fig:triHdiagrams}
	\end{center}
\end{figure}

As discussed in Section\,\ref{sec:Hcouplings} and shown in Fig.~\ref{fig:ma-tb}, the currently allowed parameter space of the NMSSM allows for large branching ratios of heavy Higgs boson decays into unique channels, beyond what can be realized in the MSSM. The generic topologies of such channels, depicting decays into scalars, vector, and fermion states, are portrayed in Fig.~\ref{fig:triHdiagrams}. Among the decays denoted by channel $(a)$, $\left(H_3 \to h_{\rm SM} h_i\right)$, $\left(H_3 \to A_1 A_1\right)$, and $\left(A_2 \to A_1 h_{\rm SM}\right)$ are generally the dominant channels. For channel $(b)$, $\left(H_3 \to Z A_1\right)$ and $\left(A_2 \to Z h_i\right)$ tend to dominate, where $h_i$ stands for the light non-SM like CP-even Higgs mass eigenstate. Recall that due to the SM-like nature of the 125\,GeV Higgs boson, approximate alignment conditions must be fulfilled (with or without decoupling); hence, several couplings, such as $H h_{\rm SM} h_{\rm SM}$, are suppressed. Regarding channel $(c)$, both $H_3$ and $A_2$ contribute to $\chi_1\chi_j$ production; in general we will be interested in $j=3$ due to kinematic phase space considerations for further decays of $\chi_j$ into $\chi_1$ and lighter Higgs or $Z$ bosons.

\begin{figure}
	\begin{center}
		\includegraphics[width=.49\linewidth]{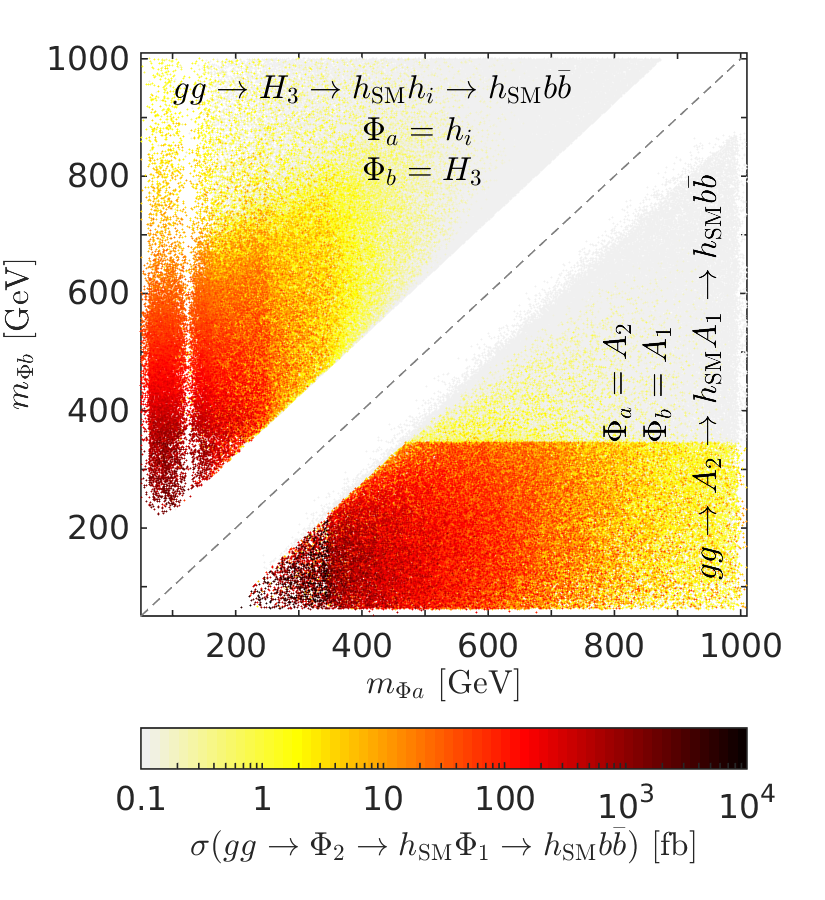}
		\includegraphics[width=.49\linewidth]{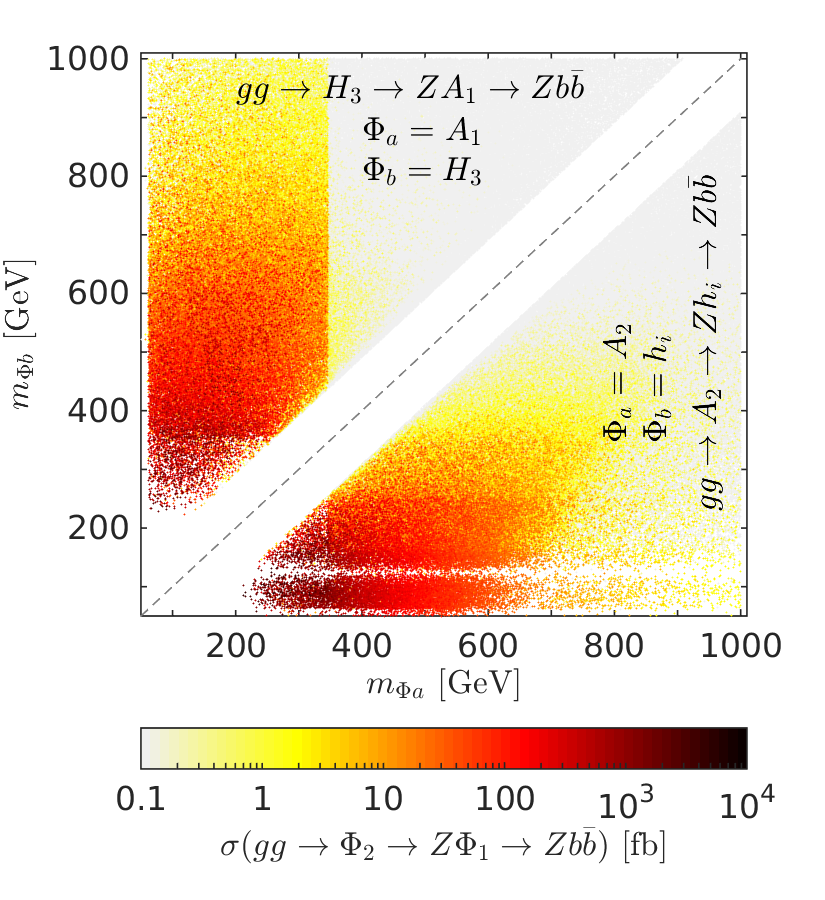}

		\includegraphics[width=.49\linewidth]{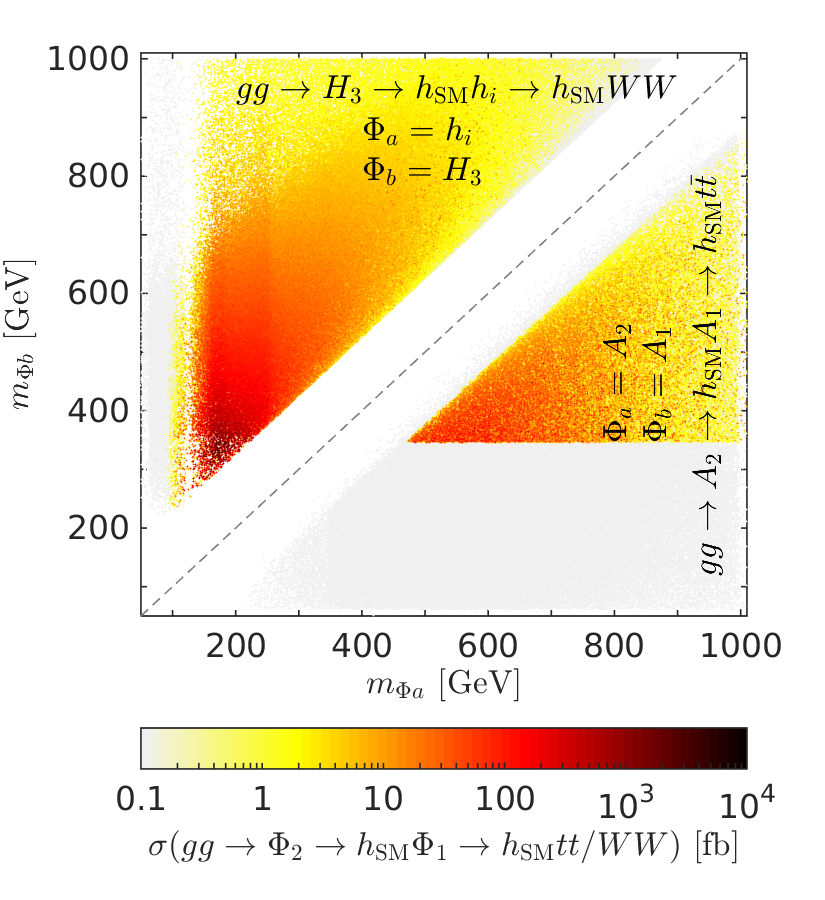}
		\includegraphics[width=.49\linewidth]{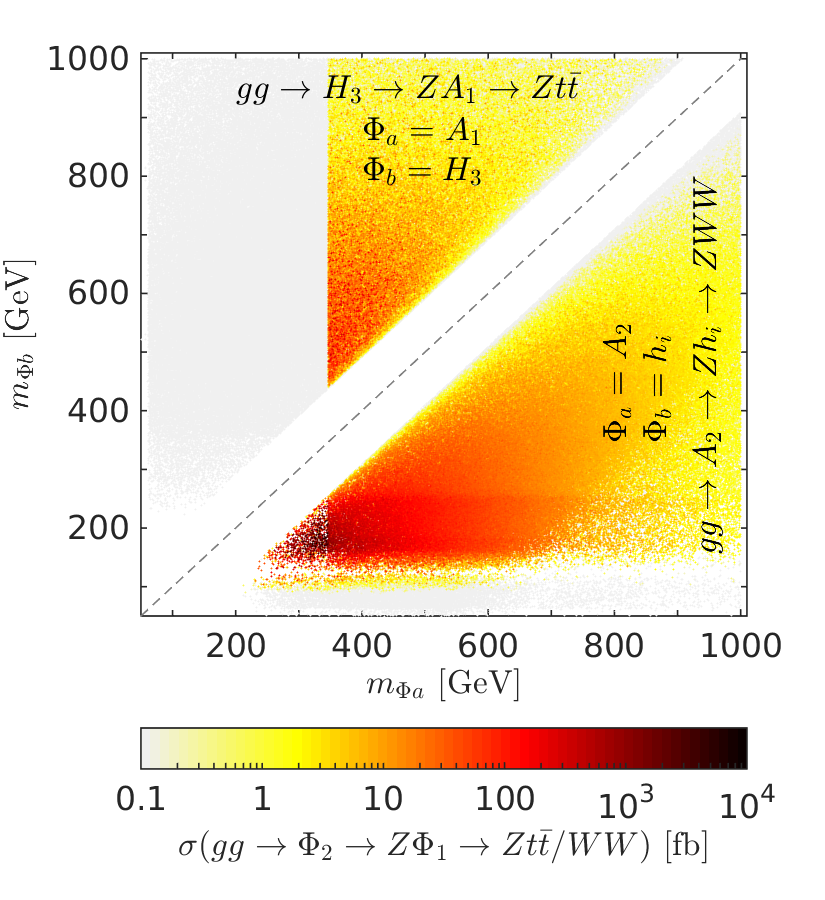}
		\caption{Cross sections for NMSSM specific Higgs search channels at $\sqrt{s} = 13\,$TeV with visible final states. The upper left panel shows $\sigma(ggA_2 \to h_{\rm SM} A_1 \to h_{\rm SM} b\bar{b})$ in the lower and $\sigma(ggH_3 \to h_{\rm SM} h_i \to h_{\rm SM} b\bar{b})$ in the upper triangles. The upper right panel shows $\sigma(ggA_2 \to Z {h_i} \to Z b\bar{b})$ and $\sigma(ggH_3 \to Z A_1 \to Z b\bar{b})$ in the lower and upper triangles. The lower panels show the same processes for $h_i \to WW$ and $A_1 \to t\bar{t}$ final states. The gap around $h_i = 125\,$GeV (visible in upper triangles in the left panels, lower triangles in the right panels) is due to the presence of the 125\,GeV SM-like Higgs.}
		\label{fig:xSec_SM}
	\end{center}
\end{figure}

\begin{figure}
	\begin{center}
		\includegraphics[width=.49\linewidth]{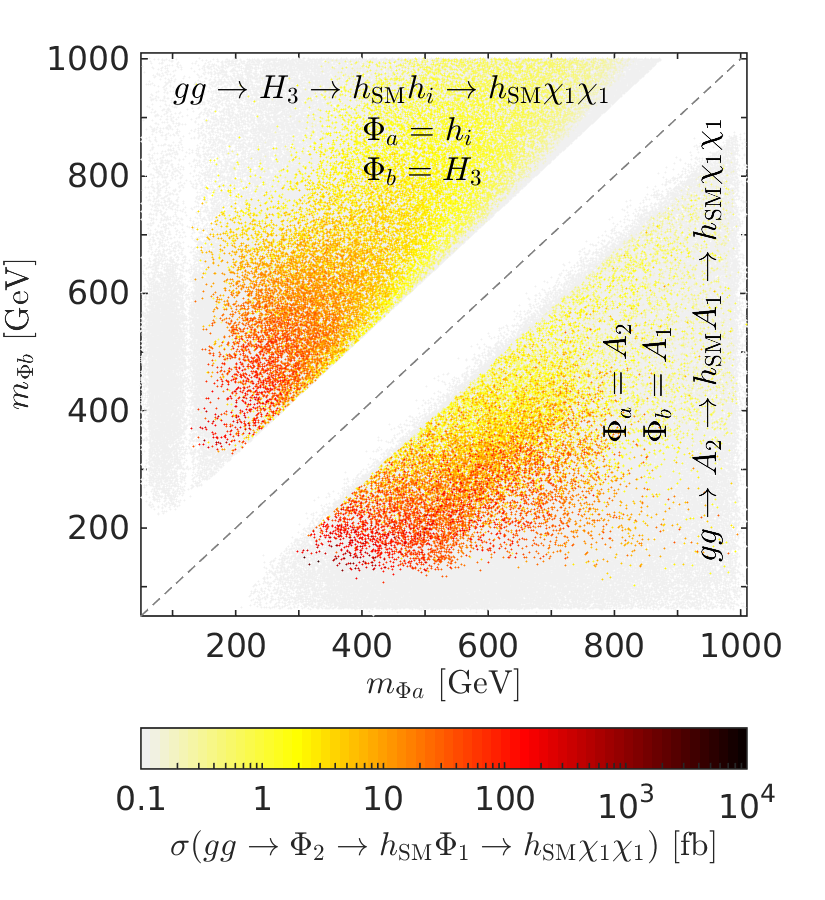}
		\includegraphics[width=.49\linewidth]{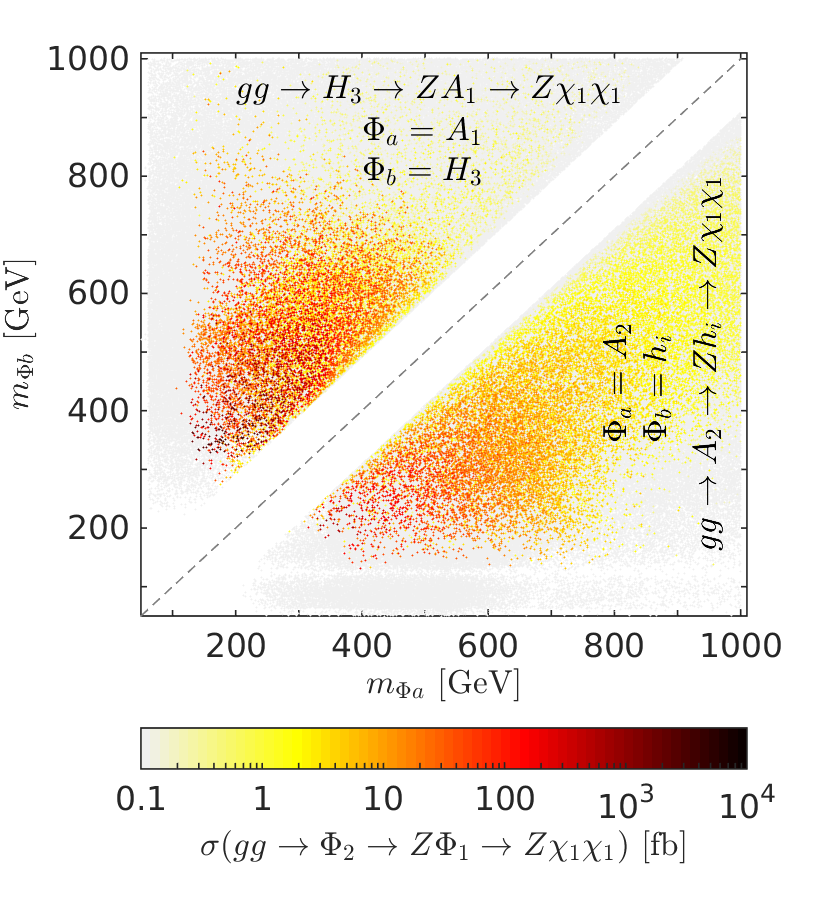}
		\caption{Same as Fig.~\ref{fig:xSec_SM} but for $h_{\rm SM}/Z + \chi_1 \chi_1$ final state.}
		\label{fig:xSec_mono}

		\includegraphics[width=.49\linewidth]{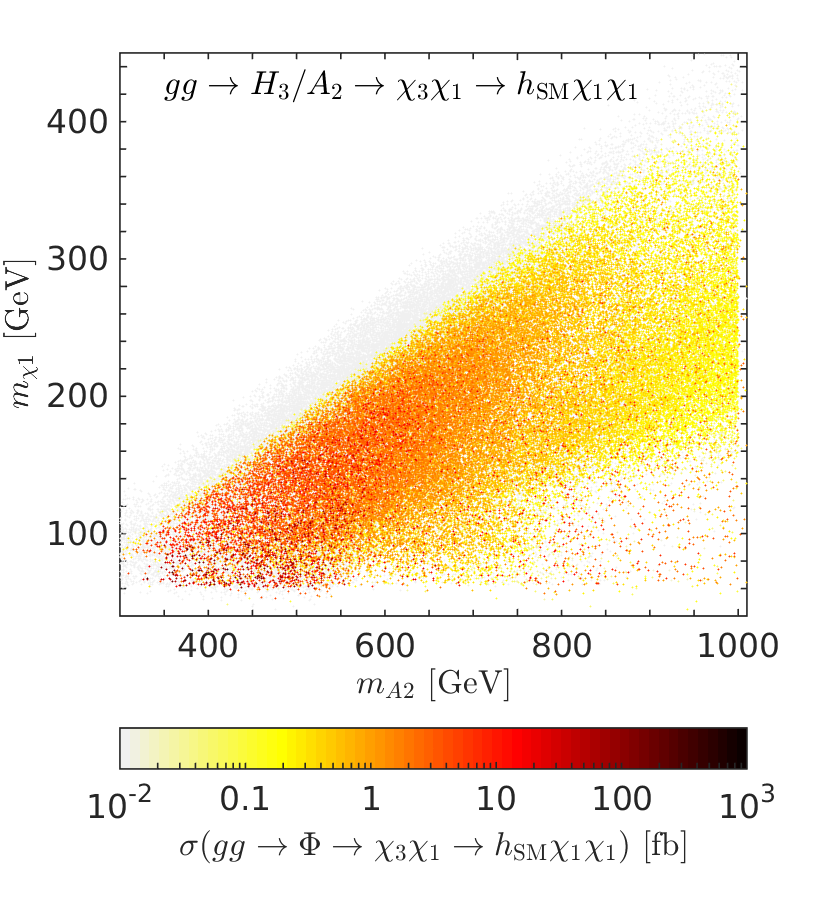}
		\includegraphics[width=.49\linewidth]{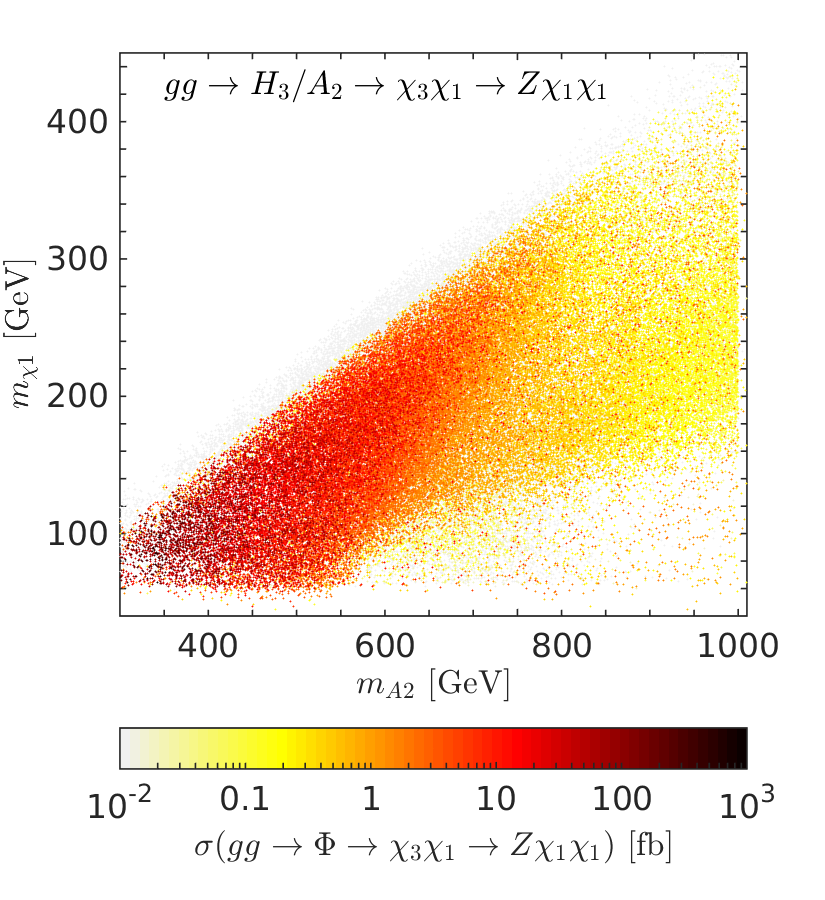}
		\caption{Cross section for NMSSM specific Higgs search channels $\sigma(ggH_3/A_2 \to \chi_3 \chi_1 \to h_{\rm SM} \chi_1 \chi_1)$ (left panel) and $\sigma(ggH_3/A_2 \to \chi_3 \chi_1 \to Z \chi_1 \chi_1)$  (right panel)  at $\sqrt{s} = 13\,$TeV. For ${(m_{H_3} - m_{A_2})/m_{A_2} < 30\,\%}$ we add production via both $H_3$ and $A_2$, otherwise we show the larger of the two production cross sections. Note that the color scale is different from that of Figs.~\ref{fig:xSec_SM}--\ref{fig:xSec_mono}.}
		\label{fig:xSec_neu}
	\end{center}
\end{figure}

A number of specific final states can be employed to search for all these processes. In particular, channels $(a)$ and $(b)$ generally result in the production of the singlet-like (pseudo)~scalar along with either $h_{\rm SM}$ or a $Z$. These NMSSM Higgs bosons $h_i/A_1$ decay with branching ratios similar to MSSM Higgs bosons of the same mass if no other decay channels are kinematically allowed. The CMS collaboration has carried out such a search at $\sqrt{s} = 8\,$TeV for channel $(b)$ in the $\left(Z h_i/ Z A_1 \to Z b\bar{b}\right)$ or $\left(Z h_i/ Z A_1 \to Z \tau^+ \tau^-\right)$ final states \cite{Khachatryan:2016are}; we show the resulting constraints on our scan  in Fig.~\ref{fig:HZA_AZH_excluded} in Appendix \ref{app:constraints}.

When $\left(h_i/A_1 \to t\bar{t}\right)$ decays are kinematically accessible, they typically dominate among the visible final states. Below the top threshold, the ($h_i \to WW$) channels become the dominant SM decay modes for CP-even Higgs bosons; such decays are forbidden at tree-level for CP-odd Higgs bosons. For $m_{A_1} \lesssim 2 m_t$ / $m_{h_i} \lesssim 2 m_W$, $\left(h_i/A_1 \to b \bar{b}\right)$ is typically the dominant decay mode among the SM final states. In the upper panels of Fig.~\ref{fig:xSec_SM}, we show the cross sections  for $\left(h_{\rm SM}/Z + b\bar{b}\right)$ final states at the $\sqrt{s} = 13\,$TeV LHC for points from our scan passing all constraints. As expected, there is a drastic reduction in the cross section above the $WW$/$t\bar{t}$ threshold. In the lower panels of Fig.~\ref{fig:xSec_SM} we show the cross section for processes arising through channels $(a)$ and $(b)$ for ($h_i \to WW$) and ($A_1 \to t\bar{t}$) final states  as examples of promising search channels for heavier NMSSM Higgs bosons with $m_{h_i} \gtrsim 2 m_W$ or $m_{A_1} \gtrsim 2m_t$.

In addition to the visible decay channels discussed above, large values of NMSSM couplings $\lambda$ and $\kappa$ lead to significant branching ratios for $\left(h_i/A_1 \to \chi_1 \chi_1\right)$ decays much larger than what is possible in the MSSM. Such invisible decay modes lead to mono-Higgs and mono-$Z$ signatures from the diagrams in Fig.~\ref{fig:triHdiagrams}$(a)$ and Fig.~\ref{fig:triHdiagrams}$(b)$ respectively, and both mono-Higgs and mono-$Z$ signatures from the diagram in Fig.~\ref{fig:triHdiagrams}$(c)$. Note that the visible and invisible decay modes of the light Higgs bosons are complementary, since these branching ratios are mainly determined by which decay channels are kinematically open. We show the invisible cross sections arising from channels $(a)$ and $(b)$ in the $h_{\rm SM}/Z \chi_1 \chi_1$ final state in Fig.~\ref{fig:xSec_mono}. The corresponding production cross sections for these final states through channel $(c)$ are shown in Fig.~\ref{fig:xSec_neu}.

Excluding the $(h_{\rm SM}/Z/W \to {\rm final \ state})$ branching ratios, cross sections up to $\mathcal{O}(10\,{\rm pb})$ for the processes shown in Figs.~\ref{fig:xSec_SM},~\ref{fig:xSec_mono} and up to $\mathcal{O}(1\,{\rm pb})$ for those shown in Fig.~\ref{fig:xSec_neu} are possible at the LHC at $\sqrt{s} = 13\,$TeV, making these channels very promising for collider searches. The focus of this paper is the mono-Higgs signal, and we leave the investigation of the other channels outlined above to future work.

%%%%%%%%%%%%%%%%%%%%%%%%%%%%%%%%%%%%%%%%
\section{Mono-Higgs signatures} \label{sec:monoH}

The mono-Higgs signature was first proposed in Refs.~\cite{Petrov:2013nia, Carpenter:2013xra} to search for dark matter pair production in association with a SM-like Higgs boson. Those works considered simplified models with additional singlets or a $Z'$ boson as well as an effective field theory approach with $H h_{\rm SM} \chi_1 \chi_1$ contact interactions, and found that cross sections $\sigma(pp \rightarrow h_{\rm SM}\chi\chi) > \mathcal{O}(100\,{\rm fb})$ could be ruled out with $300\,$fb$^{-1}$ of data at the 14 TeV LHC \cite{Carpenter:2013xra}. Ref.~\cite{Carpenter:2013xra} found the $\gamma\gamma + {\rm missing \ transverse \ energy \ } (E_T^{\rm miss})$ final state to give the best reach (despite the small branching fraction of $h_{\rm SM}\rightarrow \gamma\gamma$) due to smaller backgrounds and well-measured objects allowing for good $E_T^{\rm miss}$ reconstruction, reporting projected exclusion limits typically one order of magnitude or more stronger than from $(h_{\rm SM} \to b\bar{b}, 4\ell, 2\ell 2j)$. We therefore focus on the $h_{\rm SM}\rightarrow \gamma\gamma$ mode for our analysis. See also \cite{Berlin:2014cfa, No:2015xqa, Basso:2015aee, Abdallah:2016vcn, Liew:2016oon, Bauer:2017ota} for phenomenological studies of the mono-Higgs signature.

The ATLAS collaboration has searched for mono-Higgs signatures in the $\gamma\gamma + E_T^{\rm miss}$ final state at $\sqrt{s} = 8\,$TeV \cite{Aad:2015yga} and at $\sqrt{s} = 13\,$TeV \cite{ATLAS-CONF-2016-011,ATLAS-CONF-2016-087}. The CMS collaboration has conducted a search for the same signature at $\sqrt{s} = 13\,$TeV \cite{CMS-PAS-EXO-16-011}. Searches for mono-Higgs in the $b\bar{b} + E_T^{\rm miss}$ final states have been carried out by the ATLAS collaboration at $\sqrt{s} = 8\,$TeV \cite{Aad:2015dva} and by both the ATLAS and CMS collaborations at $\sqrt{s} = 13\,$TeV \cite{ATLAS-CONF-2016-019, CMS-PAS-EXO-16-012}.

\begin{figure}
	\begin{center}
		\includegraphics[width=1\linewidth]{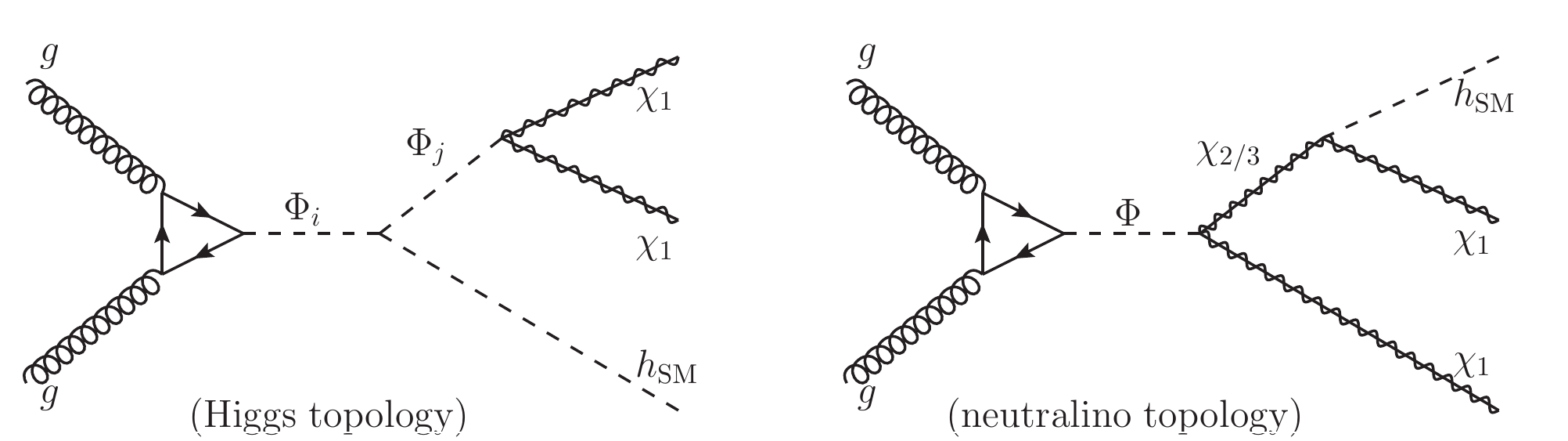}
		\caption{Illustration of the two channels we consider for producing mono-$h_{\rm SM}$ signatures in the NMSSM. For the left diagram~(the ``Higgs topology''), $\Phi_i = A_2$ and $\Phi_j = A_1$ or $\Phi_i = H_3$ and $\Phi_j = h_i$. For the right diagram (the ``neutralino topology''), $\Phi$ can be either $A_2$ or $H_3$.}
		\label{fig:triHNMSSM}
	\end{center}
\end{figure}

The decay chains that give mono-Higgs signatures in the NMSSM are shown in Fig.~\ref{fig:triHNMSSM}, where we label them as the ``Higgs topology" or the ``neutralino topology". The Higgs topology can be realized with $\Phi_i = A_2$, $\Phi_j = A_1$ or $\Phi_i = H_3$, $\Phi_j = h_i$. For the neutralino topology, $\Phi$ can be either $H_3$ or $A_2$, while the intermediate neutralino $\chi_j$ is $\chi_3$ for most of our points, since it is hard to realize mass splittings $(m_{\chi_2} - m_{\chi_1}) > m_{h_{\rm SM}} \approx 125\,$GeV required for the mono-Higgs signature, and the bino- and wino-like $\chi_4$ and $\chi_5$ are decoupled in our study. 
For our analysis, we run Monte Carlo (MC) simulations of these decay chains at the 13 TeV LHC. We use the event generator \texttt{MadGraph5\_v2.3.3} \cite{Alwall:2014hca} for the MC simulation of the hard event, \texttt{pythia6} \cite{Sjostrand:2006za} for hadronization, and \texttt{Delphes} \cite{deFavereau:2013fsa} for detector simulation. Following the analysis in Ref.~\cite{ATLAS-CONF-2016-087}, we employ the following cuts on our simulated signal events\footnote{Note that our MC configuration differs from that used in Ref.~\cite{ATLAS-CONF-2016-087}; in particular, we use the fast detector simulation \texttt{Delphes} (with the default \texttt{Delphes} card in \texttt{MadGraph5\_v2.3.3}) instead of a full detector simulation. This may have numerical impact, although beyond the scope of this work.}:
\begin{itemize}
	\item two photons with transverse momenta $p_T > 25\,$GeV and pseudorapidity $\left|\eta\right| < 2.37$, excluding the barrel-end cap transition region $1.37 < \left|\eta\right| < 1.52$,
	\item the invariant mass of the two-photon system satisfies $105\,{\rm GeV} < m_{\gamma\gamma} < 160\,$GeV,
	\item the (sub-)leading photon has $p_T^\gamma/m_{\gamma\gamma} > 0.35 \ (0.25)$.
\end{itemize}
To discriminate the signal from background events, Ref.~\cite{ATLAS-CONF-2016-087} uses the $E_T^{\rm miss}$ significance variable, defined as 
\be
S_{E_T^{\rm miss}} \equiv E_T^{\rm miss}/\sqrt{\sum E_T}, \label{eq:SMET}
\ee
where the sum in the denominator is the total transverse energy deposited in the calorimeters in the event. For our simulated event samples, we use the scalar sum of the $p_T$ of all visible objects from the \texttt{Delphes} output as a proxy for $\sum E_T$. 

\begin{figure}
	\includegraphics[width=.7\linewidth]{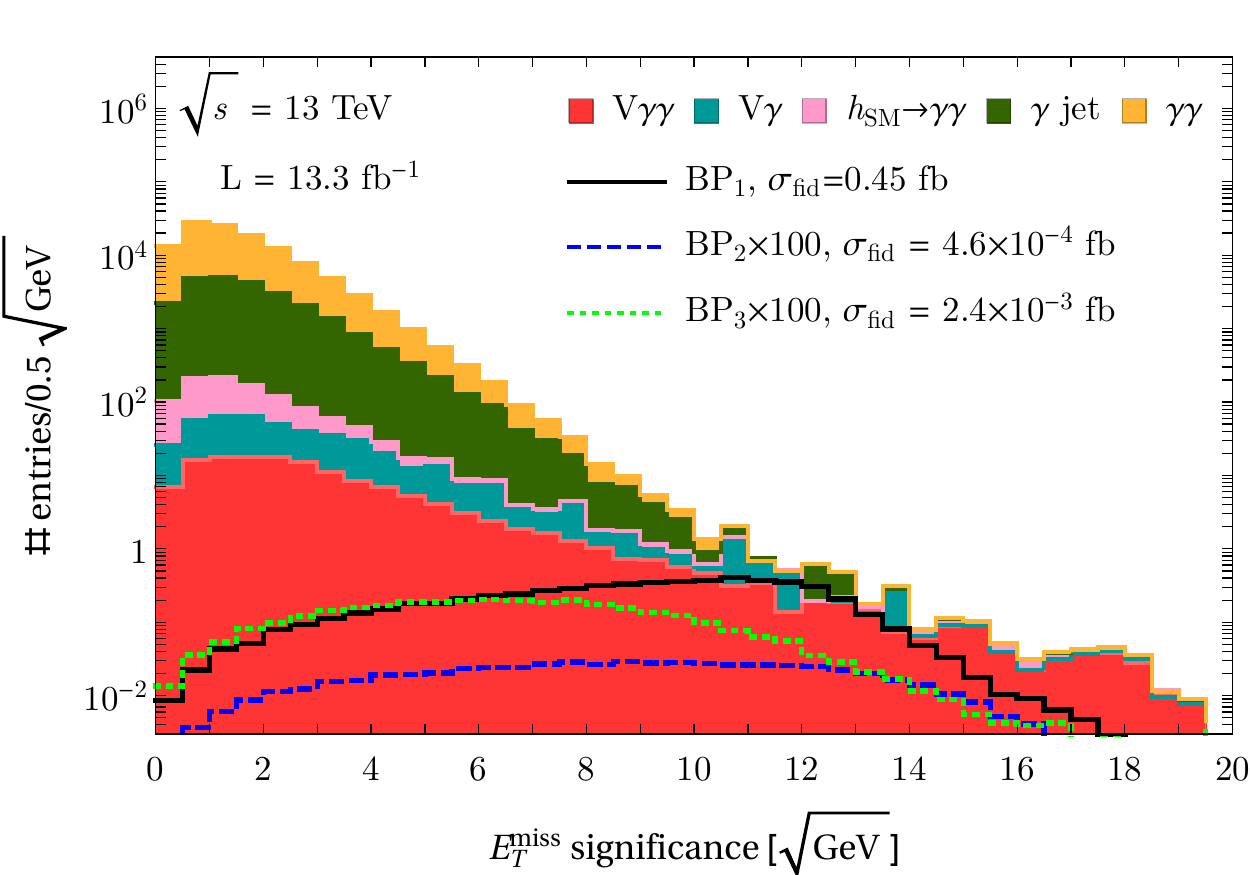}
	\caption{Simulated $E_T^{\rm miss}$ significance ($S_{E_T^{\rm miss}}$, Eq.~\eqref{eq:SMET}) distribution for three benchmark points (cf.\,Table\,\ref{tab:pointsForS_B_plots}) against background taken from Ref.\,\cite{ATLAS-CONF-2016-087} at $\sqrt{s} = 13\,$TeV and $L = 13.3\,{\rm fb}^{-1}$. The $S_{E_T^{\rm miss}}$ distributions for benchmark points BP$_2$ and BP$_3$ have been multiplied $\times 100$ for visibility. The stacked background-histogram shows different SM contributions indicated in the legend; see Ref.~\cite{ATLAS-CONF-2016-087} for details.}
	\label{fig:S_B_plots}
\end{figure}

\begin{table}
	\begin{center}
		\begin{tabular}{|c|c|c|c|}
		\hline
			& BP$_1$ & BP$_2$ & BP$_3$ \\ \hline 
			$\tan\beta$ & 2.17 & 2.16 & 2.24 \\ \hline 
			$\lambda$ & 0.60 & 0.55 & 0.55 \\ \hline 
			$\kappa$ & -0.38 & -0.33 & -0.45 \\ \hline 
			$A_\lambda$\,[GeV] & -554 & -859 & -539 \\ \hline 
			$A_\kappa$\,[GeV] & -254 & - 195 & -497 \\ \hline 
			$\mu$\,[GeV] & -144 & -222 &  -123 \\ \hline 
			$M_{Q_3}$\,[TeV] & 2.55 & 4.46 & 8.48 \\ \hline 
			$m_{h_{\rm SM}}$\,[GeV] & 122 & 123 & 126 \\ \hline 
			$m_{h_i}$\,[GeV] & 157 & 238 & 77.6 \\ \hline 
			$m_{H_3}$\,[GeV] & 421 & 650 & 390 \\ \hline 
			$m_{A_1}$\,[GeV] & 184 & 232 & 295 \\ \hline 
			$m_{A_2}$\,[GeV] & 457 & 669 & 464 \\ \hline 
			$m_{\chi_1}$\,[GeV] & 69.5 & 156 & 73.1 \\ \hline 
			$m_{\chi_2}$\,[GeV] & 158 & 238 & 139 \\ \hline  
			$m_{\chi_3}$\,[GeV] & 268 & 343 & 270 \\ \hline 
			BR($A_2 \to A_1 h_{\rm SM}$) & 18\,\% & 31\,\% & 0.10\,\% \\ \hline 
			BR($A_1 \to \chi_1 \chi_1$) & 99\,\% & -- & 69\,\% \\ \hline 
			BR($H_3 \to h_i h_{\rm SM}$) & 9.3\,\% & 5.0\,\% & 14\,\% \\ \hline
			BR($h_i \to \chi_1 \chi_1$) & 98\,\% & -- & -- \\ \hline
			BR($A_2 \to \chi_3 \chi_1$) & 0.71\,\% & 0.80\,\% & 0.34\,\% \\ \hline  
			BR($H_3 \to \chi_3 \chi_1$) & 0.57\,\% & 0.28\,\% & 1.1\,\% \\ \hline 
			BR($\chi_3 \to \chi_1 h_{\rm SM}$) & 3.2\,\% & 6.1\,\% & 11\,\% \\ \hline 
		\end{tabular}
		\caption{NMSSM parameters, mass spectra, and relevant branching ratios for the three benchmark points shown in Fig.~\ref{fig:S_B_plots}}
		\label{tab:pointsForS_B_plots}
	\end{center}
\end{table}

We show the $S_{E_T^{\rm miss}}$ distribution for three benchmark points from our scan in Fig.~\ref{fig:S_B_plots}; the associated NMSSM parameters and Higgs spectra are shown in Table~\ref{tab:pointsForS_B_plots}. Benchmark point BP$_1$ has a production cross section close to the background but peaks at a significantly high $S_{E_T^{\rm miss}}$, which makes it detectable at the LHC with $L = 300\,$fb$^{-1}$. The other benchmarks BP$_2$ and BP$_3$ have mono-Higgs cross sections too small to be detectable, and are shown to illustrate how points close in parameter space to BP$_1$ can fail to produce sizable mono-Higgs signals for various reasons. BP$_2$ has $2 m_{\chi_1} > m_{A_1}, m_{h_i}$, such that only the neutralino topology in  Fig.~\ref{fig:triHNMSSM} can be realized. Similarly, BP$_3$ has small branching ratios into $(h_{\rm SM} + A_1)$ and $(h_{\rm SM} + h_i)$ due to its coupling parameters and phase space suppression, hence mono-Higgs signatures are again dominantly produced via $(A_2 + H_3 \to \chi_1 \chi_3 \to \chi_1 \chi_1 h_{\rm SM})$. The decay chain from the neutralino topology generically results in smaller $E_T^{\rm miss}$ and consequently softer $S_{E_T^{\rm miss}}$ distributions, as the visible $h_{\rm SM}$ is produced via a secondary decay (in contrast with the Higgs topology from Fig.~\ref{fig:triHNMSSM}, where the visible and invisible Higgs bosons are produced back to back at the primary vertex). When combined with the smaller cross sections, this puts BP$_2$ and BP$_3$ out of reach of the LHC. 

We evaluate the reach of mono-Higgs searches with $\gamma\gamma + E_{T}^{\rm miss}$ final state for a range of the involved masses, $\{m_{\Phi_2}, m_{\Phi_1}, m_{\chi_1}\}$ for the Higgs topology and $\{m_{\Phi}, m_{\chi_3}, m_{\chi_1}\}$ for the neutralino topology, at the 13 TeV LHC with $300\,{\rm fb}^{-1}$ of data. We compare the simulated signals to the background taken from Ref.~\cite{ATLAS-CONF-2016-087} scaled up to $300\,{\rm fb}^{-1}$ of data. For each combination of involved masses, we optimize the $S_{E_T^{\rm miss}}$ cut and find the minimal detectable cross section $\sigma_{\rm min}$, where we define detectability as $S>5$ and $S/\sqrt{B+\Delta^2 B^2} > 2$, with $S$ and $B$ the number of signal and background events after the relevant cuts. In the latter condition, the two terms in the denominator represent the statistical and systematic uncertainties respectively in the background. We assume a systematic uncertainty of $10\%$, $\Delta=0.1$. We find that the optimal $S_{E_T^{\rm miss}}$ cut is typically $S_{E_T^{\rm miss}} \gtrsim 10\,\sqrt{\rm GeV}$ for small splittings of the involved masses, increasing to $S_{E_T^{\rm miss}} \gtrsim 15\,\sqrt{\rm GeV}$ for the largest mass splittings considered.

\begin{figure}
	\begin{center}
		\includegraphics[width=.5\linewidth]{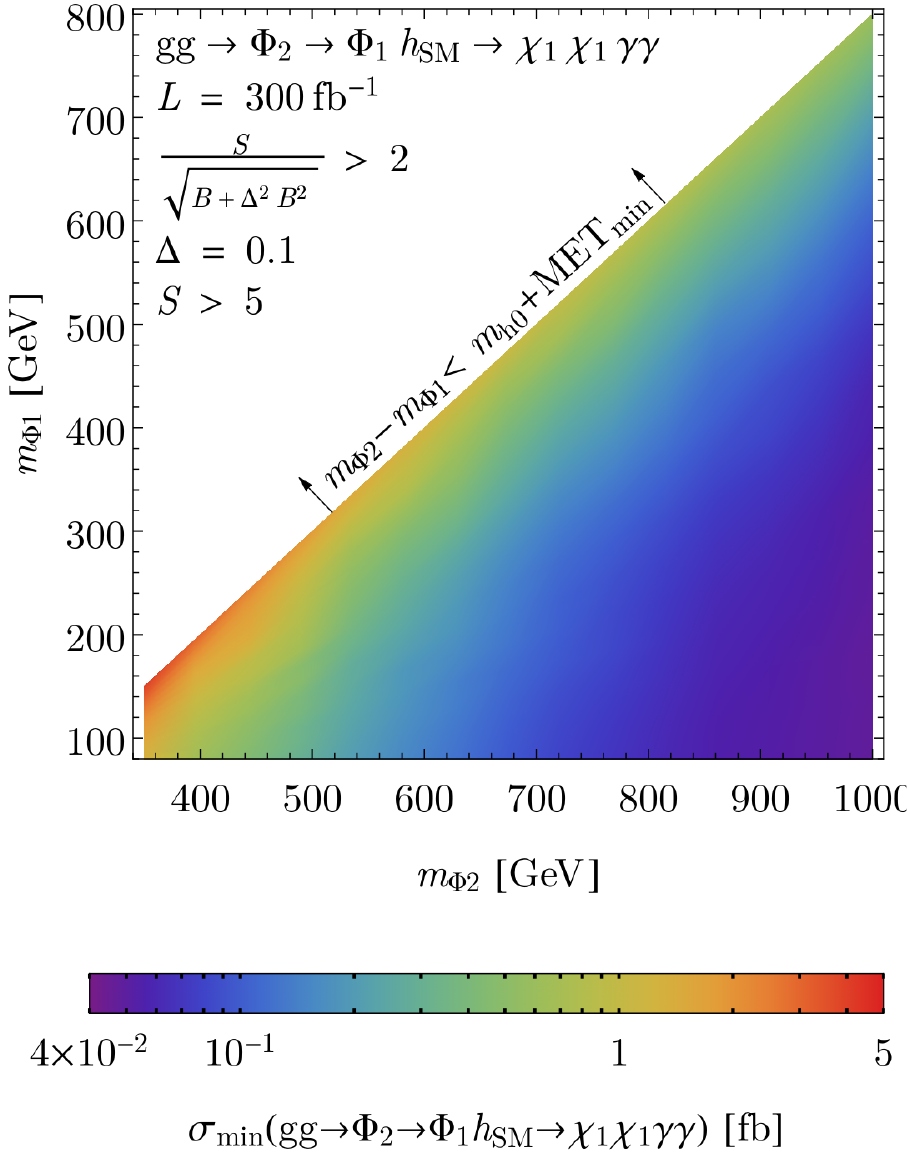}
		\caption{Mono-Higgs reach for the Higgs topology from Fig.~\ref{fig:triHNMSSM} with $\gamma\gamma+E_T^{\rm miss}$ final state for $\sqrt{s} = 13\,$TeV and $L = 300\,{\rm fb}^{-1}$, search criteria as indicated in the legend and discussed in the text, and assuming the backgrounds from Ref.~\cite{ATLAS-CONF-2016-087} rescaled to $L = 300\,{\rm fb}^{-1}$. The color coding shows the minimum signal cross section $\sigma_{\rm min}(gg \to \Phi_2 \to \Phi_1 h_{\rm SM} \to \chi_1\chi_1 \gamma\gamma)$ to which we project the LHC to be sensitive for decay topologies similar to the Higgs topology from Fig.~\ref{fig:triHNMSSM}. These results can be used to estimate the reach for any model with similar decay topologies by computing the corresponding signal cross section. In the NMSSM, $\Phi_2 = A_2$ and $\Phi_1 = A_1$ or $\Phi_2 = H_3$ and $\Phi_1 = h_i$. We use $m_{h_{\rm SM}} = m_{h_0} = 125\,$GeV and MET$_{\rm min} \equiv \left(E_T^{\rm miss}\right)_{\rm min} = 75\,$GeV.}
		\label{fig:reach_Higgs bosons}
	\end{center}
\end{figure}

\begin{figure}
	\begin{center}
		\includegraphics[width=1\linewidth]{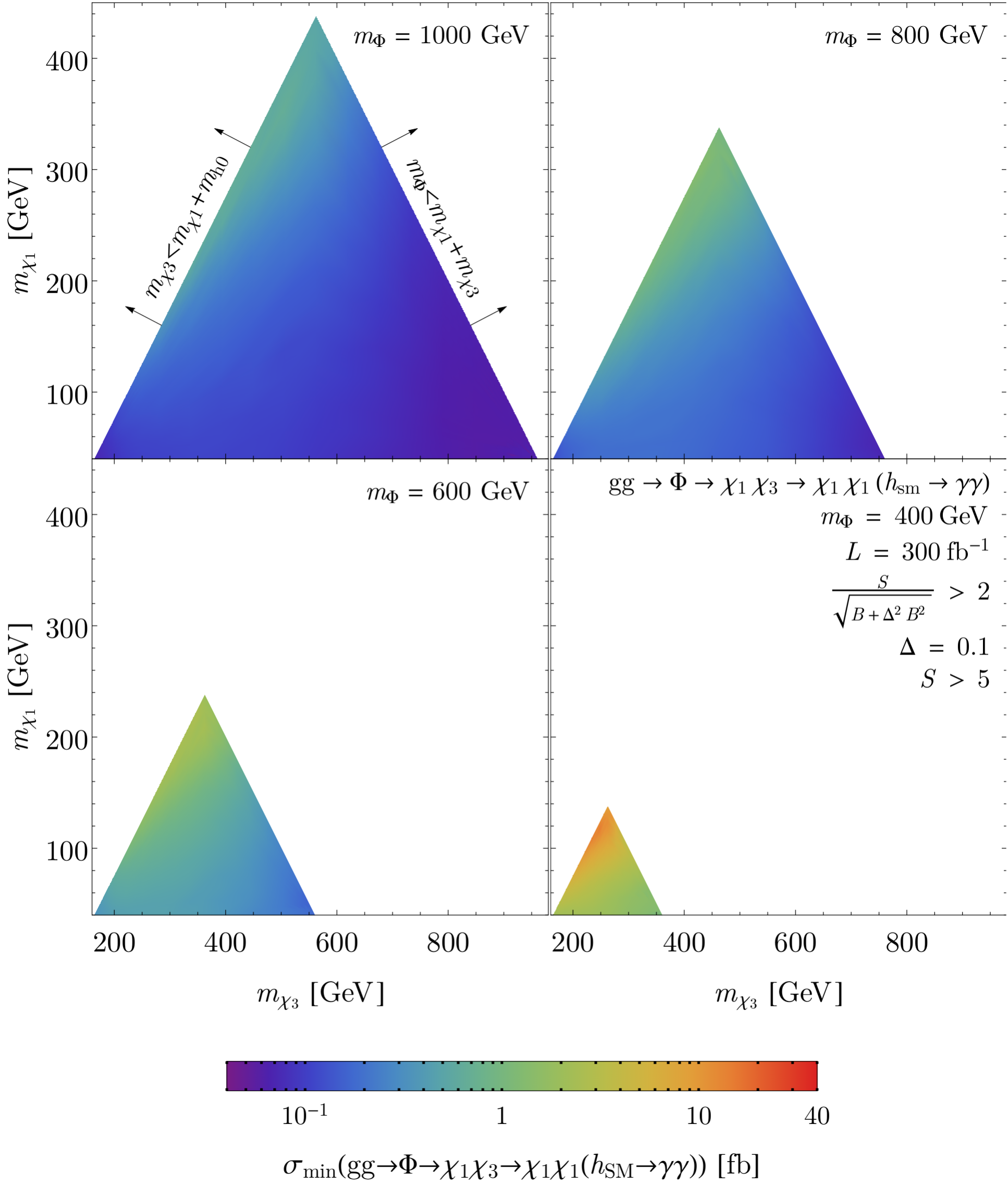}
		\caption{Same as Fig.~\ref{fig:reach_Higgs bosons} but for the neutralino topology from Fig.~\ref{fig:triHNMSSM}. The color coding shows $\sigma_{\rm min}(gg \to \Phi \to \chi_1 \chi_3 \to \chi_1 \chi_1 (h_{\rm SM} \to \gamma\gamma))$. The different panels are for different values of $m_\Phi$ indicated in the respective panel. In the NMSSM, $\Phi$ can  be either $A_2$ or $H_3$, and $m_{h_{\rm SM}} = m_{h_0} = 125\,$GeV. Note that the color scale is different from that of Fig.~\ref{fig:reach_Higgs bosons}.}
		\label{fig:reach_neutralinos}
	\end{center}
\end{figure}

The resulting reach is shown in Fig.~\ref{fig:reach_Higgs bosons} for the Higgs topology and in Fig.~\ref{fig:reach_neutralinos} for the neutralino topology, where we portray our results as color coding for the minimum signal cross section that satisfies the above criteria. We present our results in this manner so that they can be used to interpret the reach for any model with topologies similar to those shown in Fig.~\ref{fig:triHNMSSM}, and not just the NMSSM.  

For the Higgs topology $\left(gg \to \Phi_2 \to \Phi_1 h_{\rm SM} \to \chi_1 \chi_1 \gamma \gamma\right)$, the reach depends primarily on $m_{\Phi_2}$, which controls the overall energy scale, and the mass splitting $[m_{\Phi_2} - (m_{\Phi_1} + m_{h_ {\rm SM}})]$, which sets the maximal $E_T^{\rm miss}$ in the process. In Fig.~\ref{fig:reach_Higgs bosons} we present results for mass spectra that satisfy $300\,{\rm GeV} \leq m_{\Phi_2} \leq 1\,$TeV, $[m_{\Phi_2} - (m_{\Phi_1} + m_{h_{\rm SM}})] \geq 25\,$GeV, and $m_{\Phi_1} \geq 2 m_{\chi_1}$. We see that cross sections as low as $4\times 10^{-2}\,$fb can be probed. The reach $\sigma_{\rm min}(gg \to \chi_1 \chi_1 h_{\rm SM}) \lesssim 100\,$fb is maintained over most of the parameter space where $(m_{\Phi_2} - m_{\Phi_1}) > 300\,$GeV, and deteriorates sharply as this mass splitting decreases. 

For the neutralino topology $\left(gg \to \Phi \to \chi_1 \chi_3 \to \chi_1 \chi_1 h_{\rm SM}\right)$, the reach shown in Fig.~\ref{fig:reach_neutralinos} depends on $m_\Phi$, as it controls the overall energy scale, and the mass splittings at the two vertices, $[m_{\Phi} - (m_{\chi_1} + m_{\chi_3})]$ and $[m_{\chi_3} - (m_{\chi_1} + m_{h_{\rm SM}})]$. Since three independent masses are involved, we show our results in the $(m_{\chi_1}-m_{\chi_3})$ plane for several values of $m_{\Phi}$, for mass spectra satisfying $300\,{\rm GeV} \leq m_\Phi \leq 1\,$TeV, $[m_{\Phi} - (m_{\chi_1} + m_{\chi_3})] \geq 50\,$GeV, and $m_{\chi_3} \geq (m_{\chi_1} + m_{h_{\rm SM}})$. The reach is more sensitive to the $(m_{\chi_3} - m_{\chi_1})$ mass splitting than to the $[m_{\Phi} - (m_{\chi_1} + m_{\chi_3})]$ splitting, and is considerably weaker than the corresponding reach for the Higgs topology (Fig.~\ref{fig:reach_Higgs bosons}), where the 125\,GeV Higgs is produced at the primary decay vertex and back-to-back with missing energy. 

Overall, our computed reach is roughly one order of magnitude better than that estimated in Ref.~\cite{Carpenter:2013xra} which used simplified models and contact interaction terms. The reach is significantly augmented by the SM-like Higgs and the $E_T^{\rm miss}$ being produced back-to-back in the Higgs topology in addition to the improved background modeling provided by the ATLAS collaboration.

\section{NMSSM Interpretation} \label{sec:NMSSMinterpretation}

In this section, we apply the results from Figs.~\ref{fig:reach_Higgs bosons} and \ref{fig:reach_neutralinos} to our \texttt{NMSSMTools} scan, and study the implications for the NMSSM parameter space. In all the plots in this section, we show the reach for both 300 and 3000 fb$^{-1}$ of data at the 13 TeV LHC using the mono-Higgs signal, with points color coded to show the cross section in terms of the corresponding minimum cross section $\sigma_{\rm min}$ needed to probe the signal as discussed in the previous section. In all plots, the color spectrum legend spans $10^{-2}\leq \sigma/\sigma_{\rm min} \leq 1$ going from light grey to black, but it should be understood that light grey points include $\sigma/\sigma_{\rm min}\leq 10^{-2}$ and black points include $\sigma/\sigma_{\rm min}\geq 1$. 

\begin{figure}
	\begin{center}
		\includegraphics[width=.49\linewidth]{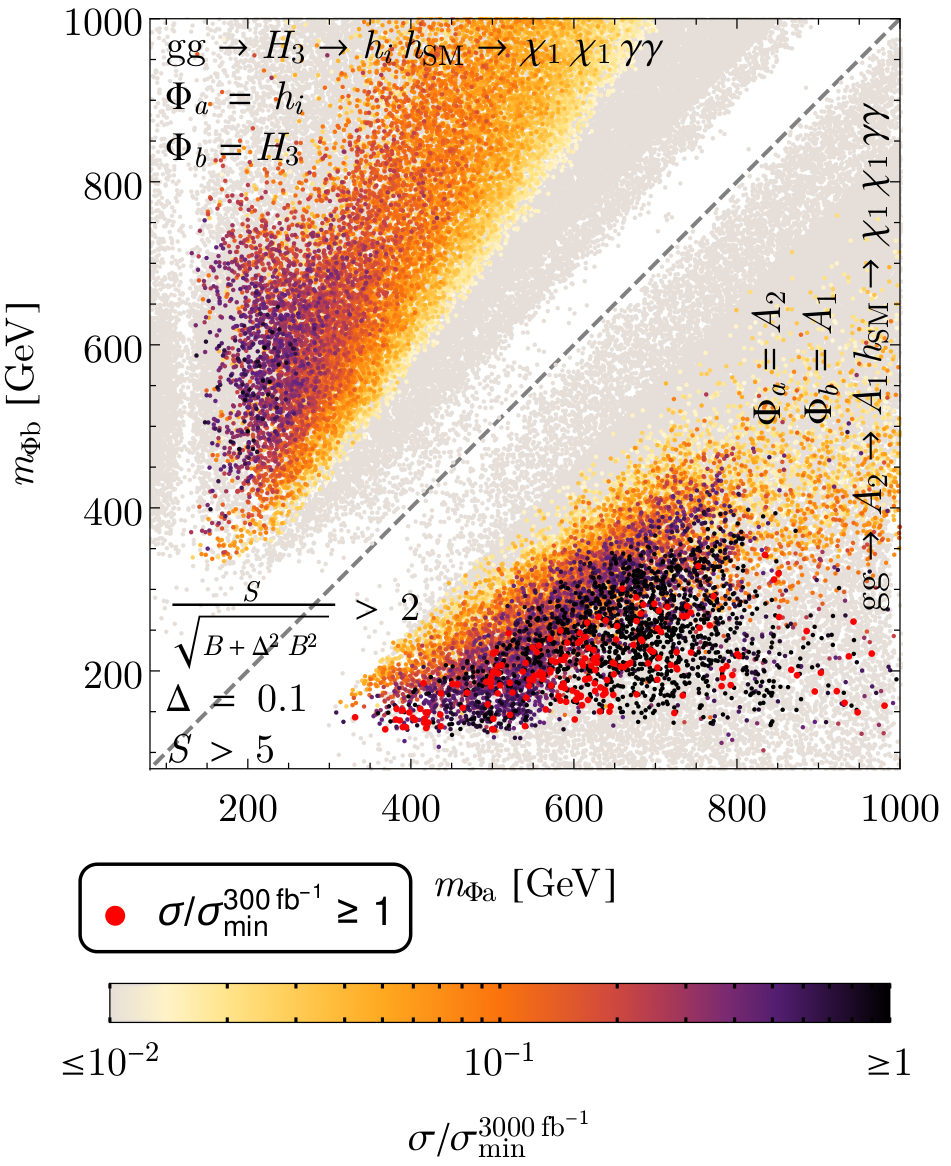}
		\includegraphics[width=.49\linewidth]{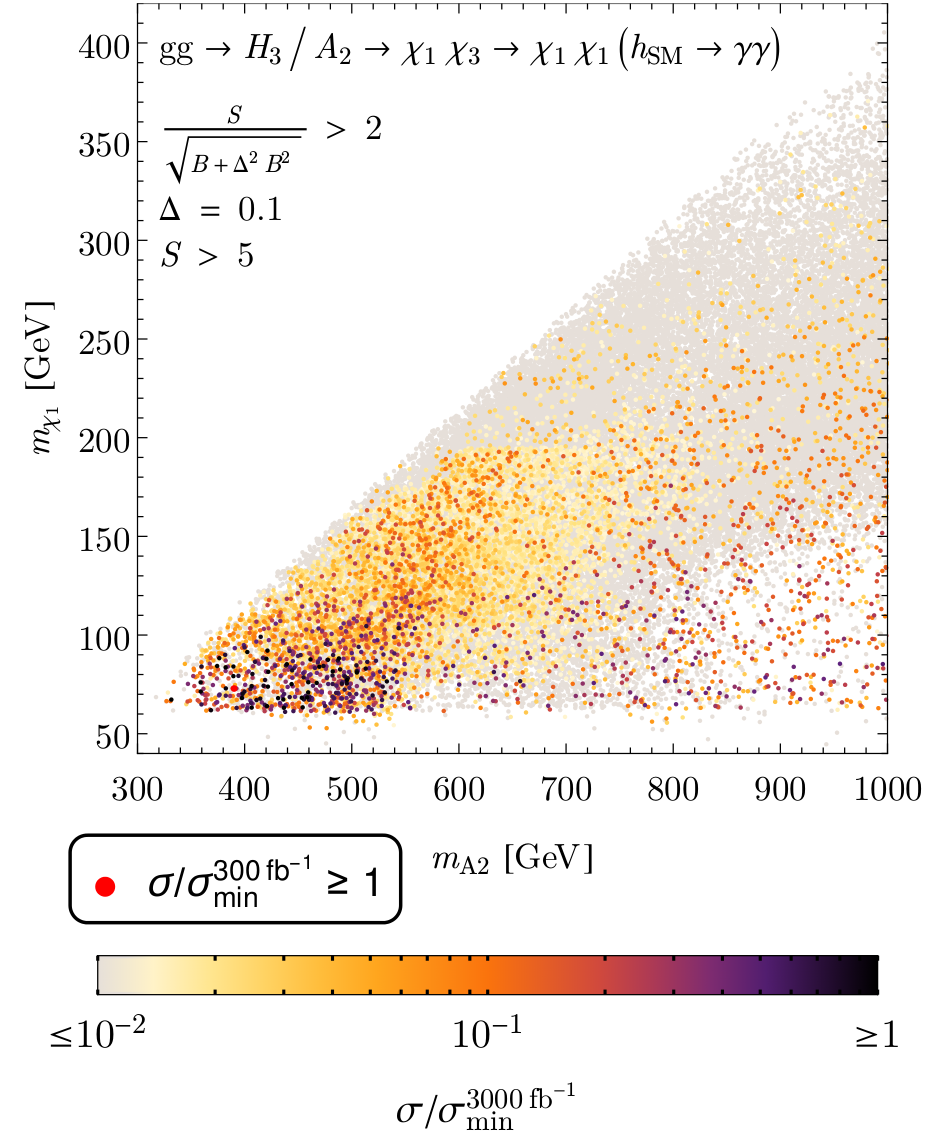}
		\caption{Distribution of scan points within reach of the LHC for the Higgs topology (left panel) and neutralino topology from Fig.~\ref{fig:triHNMSSM} (right panel). The color coding shows the cross sections in terms of the reach $\sigma_{\rm min}^{3000\,{\rm fb}^-1}$ of the LHC at $\sqrt{s} = 13\,$TeV and $L = 3000\,$fb$^{-1}$. Thick red dots indicate points within reach with $L = 300\,$fb$^{-1}$. See the text for definitions of the reach.}
		\label{fig:detectable_points}
	\end{center}
\end{figure}

In Fig.~\ref{fig:detectable_points}, we show the distribution of points from our \texttt{NMSSMTools} scans accessible to the LHC. The left and right panels show the results for the Higgs and neutralino topologies respectively. In the left panel, we show the reach via $\left(gg \to H_3 \to h_i h_{\rm SM}\right)$ and $\left(gg \to A_{2} \to A_1 h_{\rm SM}\right)$ separately in the upper and lower triangles since the mass splitting $[m_{\Phi_2} - (m_{\Phi_1} + m_{h_{\rm SM}})]$, which affects the $S_{E_T^{\rm miss}}$ distribution, is generally different for the two decays. The results for the neutralino topology are shown in the right panel in the $m_{A2}$ vs. $m_{\chi_1}$ plane. The reach in this case depends on the heavy Higgs boson mass  as well as the two neutralinos it decays into. However, as noted previously, the reach is much less sensitive to $m_{\chi_3}$ than to $m_{\chi_1}$. Additionally, since $H_3$ and $A_2$ are mostly mass degenerate, we combine the contributions from $\left(gg \to H_3 \to \chi_1 \chi_3\right)$ and $\left(gg \to A_2 \to \chi_1 \chi_3\right)$. Specifically, we add the two contributions if $|m_{A_2} - m_{H_3}|/m_{A_2} \leq 30\,\%$, which is the case for most of the points, else we use the channel with the larger cross section in terms of the reach. 

\begin{figure}
	\begin{center}
		\includegraphics[width = .49\linewidth]{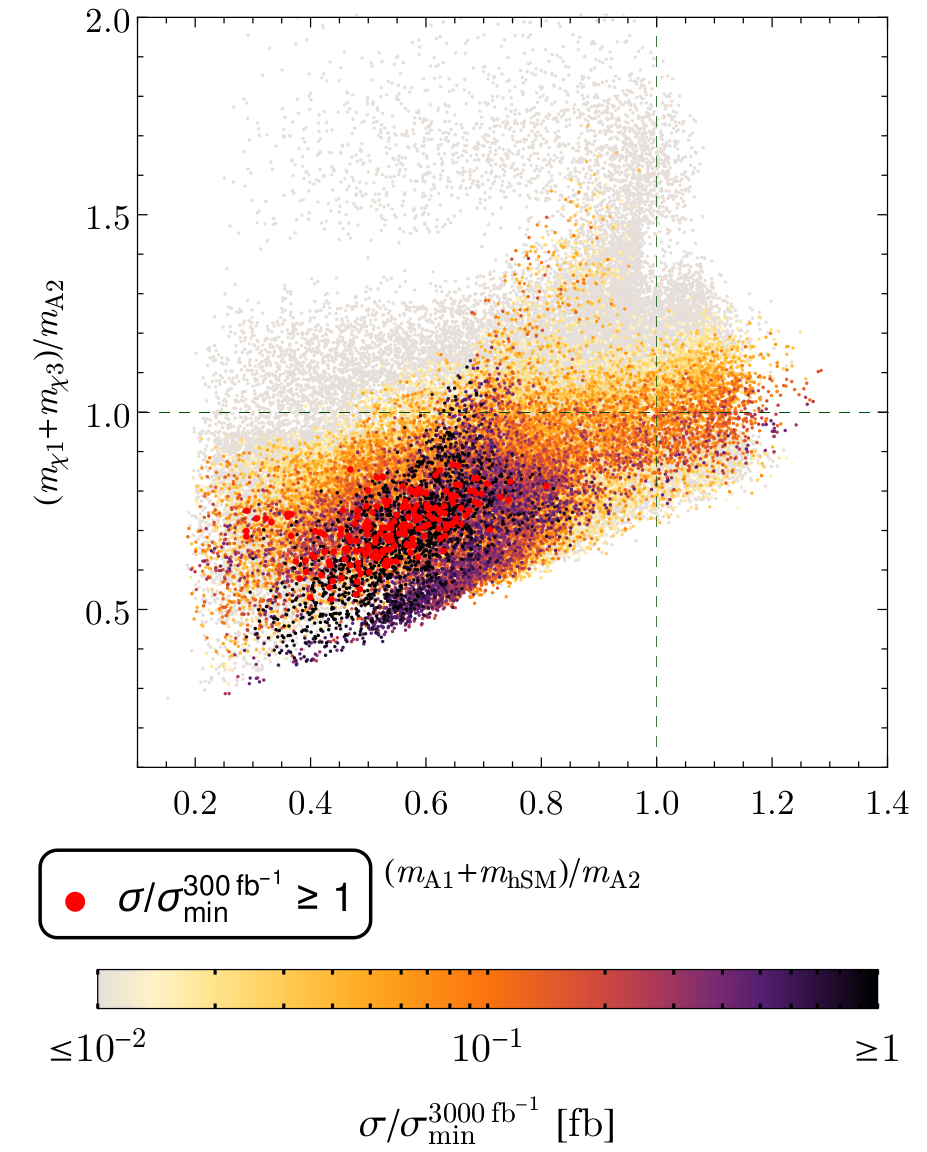}
		\includegraphics[width = .49\linewidth]{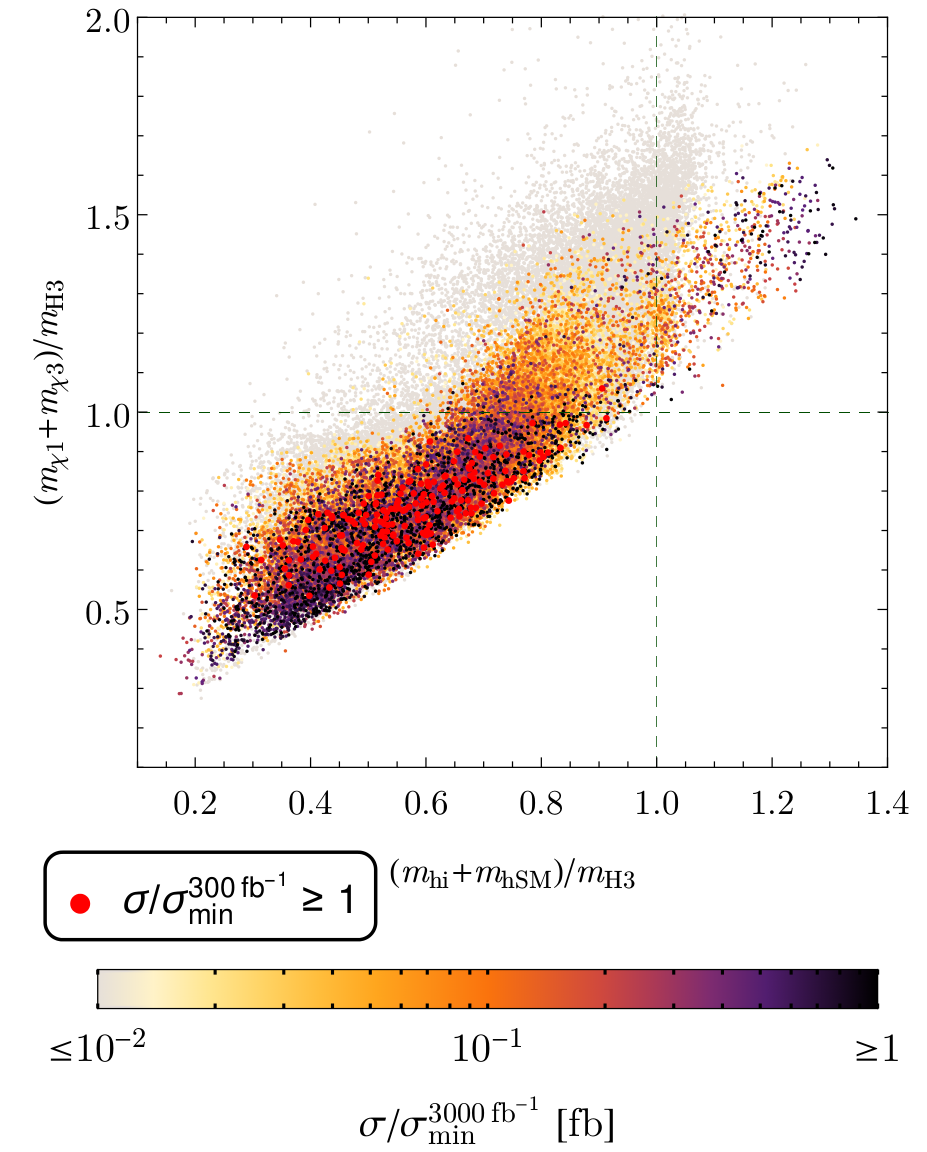}
		\caption{Relevant mass splittings for the two mono-Higgs topologies in the NMSSM for $A_2$ (left panel) and $H_3$ (right panel). Decays are kinematically allowed, if the ratio of the sum of masses of the decay products to the parent particle is smaller than 1 (indicated by the thin dashed lines). Thick red dots are within reach of the LHC at $300\,$fb$^{-1}$. The color coding shows the cross section in terms of the reach of LHC at $\sqrt{s} = 13\,$TeV and $L = 3000\,$fb$^{-1}$, where we use the best channel for each point. }
		\label{fig:detectable_massSplittings}
	\end{center}
\end{figure}

\begin{figure}
	\begin{center}
		\includegraphics[width=.49\linewidth]{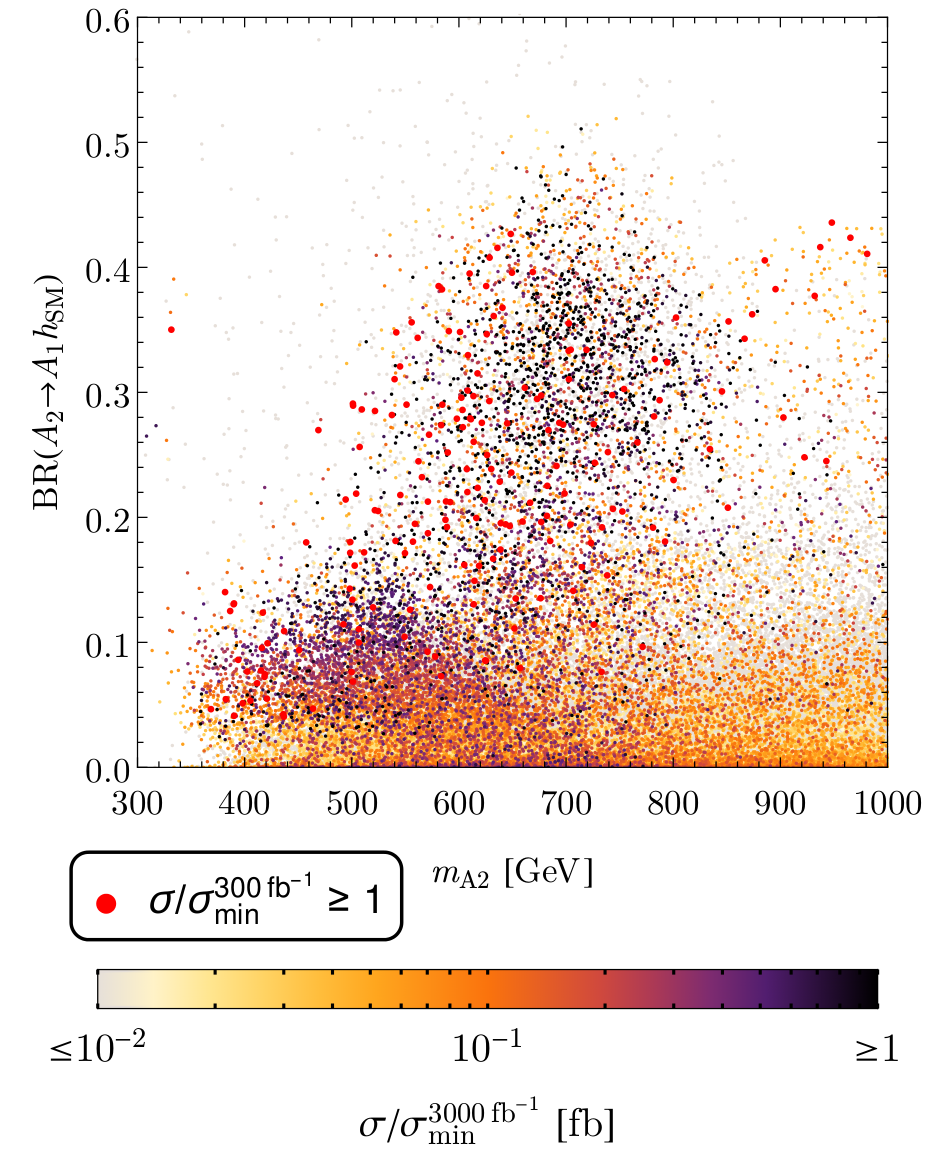}
		\includegraphics[width=.49\linewidth]{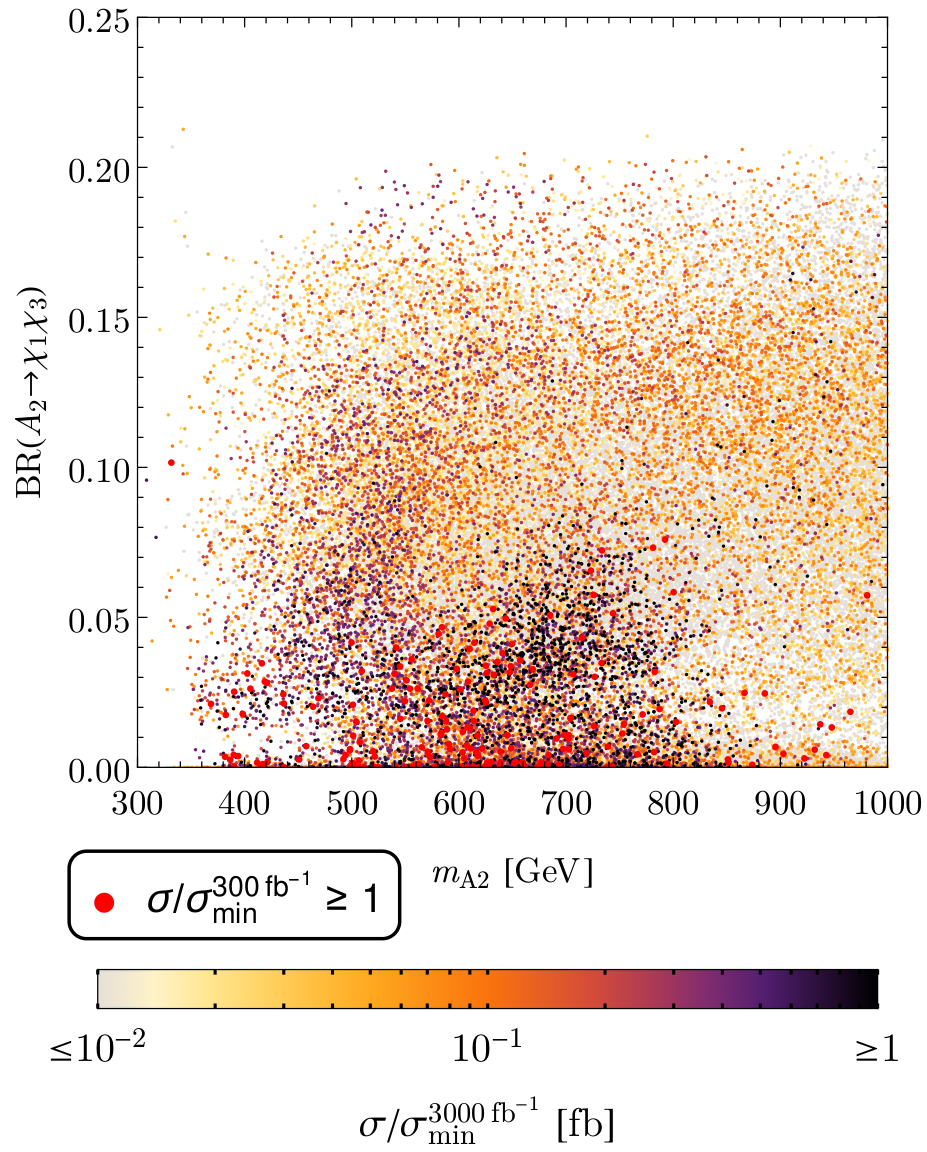}
		\caption{Branching ratios BR$(A_2 \to A_2 h_{\rm SM})$ (left panel) and BR$(A_2 \to \chi_1 \chi_3)$ (right panel) giving rise to mono-Higgs signals in the Higgs and neutralino topologies from $s$-channel production of an $A_2$ Higgs boson. Note that the scales on the $y$-axes differ between the panels. The color coding is the same as in Fig.~\ref{fig:detectable_massSplittings}. The branching ratios BR$(H_3 \to h_i h_{\rm SM})$ and BR$(H_3 \to \chi_1 \chi_3)$ are similar; hence, we do not plot them separately. }
		\label{fig:detectable_BRs}
	\end{center}
\end{figure}

Fig.~\ref{fig:detectable_points} shows that the Higgs topology $\left(gg \to \Phi_2 \to \Phi_1 h_{\rm SM}\right)$ (left panel) has significantly better prospects of being observed at the LHC than the neutralino topology $(gg \to \Phi \to \chi_1 \chi_3 \to \chi_1 \chi_1 h_{\rm SM})$ (right panel), as the SM-like Higgs is produced from the primary decay of $\Phi_2$, and back to back with the invisibly decaying particle. Decays of the pseudoscalar $A_2$ (left panel, lower triangle) are more promising than those of the scalar $H_3$ (left panel, upper triangle), as the gluon fusion production cross section for the former can be approximately a factor of 2 larger than that of the latter at the same mass. Furthermore, $A_1$ has a larger branching ratios into neutralinos than $h_i$ because the decay of $A_1$ into pairs of vector bosons is forbidden at tree level. For $\left(gg \to A_2 \to A_1 h_{\rm SM}\right)$, we find that TeV scale pseudoscalars can already be probed at the 13\,TeV LHC with 300\,fb$^{-1}$ of data. We also find that a significant part of parameter space has mono-Higgs cross sections within $\mathcal{O}(1)$ of the LHC reach with 3000\,fb$^{-1}$ of data, implying that improved search strategies or an improvement in background rejection can render them accessible. In the remainder of this paper, we combine the reach from all topologies, such that the color coding for each point shows the more promising of the two reaches in the Higgs or neutralino topology from Fig.~\ref{fig:triHNMSSM}.

In Fig.~\ref{fig:detectable_massSplittings}, we show the relevant mass splitting ratios for the primary decays of $A_2$ (left panel) and $H_3$ (right panel) Higgs bosons. The $x$-axis corresponds to the respective decay mode giving rise to the Higgs topology, while the $y$-axis corresponds to the decay mode of the neutralino topology. Decays are kinematically allowed if the ratio of the sum of masses of the decay products to the parent particle is smaller than 1. We observe that the relevant mass splittings for the $H_3$ decays into the two channels (right panel) are more correlated than the corresponding ones for the $A_2$ decays (left panel); cf. the discussion of the correlation of Higgs and neutralino masses in Section \ref{sec:Hcouplings}. Recalling the sizable couplings between the Higgs bosons and between Higgs bosons and neutralinos, we find that large branching ratios into these channels are indeed generic in the currently allowed NMSSM parameter space. In Fig.~\ref{fig:detectable_BRs} we show the branching ratios of the CP-odd Higgs boson $A_2$ giving rise to the Higgs topology (left panel) and the neutralino topology (right panel); branching ratios for $H_3$ are similar, hence we do not plot them separately. These plots together illustrate that the most promising points (large red dots) are driven by large mass splittings $[m_{A_2} - (m_{A_1} + m_{h_{\rm SM}})]$, giving rise to significant branching ratios in the $(A_2 \to A_1 h_{\rm SM})$ mode and large $E_T^{\rm miss}$.

\begin{figure}
	\begin{center}
		\includegraphics[width=.47\linewidth]{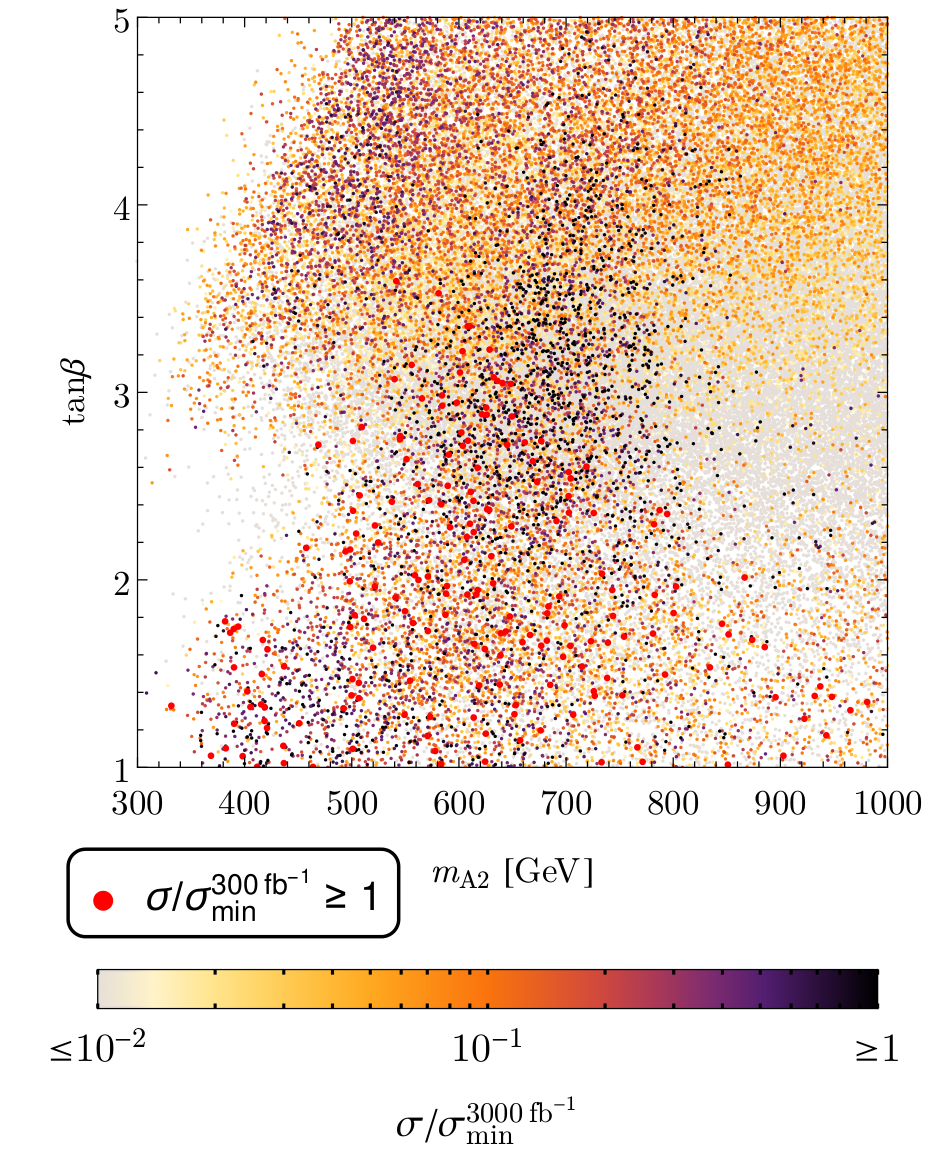}
		\caption{Potentially detectable points in the $m_{A_2} - \tan\beta$ plane.}
		\label{fig:detectable_ma-tb}
	\vskip1cm
		\includegraphics[width = .47\linewidth]{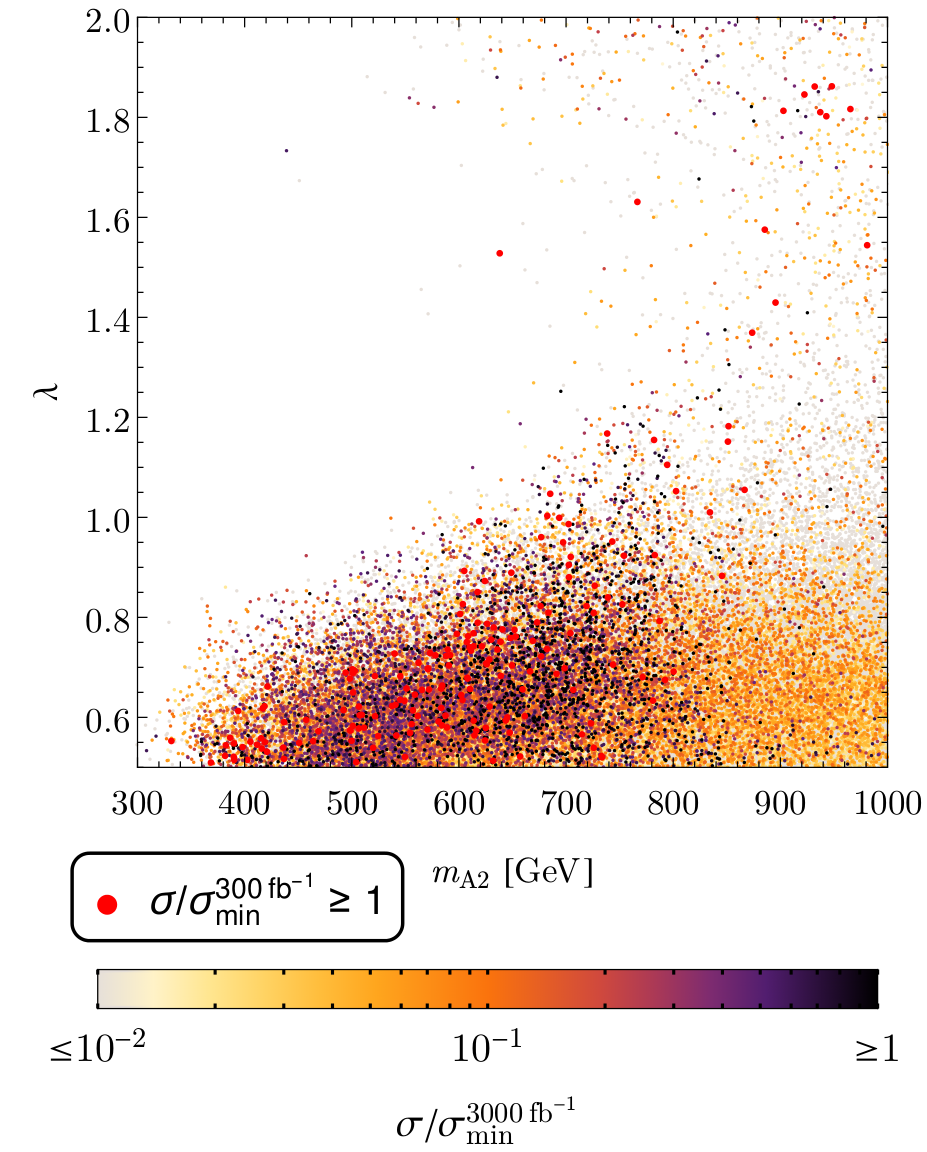}
		\includegraphics[width = .47\linewidth]{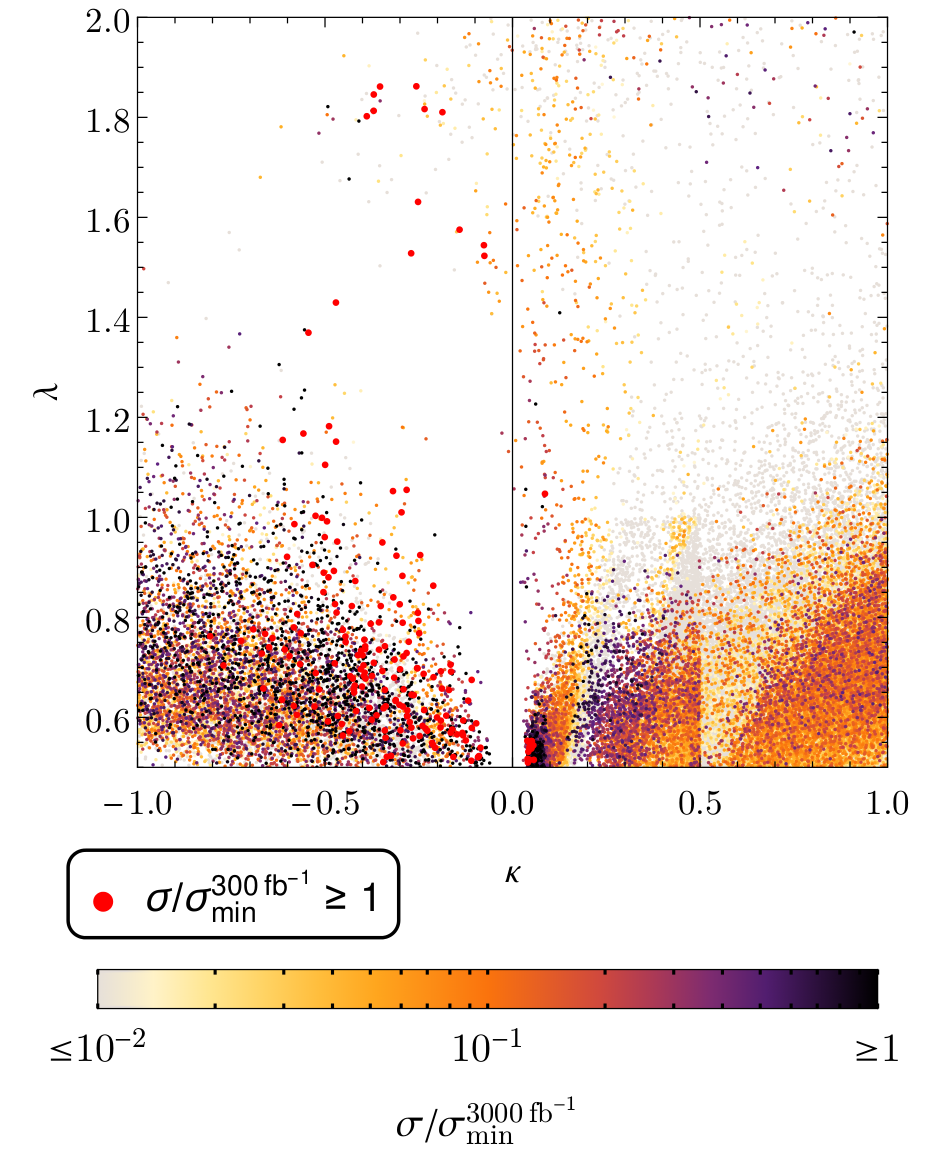}
		\caption{Potentially detectable points in the in the $m_{A_2}-\lambda$ (left) and $\kappa - \lambda$ plane (right).}
		\label{fig:detectable_ma-lam_lam-kap}
	\end{center}
\end{figure}

For comparison with the MSSM heavy Higgs searches, Fig.~\ref{fig:detectable_ma-tb} shows the distribution of points in the traditional $m_{A_2} - \tan\beta$ plane. Note that the most promising points populate the large $m_{A_2} (> 2m_t)$ and small $\tan\beta$ region, which is traditionally dominated by $t\bar{t}$ decays and therefore difficult to probe. The mono-Higgs channel provides a particularly clean and powerful probe of this theoretically well-motivated region of the NMSSM parameter space. 

It is instructive to plot the distribution of points in terms of the NMSSM couplings $\lambda$ and $\kappa$; these are shown in Fig.~\ref{fig:detectable_ma-lam_lam-kap}. The left panel shows the distribution in the $\lambda-m_{A_2}$ plane. The most promising points are clustered around $\lambda\sim 0.65, m_{A_2}\sim$ few hundred GeV, consistent with the discussion about alignment without decoupling in Section \ref{sec:Hcouplings}. Beyond this cluster, one can also see that there are points closer to $m_{A_2}\sim 1\,$TeV that are still promising, because they feature large values $\lambda\gsim 1.2$, representing a qualitatively different region of parameter space that is consistent with a 125\,GeV Higgs and other collider constraints. The right panel in Fig.~\ref{fig:detectable_ma-lam_lam-kap} shows the interplay between $\lambda$ and $\kappa$, where one sees a preference for negative values of $\kappa$ for points that are promising for LHC searches. This can be understood from noting that this choice of sign leads to large $(H^{\rm SM} A^{\rm NSM} A^S)$ couplings, cf. Table~\ref{tab:triH}. 

%%%%%%%%%%%%%%%%%%%%%%%%%%%%%%%%%%%%
\section{Conclusions} \label{sec:conclusions}
In this paper, we have studied LHC probes of the Higgs sector of the NMSSM, focusing on phenomenology arising from a large coupling $\lambda$ between the NMSSM singlet and the Higgs doublets, which is characteristic of the region of parameter space most compatible with a 125\,GeV SM-like Higgs boson, null results for sparticle searches, and naturalness considerations. We have considered the range $0.5\leq\lambda\leq 2$, which includes both the $\lambda\sim 0.65$ region favored by alignment without decoupling as well as the $\lambda\gsim 1$ values favored by natural electroweak symmetry breaking with a heavier supersymmetric scale. Such large values of $\lambda$ lead to large couplings among the CP-even and CP-odd Higgs boson as well as between the Higgs bosons and (Higgsinos and singlino) neutralinos. These couplings can provide the dominant decay modes of heavy Higgs bosons, significantly modifying collider phenomenology and evading the LHC bounds on heavy Higgs bosons as interpreted in the MSSM framework.

We performed a parameter scan of the NMSSM (Section \ref{sec:scan}), demanding a 125\,GeV Higgs boson with couplings compatible with LHC measurements. As expected, this picks out the region of parameter space favored by alignment. We then subject this data set to a number of direct Higgs searches at the LHC (Section \ref{sec:lhcconstraints}). Points evading LHC bounds generically show large branching ratios into decay modes beyond the MSSM, calling for NMSSM-specific search strategies to target these regions (Section\,\ref{sec:NMSSMsearch}). While various signatures are possible, we focused on the mono-Higgs channel in the $\left({h_{\rm SM} \to \gamma\gamma}\right) + E_T^{\rm miss}$ final state in this paper. 

We studied the prospects of probing topologies yielding mono-Higgs signatures at the 13 TeV LHC with $300\,$fb$^{-1}$ and $3000\,$fb$^{-1}$ of data. We present out reach results in Figs.~\ref{fig:reach_Higgs bosons},~\ref{fig:reach_neutralinos} as a function of the relevant masses in the decay topologies in Fig.~\ref{fig:triHNMSSM}; these results can be directly applied to any model that allows for such decay chains by comparing our projected reach to the corresponding signal cross section in the model. In the NMSSM (Section\,\ref{sec:NMSSMinterpretation}), we found that $300\,$fb$^{-1}$ of data can probe up to TeV scale heavy Higgs bosons, and a significantly larger region of parameter space becomes accessible with $3000\,$fb$^{-1}$. In particular, this search strategy remains effective even in the heavy  $m_{A_2} (> 2\,m_t$), low tan$\beta$ regions usually overwhelmed by $t\bar{t}$ decays (Fig.~\ref{fig:detectable_ma-tb}). In addition, we have also provided an NMSSM benchmark point BP$_1$ (Table\,\ref{tab:pointsForS_B_plots}) for further study.

These results show that the mono-Higgs channel is a powerful search strategy for heavy Higgs bosons in the currently well-motivated regions of the NMSSM, and more careful treatment, both theoretical and experimental, of such signatures is crucial for discovering the NMSSM Higgs bosons at future runs of the LHC. These results can be complemented and enhanced with studies in several directions. As pointed out in Section\,\ref{sec:NMSSMsearch}, several other final states, such as mono-$Z$ or $h_{\rm SM} b\bar{b}$, can provide complementary coverage of the NMSSM parameter space. Furthermore, heavy Higgs decays can be the dominant source of neutralino and chargino production at the LHC in certain regions of parameter space, augmenting the reach from direct searches for these particles. Likewise, focusing on regions of parameter space that contain viable dark matter candidates can also sharpen the expected signatures at various detectors. We leave such directions of study for future work.

\acknowledgements{
KF and SB acknowledge support from the Swedish Research Council (Vetenskapsr\r{a}det) through the Oskar Klein Centre (Contract No. 638-2013-8993). KF and BS acknowledge support from DoE grant DE-SC007859 at the University of Michigan. BS also acknowledges support from the University of Cincinnati. NRS is supported by Wayne State University. 

SB would like to thank the University of Michigan and Wayne State University, where part of this work was conducted, for hospitality. This work was performed in part at the Aspen Center for Physics, which is
supported by National Science Foundation grant PHY-1066293. BS also thanks the CERN Theory Group, where part of this work was conducted, for hospitality. We thank Marcela Carena and Carlos Wagner for useful discussions.   
}

\bibliography{nmssmbib}

\newpage
\appendix

\section{Trilinear Higgs couplings} \label{app:triH}

\begin{table}[h!]
\centering    
\begin{tabular}{|c|c|}
\hline
&  $\mathbb{Z}_3$-invariant NMSSM \\
\hline
$H^{\rm SM} H^{\rm SM} H^{\rm SM}$  &  $m_{h_{\rm SM}}^2/\,2v$\\
\hline
$H^{\rm SM} H^{\rm SM} H^{\rm NSM}$  &   $3s^{-1}_\beta\left(m_{h_{\rm SM}}^2 c_\beta-m_Z^2c_{2\beta}c_\beta-\frac{1}{2}\lambda^2 v^2s_{2\beta}s_\beta\right)/\,2v $ \\
\hline
$H^{\rm SM} H^{\rm NSM} H^{\rm NSM}$  & $3s^{-2}_\beta\left[m_{h_{\rm SM}}^2 c^2_\beta-m_Z^2(c^2_{2\beta}-\frac{2}{3} s^2_\beta)-\lambda^2 v^2 s^2_\beta(c_{2\beta}+\frac{2}{3})\right] /\,2v $ \\
\hline
$H^{\rm NSM} H^{\rm NSM} H^{\rm NSM}$ & $s_\beta^{-3}\left[m_{h_{\rm SM}}^2 c_\beta^3+m_Z^2 c_{2\beta} c_\beta(2s^2_\beta-c^2_\beta)-\frac{1}{2}\lambda^2 v^2 s_{2\beta} s_\beta(2c^2_\beta-s^2_\beta)\right] /\,2v$ \\
\hline
$H^{\rm SM} H^{\rm SM} H^{\rm S}$  &  $\frac{\lambda \mu}{\sqrt{2}}\left[ 1- \frac{1}{2} s_{2 \beta}\left(\frac{\kappa}{\lambda}+\frac{M_A^2}{2 \mu^2} s_{2\beta}\right)\right]$\\
\hline
$H^{\rm SM} H^{\rm NSM} H^{\rm S}$  & $-\frac{\lambda \mu c_{2\beta}}{\sqrt{2}}\left(\frac{\kappa}{\lambda}+\frac{M_A^2}{2\mu^2}s_{2\beta}\right) $ \\
\hline
$H^{\rm NSM} H^{\rm NSM} H^{\rm S}$  &
 $\frac{\lambda \mu}{\sqrt{2}}\left[ 1+\frac{1}{2} s_{2\beta}\left(\frac{\kappa}{\lambda}+\frac{M_A^2}{2\mu^2}s_{2\beta}\right)\right]$ \\
\hline
$H^{\rm SM} H^{\rm S} H^{\rm S}$ & $\frac{1}{2} v\lambda(\lambda-\kappa s_{2\beta})$ \\
\hline
$H^{\rm NSM} H^{\rm S} H^{\rm S}$ & $-\frac{1}{2} v\kappa\lambda c_{2\beta}$\\
\hline
$H^{\rm S} H^{\rm S} H^{\rm S}$ & $\frac{\kappa}{3 \sqrt{2}}(A_\kappa+6\frac{\kappa \mu}{\lambda})$\\
\hline
$H^{\rm SM} A^{\rm NSM} A^{\rm NSM}$ & $s^{-2}_{\beta}\left(m_{h_{\rm SM}}^2 c_\beta^2-m_Z^2 c^2_{2\beta}-\lambda^2 v^2 c_{2\beta}s_\beta^2 \right)/\,2v$ \\
\hline
$H^{\rm NSM} A^{\rm NSM} A^{\rm NSM}$ & $s_\beta^{-3}\left[m_{h_{\rm SM}}^2 c_\beta^3+m_Z^2 c_{2\beta} c_\beta(2s^2_\beta-c^2_\beta)-\frac{1}{2}\lambda^2 v^2 s_{2\beta} s_\beta(2c^2_\beta-s^2_\beta)\right]/\,2v$ \\
\hline
$H^{\rm S} A^{\rm NSM} A^{\rm NSM}$ &
$\frac{\lambda \mu}{\sqrt{2}}\left[ 1+ \frac{1}{2} s_{2 \beta} \left(\frac{\kappa}{\lambda}+\frac{M_A^2}{2 \mu^2} s_{2 \beta}\right)\right]$  \\
\hline
$H^{\rm SM} A^{\rm NSM} A^{\rm S}$ & 
$\frac{\lambda\mu}{ \sqrt{2}}\left(\frac{M_A^2}{2 \mu^2} s_{2 \beta}-3\frac{\kappa}{\lambda}\right)$\\
\hline
$H^{\rm NSM} A^{\rm NSM} A^{\rm S}$ & 0 \\
\hline
$H^{\rm S} A^{\rm NSM} A^{\rm S}$ & $-\kappa\lambda v$ \\
\hline
$H^{\rm SM} A^{\rm S} A^{\rm S}$  &  $\frac{1}{2} v\lambda(\lambda+\kappa s_{2\beta})$\\
\hline
$H^{\rm NSM} A^{\rm S} A^{\rm S}$  & $\frac{1}{2} v\kappa\lambda c_{2\beta}$  \\
\hline
$H^{\rm S} A^{\rm S} A^{\rm S}$  & $-\kappa(A_\kappa-2\frac{\kappa\mu}{\lambda})/\sqrt{2}$\\
\hline
$H^{\rm SM}H^+H^-$ & $2m_W^2/v + 2s^{-2}_\beta\left(m_{h_{\rm SM}}^2 c_\beta^2-m_Z^2 c^2_{2\beta}-\frac{1}{2}\lambda^2 v^2s^2_{2\beta}\right)/\,2v$ \\
\hline
$H^{\rm NSM}H^+H^-$ & $2s_\beta^{-3}\left[m_{h_{\rm SM}}^2 c_\beta^3+m_Z^2 c_{2\beta} c_\beta(2s^2_\beta-c^2_\beta)-\frac{1}{2}\lambda^2 v^2 s_{2\beta} s_\beta(2c^2_\beta-s^2_\beta)\right]/\,2v$ \\
\hline
$H^{\rm S}H^+H^-$ & $\sqrt{2}\,\lambda\mu\left[ 1+\frac{1}{2}  s_{2\beta}\left(\frac{\kappa}{\lambda}+\frac{M_A^2}{2\mu^2}s_{2\beta}\right)\right]$ \\
\hline
\end{tabular}
\caption{Trilinear scalar interactions including dominant one-loop corrections, cf. Appendices of Ref.~\cite{Carena:2015moc} for details.}
\label{tab:triH}
\end{table}

\newpage
\section{Figures for LHC constraints from direct Higgs searches}
\label{app:constraints}

\twocolumngrid
\linespread{1.0}

\begin{figure}
	\begin{center}
		\includegraphics[width=1\linewidth]{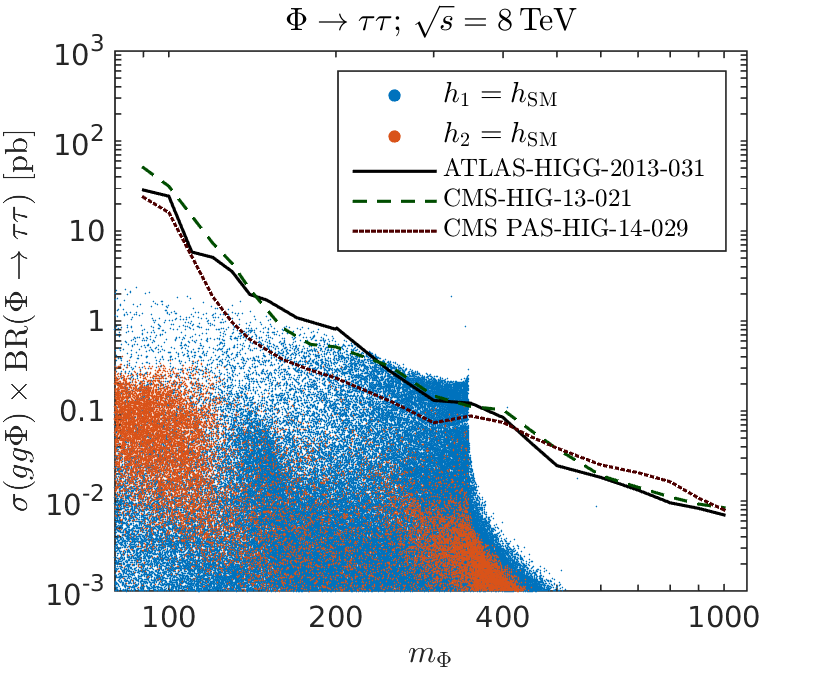}
		\caption{Upper limits from direct Higgs searches in the ${\tau^+ \tau^-}$ final state at the LHC for ${\sqrt{s} = 8}\,$TeV \cite{Khachatryan:2014wca, CMS-PAS-HIG-14-029, Aad:2014vgg} compared to NMSSM points from our \texttt{NMSSMTools}. Note, that we show points from both the ``standard'' and the ``light subset'' together. Points with ${\sigma(gg\Phi ) \times {\rm BR}(\Phi \to \tau^+ \tau^-)}$ larger than the $95\,\%\,$CL limits shown are excluded, where ${\Phi = H_3, h_i, A_2, A_1}$.}
		\label{fig:LHC_tautau_8}
	\end{center}
\end{figure}

\begin{figure}
	\begin{center}
		\includegraphics[width=1\linewidth]{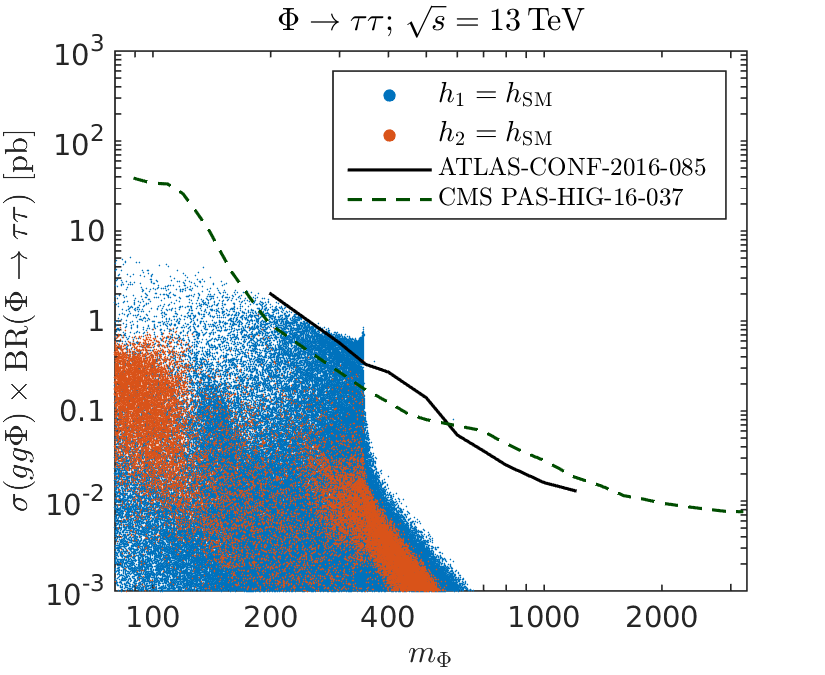}
		\caption{Same as Fig.~\ref{fig:LHC_tautau_8}, but for ${\Phi \to \tau^+ \tau^-}$ at ${\sqrt{s} = 13}\,$TeV \cite{ATLAS-CONF-2016-085, CMS-PAS-HIG-16-037}, ${\Phi = H_3, h_i, A_2, A_1}$.}
		\label{fig:LHC_tautau_13}
	\end{center}
\end{figure}

\begin{figure}
	\begin{center}
		\includegraphics[width=1\linewidth]{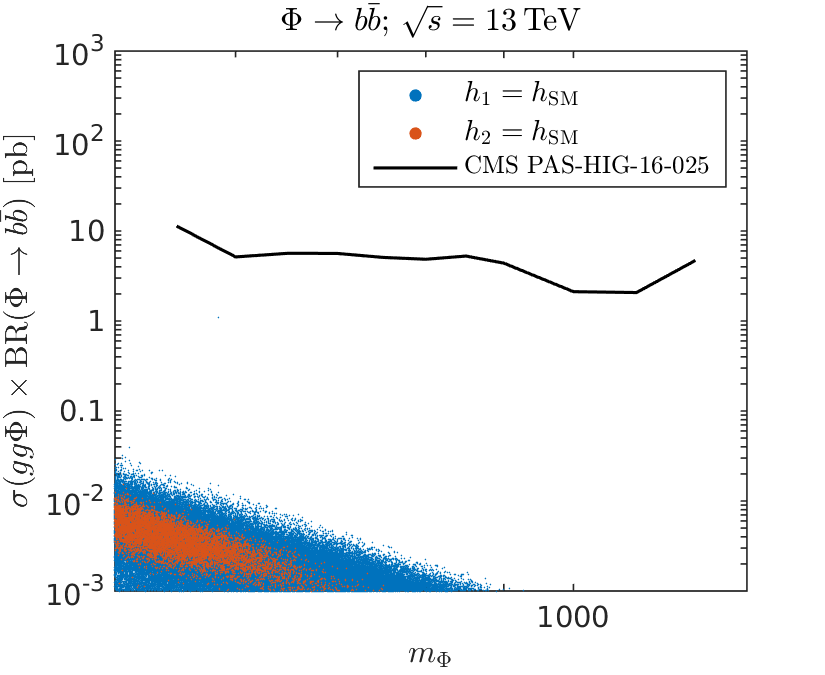}
		\caption{Same as Fig.~\ref{fig:LHC_tautau_8}, but for ${\Phi \to b\bar{b}}$ at ${\sqrt{s} = 13}\,$TeV \cite{CMS-PAS-HIG-16-025}, ${\Phi = H_3, h_i, A_2, A_1}$.}
		\label{fig:LHC_bb_13}
	\end{center}
\end{figure}

\begin{figure}
	\begin{center}
		\includegraphics[width=1\linewidth]{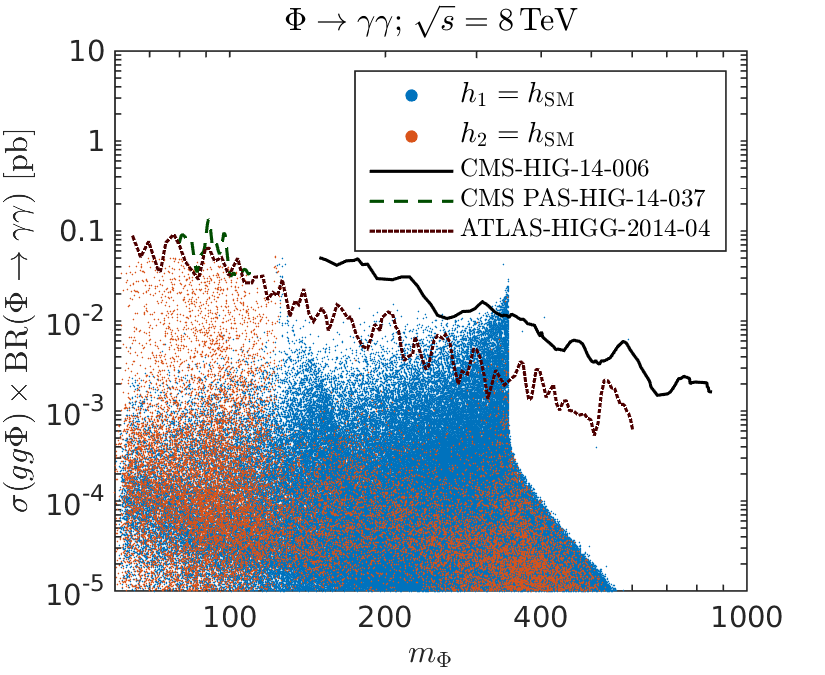}
		\caption{Same as Fig.~\ref{fig:LHC_tautau_8}, but for ${\Phi \to \gamma\gamma}$ at ${\sqrt{s} = 8}\,$TeV  \cite{Khachatryan:2015qba, CMS-PAS-HIG-14-037, Aad:2014ioa}, ${\Phi = H_3, h_i, A_2, A_1}$.}
		\label{fig:LHC_gammagamma_8}
	\end{center}
\end{figure}

\begin{figure}
	\begin{center}
		\includegraphics[width=1\linewidth]{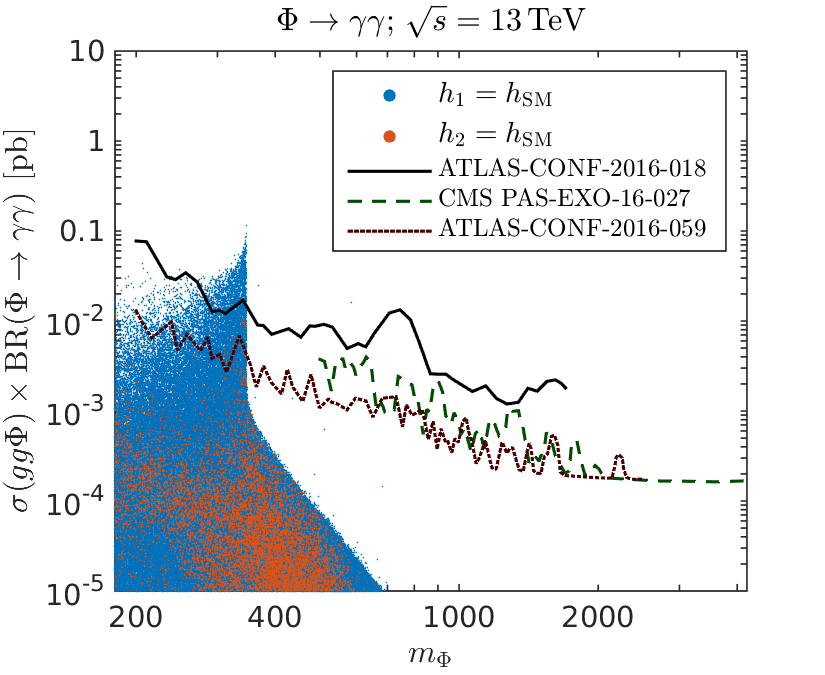}
		\caption{Same as Fig.~\ref{fig:LHC_tautau_8}, but for ${\Phi \to \gamma\gamma}$ at ${\sqrt{s} = 13}\,$TeV \cite{ATLAS-CONF-2016-018, CMS-PAS-EXO-16-027, ATLAS-CONF-2016-059}, ${\Phi = H_3, h_i, A_2, A_1}$.}
		\label{fig:LHC_gammagamma_13}
	\end{center}
\end{figure}

\begin{figure}
	\begin{center}
		\includegraphics[width=1\linewidth]{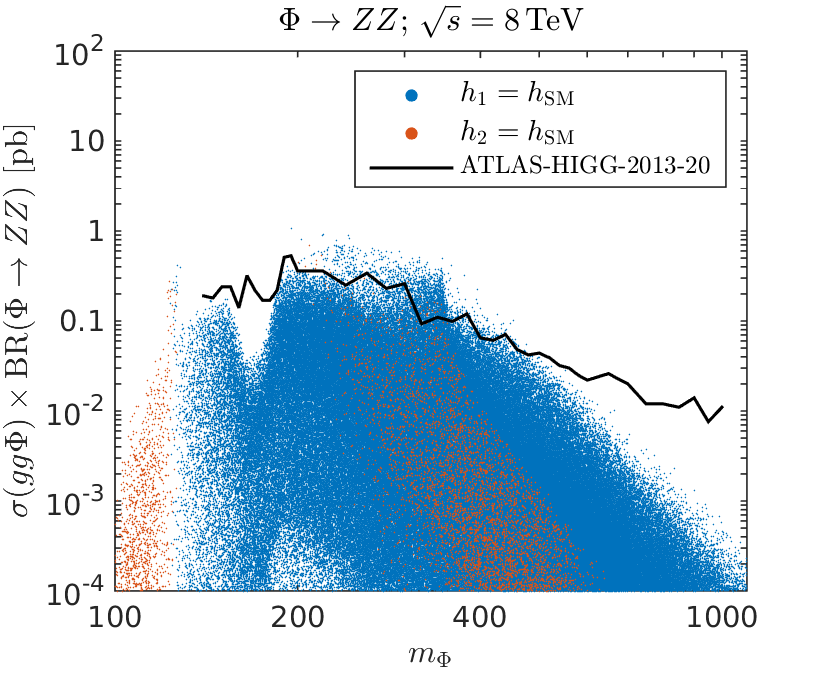}
		\caption{Same as Fig.~\ref{fig:LHC_tautau_8}, but for ${\Phi \to ZZ}$ at ${\sqrt{s} = 8}\,$TeV \cite{Aad:2015kna}, ${\Phi = H_3, h_i}$.}
		\label{fig:LHC_ZZ_8}
	\end{center}
\end{figure}

\begin{figure}
	\begin{center}
		\includegraphics[width=1\linewidth]{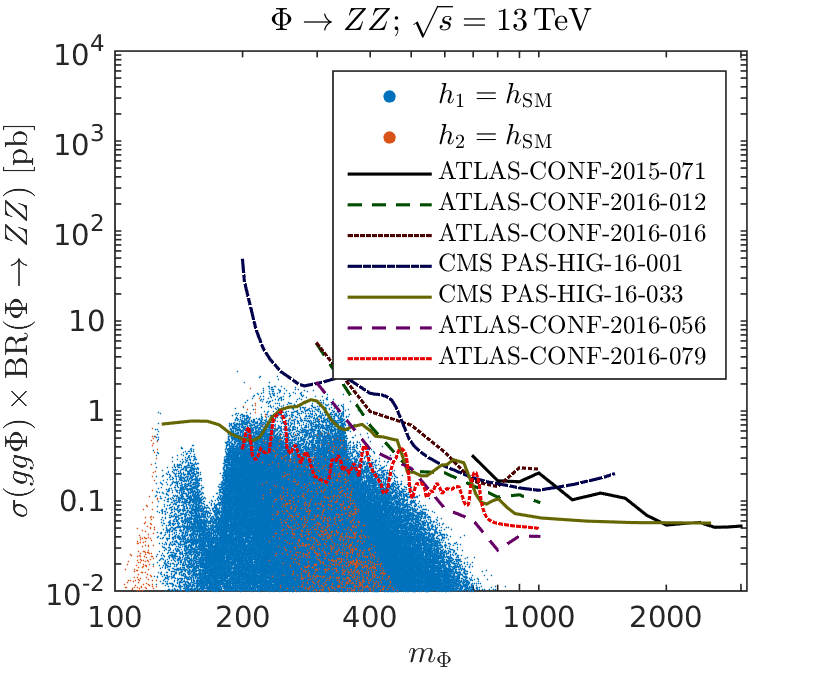}
		\caption{Same as Fig.~\ref{fig:LHC_tautau_8}, but for ${\Phi \to ZZ}$ at ${\sqrt{s} = 13}\,$TeV \cite{CMS-PAS-HIG-16-001, ATLAS-CONF-2016-012, ATLAS-CONF-2016-016, ATLAS-CONF-2015-071, CMS-PAS-HIG-16-033, ATLAS-CONF-2016-056, ATLAS-CONF-2016-079}, ${\Phi = H_3, h_i}$.}
		\label{fig:LHC_ZZ_13}
	\end{center}
\end{figure}

\begin{figure}
	\begin{center}
		\includegraphics[width=1\linewidth]{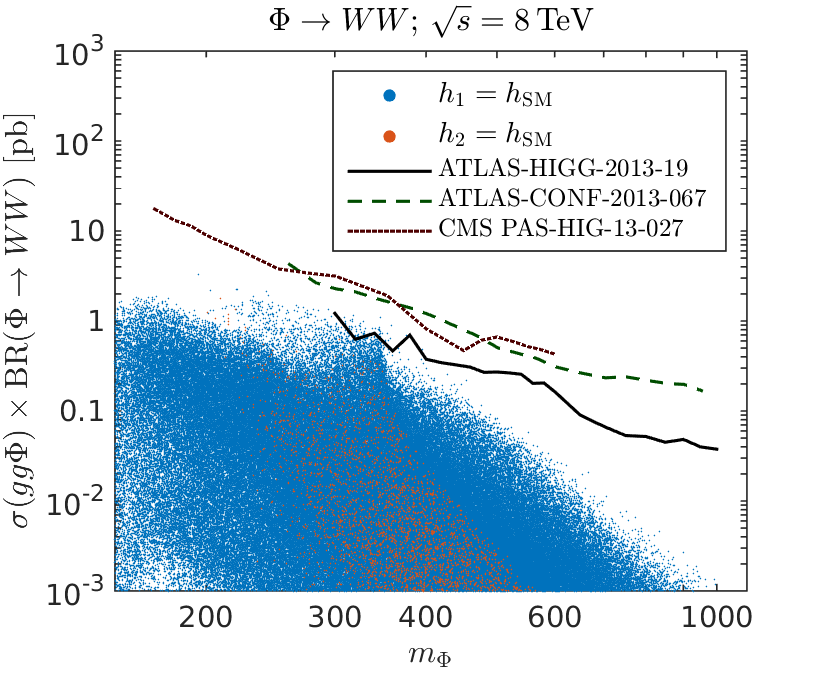}
		\caption{Same as Fig.~\ref{fig:LHC_tautau_8}, but for ${\Phi \to W^+ W^-}$ at ${\sqrt{s} = 8}\,$TeV \cite{Aad:2015agg, CMS-PAS-HIG-13-027, ATLAS-CONF-2013-067}, ${\Phi = H_3, h_i}$.}
		\label{fig:LHC_WW_8}
	\end{center}
\end{figure}

\begin{figure}
	\begin{center}
		\includegraphics[width=1\linewidth]{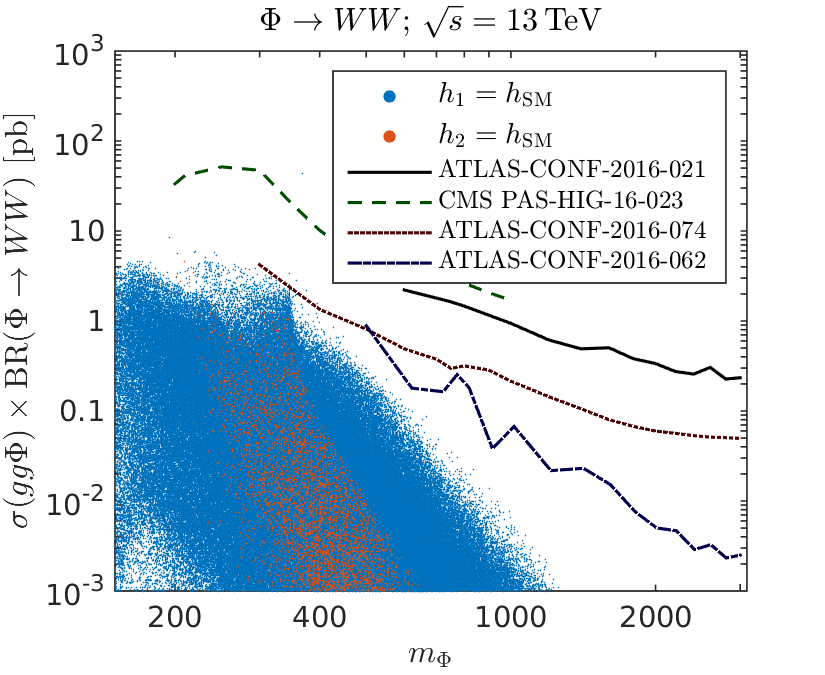}
		\caption{Same as Fig.~\ref{fig:LHC_tautau_8}, but for ${\Phi \to W^+ W^-}$ at ${\sqrt{s} = 13}\,$TeV\cite{ATLAS-CONF-2016-021, CMS-PAS-HIG-16-023, ATLAS-CONF-2016-074, ATLAS-CONF-2016-062}, ${\Phi = H_3, h_i}$.}
		\label{fig:LHC_WW_13}
	\end{center}
\end{figure}

\begin{figure}
	\begin{center}
		\includegraphics[width=1\linewidth]{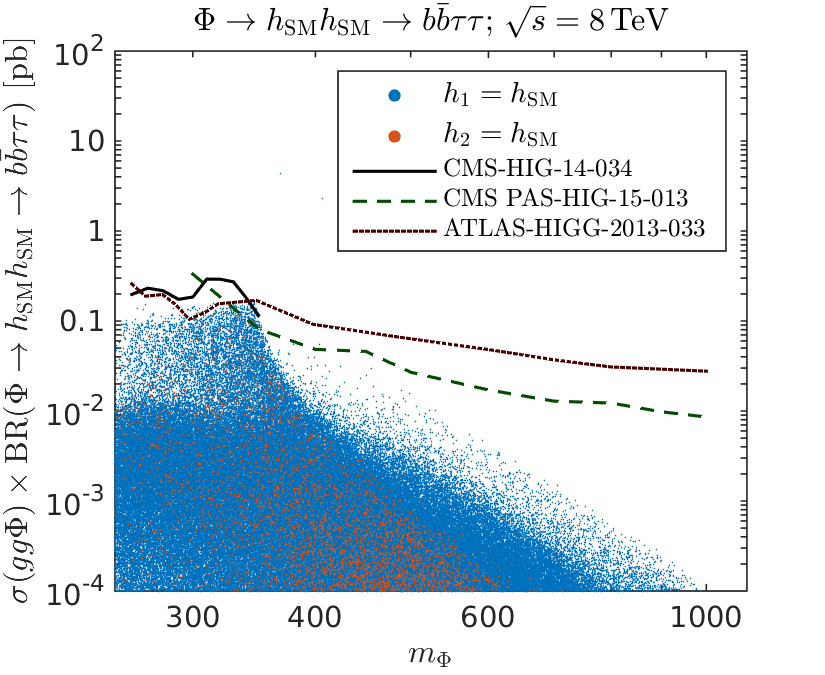}
		\caption{Same as Fig.~\ref{fig:LHC_tautau_8}, but for ${\Phi \to h_{\rm SM} h_{\rm SM} \to b\bar{b} \tau^+ \tau^-}$ at ${\sqrt{s} = 8}\,$TeV \cite{Khachatryan:2015tha, CMS-PAS-HIG-15-013, Aad:2015xja}, ${\Phi = H_3, h_i}$.}
		\label{fig:LHC_Hhh_bbtautau_8}
	\end{center}
\end{figure}

\begin{figure}
	\begin{center}
		\includegraphics[width=1\linewidth]{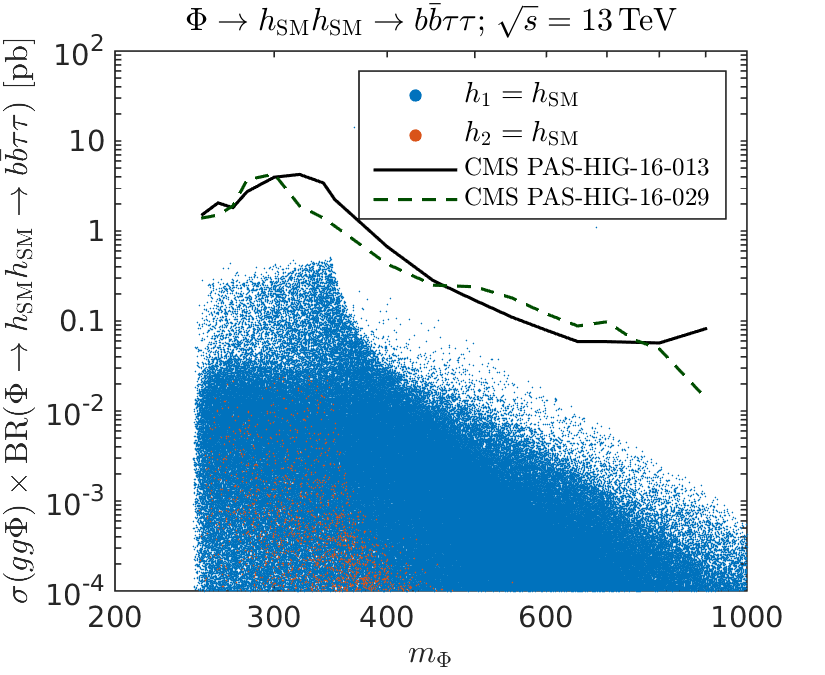}
		\caption{Same as Fig.~\ref{fig:LHC_tautau_8}, but for ${\Phi \to h_{\rm SM} h_{\rm SM} \to b\bar{b} \tau^+ \tau^-}$ at ${\sqrt{s} = 13}\,$TeV \cite{CMS-PAS-HIG-16-013, CMS-PAS-HIG-16-029}, ${\Phi = H_3, h_i}$.}
		\label{fig:LHC_Hhh_bbtautau_13}
	\end{center}
\end{figure}

\begin{figure}
	\begin{center}
		\includegraphics[width=1\linewidth]{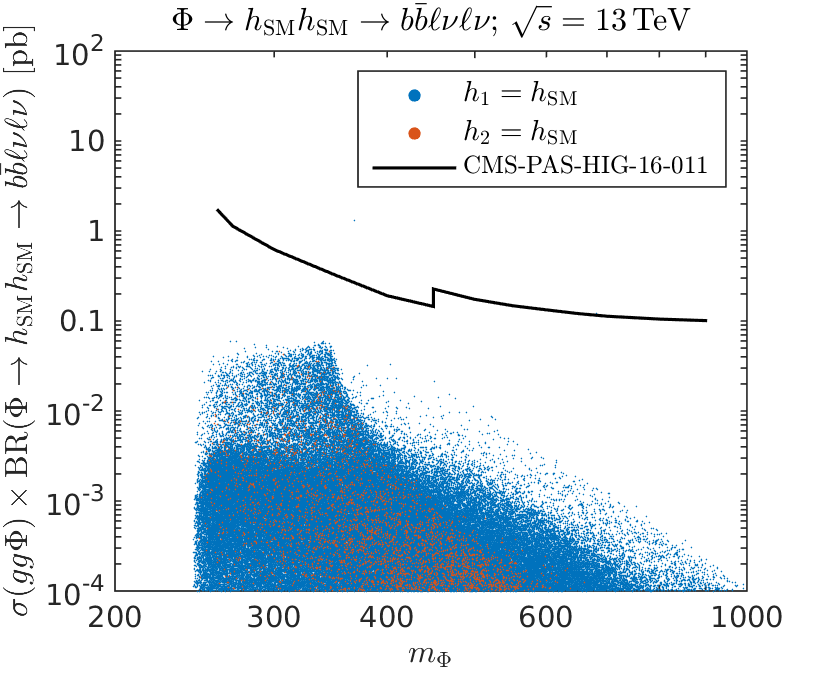}
		\caption{Same as Fig.~\ref{fig:LHC_tautau_8}, but for ${\Phi \to h_{\rm SM} h_{\rm SM} \to b\bar{b} \ell \nu_{\ell} \ell \nu_\ell}$ at ${\sqrt{s} = 13}\,$TeV \cite{CMS-PAS-HIG-16-011}, ${\Phi = H_3, h_i}$.}
		\label{fig:LHC_Hhh_bblnulnu_13}
	\end{center}
\end{figure}

\begin{figure}
	\begin{center}
		\includegraphics[width=1\linewidth]{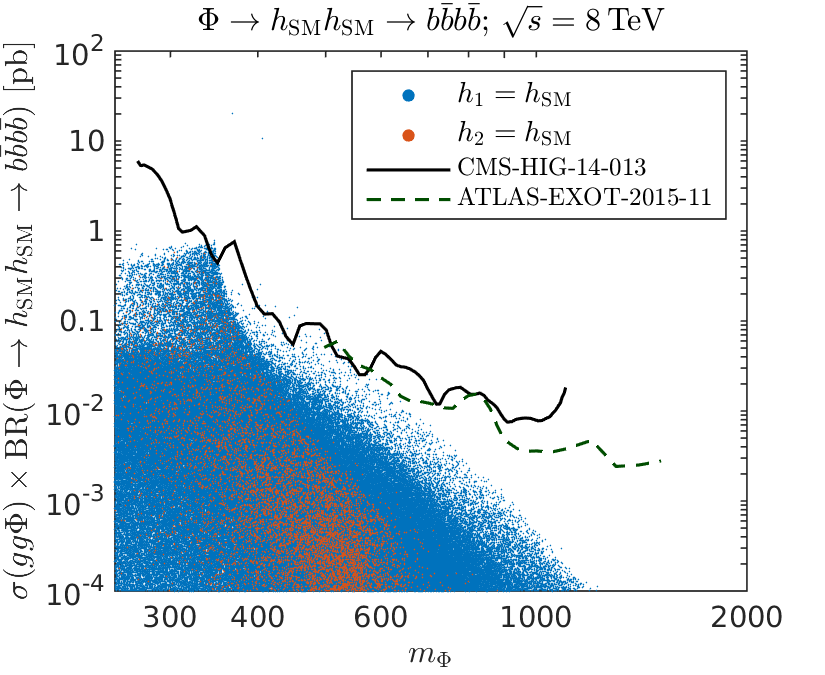}
		\caption{Same as Fig.~\ref{fig:LHC_tautau_8}, but for ${\Phi \to h_{\rm SM} h_{\rm SM} \to b\bar{b} b \bar{b}}$ at ${\sqrt{s} = 8}\,$TeV \cite{Khachatryan:2015yea, Aad:2015uka}, ${\Phi = H_3, h_i}$.}
		\label{fig:LHC_Hhh_bbbb_8}
	\end{center}
\end{figure}

\begin{figure}
	\begin{center}
		\includegraphics[width=1\linewidth]{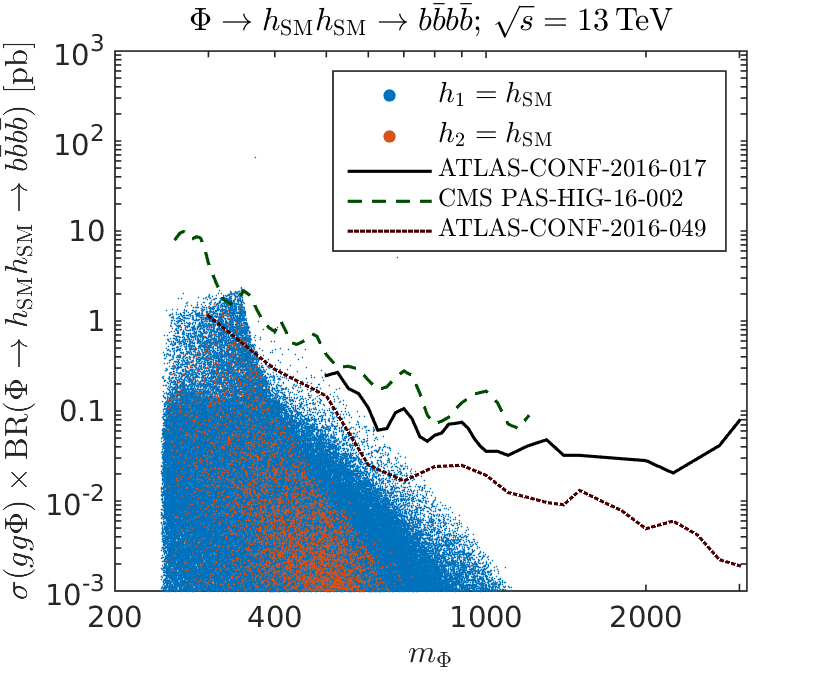}
		\caption{Same as Fig.~\ref{fig:LHC_tautau_8}, but for ${\Phi \to h_{\rm SM} h_{\rm SM} \to b\bar{b} b \bar{b}}$ at ${\sqrt{s} = 13}\,$TeV \cite{CMS-PAS-HIG-16-002, ATLAS-CONF-2016-017, ATLAS-CONF-2016-049}, ${\Phi = H_3, h_i}$.}
		\label{fig:LHC_Hhh_bbbb_13}
	\end{center}
\end{figure}

\begin{figure}
	\begin{center}
		\includegraphics[width=1\linewidth]{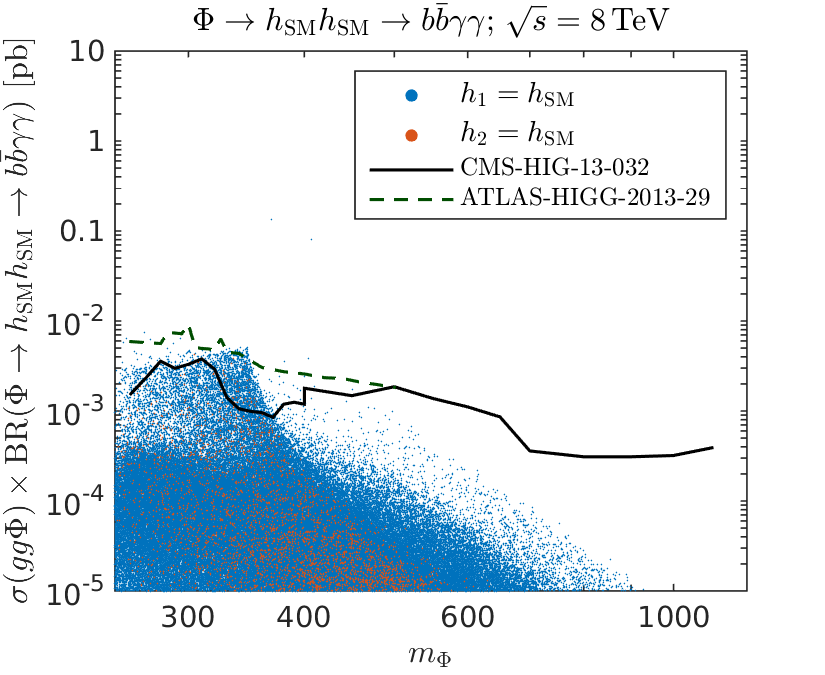}
		\caption{Same as Fig.~\ref{fig:LHC_tautau_8}, but for ${\Phi \to h_{\rm SM} h_{\rm SM} \to b\bar{b} \gamma\gamma}$ at ${\sqrt{s} = 8}\,$TeV \cite{Khachatryan:2016sey, Aad:2014yja}, ${\Phi = H_3, h_i}$.}
		\label{fig:LHC_Hhh_bbgammagamma_8}
	\end{center}
\end{figure}

\begin{figure}
	\begin{center}
		\includegraphics[width=1\linewidth]{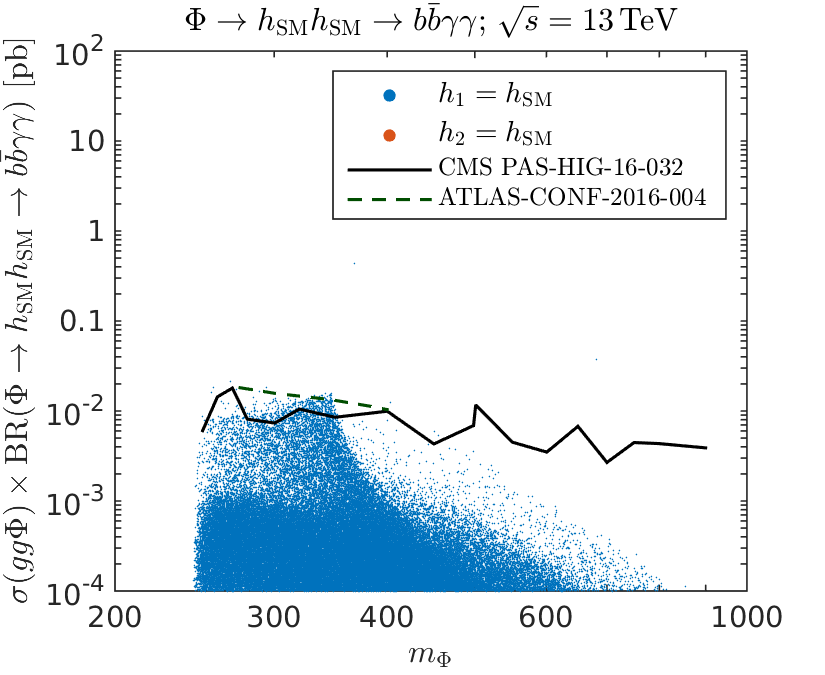}
		\caption{Same as Fig.~\ref{fig:LHC_tautau_8}, but for ${\Phi \to h_{\rm SM} h_{\rm SM} \to b\bar{b} \gamma\gamma}$ at ${\sqrt{s} = 13}\,$TeV \cite{CMS-PAS-HIG-16-032, ATLAS-CONF-2016-004}, ${\Phi = H_3, h_i}$.}
		\label{fig:LHC_Hhh_bbgammagamma_13}
	\end{center}
\end{figure}

\clearpage
\begin{figure}
	\begin{center}
		\includegraphics[width=1\linewidth]{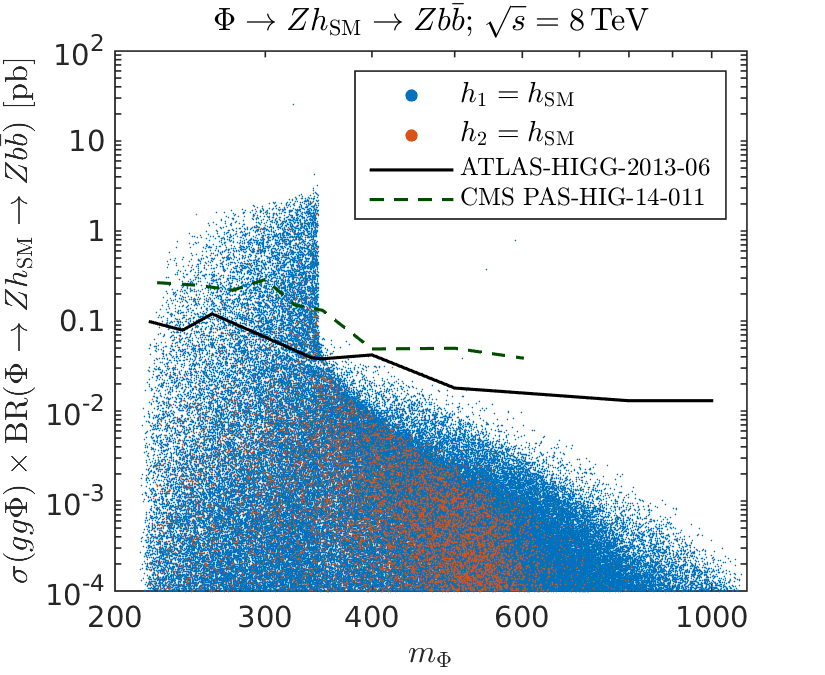}
		\caption{Same as Fig.~\ref{fig:LHC_tautau_8}, but for ${\Phi \to Z h_{\rm SM} \to Z b\bar{b}}$ at ${\sqrt{s} = 8}\,$TeV \cite{Aad:2015wra, Khachatryan:2015lba}, ${\Phi = A_2, A_1}$.}
		\label{fig:LHC_AZh_Zbb_8}
	\end{center}
\end{figure}

\begin{figure}
	\begin{center}
		\includegraphics[width=1\linewidth]{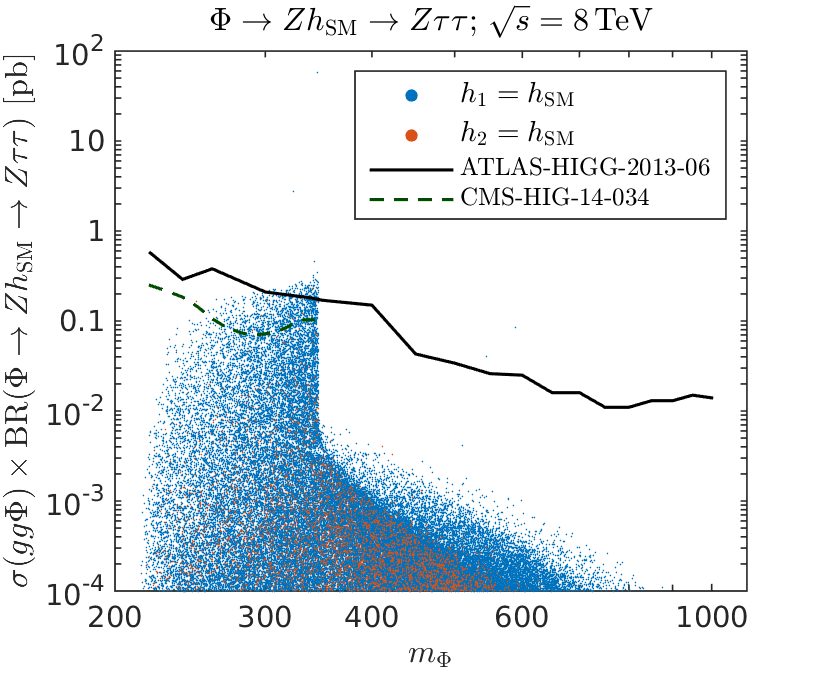}
		\caption{Same as Fig.~\ref{fig:LHC_tautau_8}, but for ${\Phi \to Z h_{\rm SM} \to Z \tau^+ \tau^-}$ at ${\sqrt{s} = 8}\,$TeV \cite{Khachatryan:2015tha, Aad:2015wra}, ${\Phi = A_2, A_1}$.}
		\label{fig:LHC_AZh_Ztautau_8}
	\end{center}
\end{figure}

\begin{figure}
	\begin{center}
		\includegraphics[width=1\linewidth]{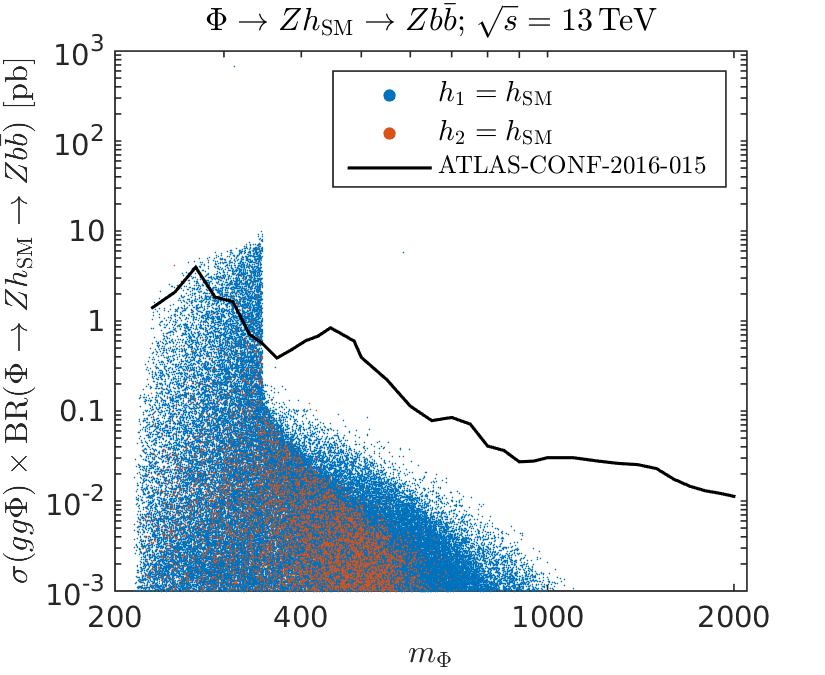}
		\caption{Same as Fig.~\ref{fig:LHC_tautau_8}, but for ${\Phi \to Z h_{\rm SM} \to Z b\bar{b}}$ at ${\sqrt{s} = 13}\,$TeV \cite{ATLAS-CONF-2016-015}, ${\Phi = A_2, A_1}$.}
		\label{fig:LHC_AZh_Zbb_13}
	\end{center}
\end{figure}

\begin{figure}
	\begin{center}
		\includegraphics[width=1\linewidth]{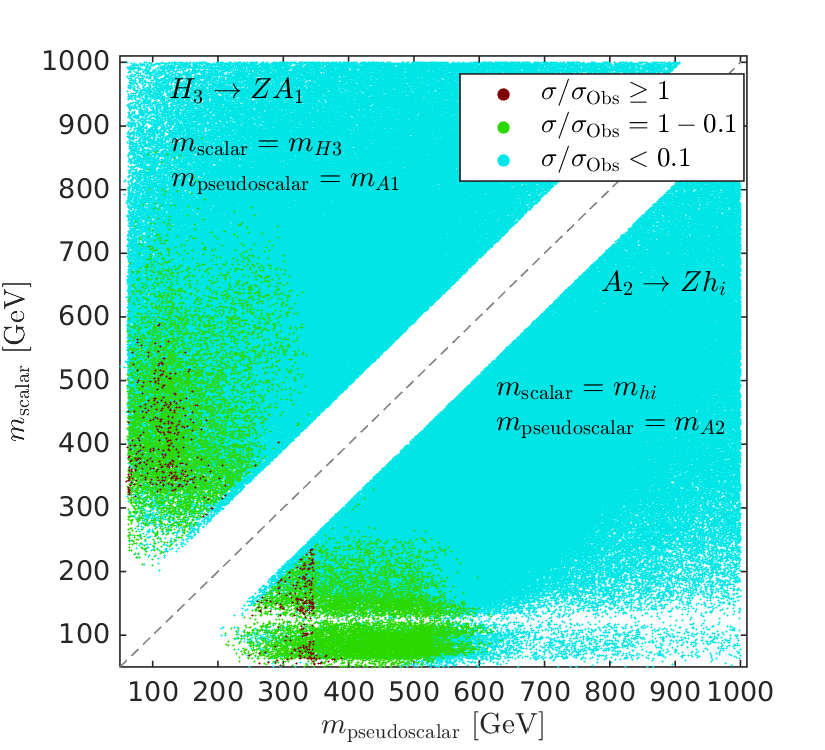}
		\caption{Constraints from direct Higgs searches at the LHC in the ${A_2/H_3 \to Z h_i/A_1 \to b\bar{b} \ell^+\ell^-}$ channel at ${\sqrt{s} = 8}\,$TeV \cite{Khachatryan:2016are} compared to NMSSM points from our \texttt{NMSSMTools}. The color coding shows the signal cross section of our scan points $\sigma$ in terms of the experimental limit $\sigma_{\rm Obs}$ as indicated in the legend.}
		\label{fig:HZA_AZH_excluded}
	\end{center}
\end{figure}

\end{document}